\documentclass[%
reprint,
superscriptaddress,
amsmath,amssymb,
aps,
prl,
longbibliography,
]{revtex4-1}

\usepackage{pdfpages} % adding supp part pdf
\usepackage{dirtytalk}
\usepackage{mathtools}
\usepackage{pdfpages}
\usepackage{mhchem}
\usepackage{graphicx}% Include figure files
\usepackage{dcolumn}% Align table columns on decimal point
\usepackage{bm}% bold math

\usepackage{hyperref}% add hypertext capabilities
\hypersetup{
colorlinks = true,   % Colours links instead of ugly boxes
urlcolor = blue,    % Colour for external hyperlinks
linkcolor = magenta,   % Colour of internal links
citecolor = blue  % Colour of citations
}
\usepackage{float}
\usepackage{color}

\makeatletter
\AtBeginDocument{\let\LS@rot\@undefined}
\makeatother

\begin{document}
\preprint{APS/123-QED}
\title{Vortex-line topology in iron-based superconductors\\with and without second-order topology}
\author{Majid Kheirkhah}
\affiliation{Department of Physics, University of Alberta, Edmonton, Alberta T6G 2E1, Canada}
\author{Zhongbo Yan}
\email{yanzhb5@mail.sysu.edu.cn}
\affiliation{School of Physics, Sun Yat-Sen University, Guangzhou 510275, China}
\author{Frank Marsiglio}
\affiliation{Department of Physics, University of Alberta, Edmonton, Alberta T6G 2E1, Canada}
\date{\today}
%======================
\begin{abstract}  
The band topology of a superconductor is known to have profound impact on the existence of Majorana zero modes
in vortices. As iron-based superconductors with band inversion and $s_{\pm}$-wave pairing
can give rise to time-reversal invariant second-order topological superconductivity, manifested by the presence of helical Majorana hinge states in three dimensions, we are motivated to investigate the interplay between the second-order topology
and the vortex lines in both weak- and strong-Zeeman-field regimes. In the weak-Zeeman-field regime, we find that vortex lines far away from the hinges are topologically nontrivial in the weakly doped regime, regardless of whether the second-order topology is present or not. However, when the superconductor falls into the second-order topological phase and a topological vortex line is moved close to the helical Majorana hinge states, we find that their hybridization will trivialize the vortex line and transfer robust Majorana 
zero modes to the hinges. Furthermore, when the Zeeman field is large enough, we find that the helical Majorana hinge 
states are changed into chiral Majorana hinge modes and thus a chiral second-order topological superconducting phase is realized. In this regime, the vortex lines are always topologically trivial, no matter how far away they are from the chiral Majorana hinge modes. By incorporating a realistic assumption of inhomogeneous
superconductivity, our findings can explain the recent experimental observation of the peculiar coexistence
 and evolution of topologically nontrivial and trivial vortex lines in iron-based superconductors.
\end{abstract}

\pacs{Valid PACS appear here}
%Use showkeys class option if keyword
\maketitle
%\tableofcontents
%======================
{\it Introduction.---}
Topological superconductors (TSCs) and iron-based superconductors (FeSCs) have been two mainstreams of the superconducting field
for more than one decade. The great interest in TSCs lies in the various kinds of Majorana modes, like  one-dimensional propagating helical and chiral Majorana modes and zero-dimensional localized Majorana zero modes (MZMs),  which hold promising applications
in topological quantum computation \cite{read2000,ivanov2001,kitaev2001unpaired,nayak2008review,fu2008,lutchyn2010,oreg2010helical,Alicea2011non, zhang2013kramers, Zhang2013mirror}. For FeSCs, the great interest lies in their high superconducting transition temperature ($T_{c}$),
strongly correlated nature, and pairing
mechanisms, leading to the emergence of unconventional pairings \cite{Paglione2010,Hirschfeld2011,Stewart2011review,Chubukov2012review,Dagotto2013,Dai2015review}, such as the widely known $s_{\pm}$-wave pairing \cite{Mazin2008,Wang2011iron}.
As FeSCs commonly have multiple bands near the Fermi energy, the possibility of the occurrence of
band inversion with a realization of topological superconductivity through the Fu-Kane mechanism \cite{fu2008} has attracted
considerable theoretical interest in the past few years \cite{Wang2015iFeSC,Wu2016iron,Xu2016FeSC,Hao2019topological,Kreise2020review}.

Recently, several groups
have experimentally observed that above $T_{c}$, band inversion occurs between the bands
near the Fermi energy in a series of FeSCs, and below $T_{c}$,
the surface Dirac cones associated with the inverted band structure
are gapped by the bulk superconductivity \cite{zhang2018iron,zhang2019multiple,wang2018evidence,kong2019half,machida2019zero,Liu2018MZM,chen2019quantized,Zhu2019MZM}.
The Fu-Kane mechanism is thus fulfilled in a single-material platform, unlike
the various actively studied heterostructures composed of different pieces of materials \cite{Mourik2012MZM,Nadj2014MZM,Sun2016Majorana,Zhang2018quantized,Fornieri2019}. Therefore,
the observation of compelling experimental evidence for vortex MZMs in FeSCs, like a
zero-bias peak with nearly quantized height and other ordered
discrete peaks in the scanning tunneling spectroscopy \cite{wang2018evidence,kong2019half,machida2019zero,Liu2018MZM,chen2019quantized,Zhu2019MZM}, has generated
tremendous interest \cite{Jiang2019vortex,Peng2019iron,Chiu2020vortex,Wang2020helical,Chen2020MZM,PhysRevX.11.011041,Wu2020ironline,Ghazaryan2020vortex,Qin2019vortex,Qin2019vortex2,Konig2019votex,Yan2020vortex}. Notably, in these FeSCs, trivial vortex lines without MZMs are observed to coexist with
topological vortex lines with MZMs \cite{wang2018evidence,kong2019half,machida2019zero,Chen2018iron}. While mechanisms like pairing change induced by a Zeeman field or spatially dependent surface states have been proposed to
explain this peculiar experimental finding \cite{Ghazaryan2020vortex,kong2019half},  a conclusive understanding is still far from established.

\begin{figure}[t!]
\centering
\includegraphics[scale=0.4]{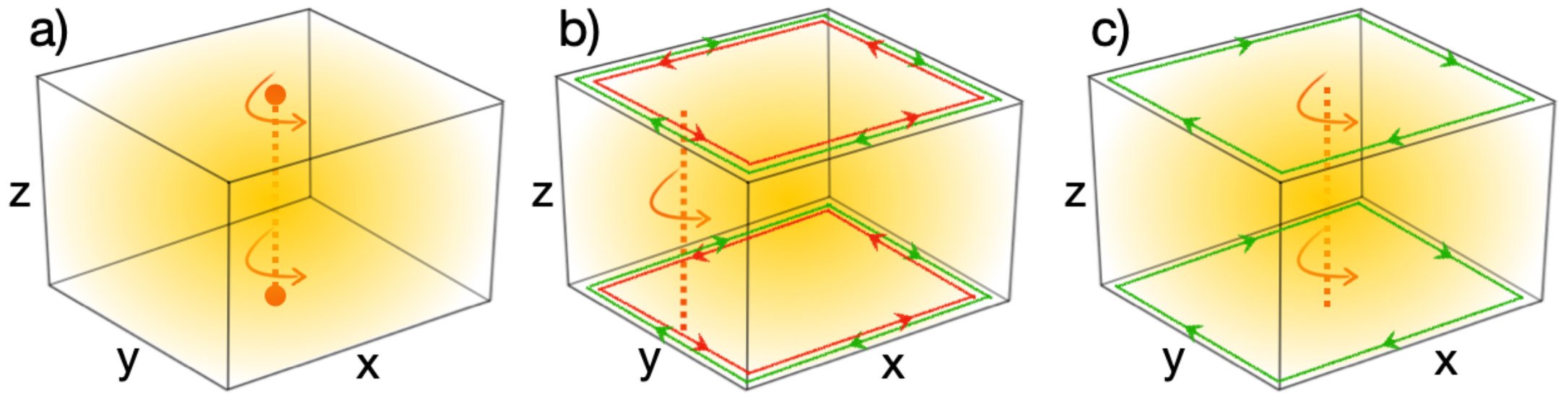}
\caption{(Color online) Schematic diagrams. (a) 
For a superconductor with a single band inversion, when the chemical potential crosses the surface bands in the normal state and there are no gapless modes on the boundary in the superconducting state, the vortex line will bind one robust 
MZM at each end, regardless of its position. 
(b) When the superconductor is a second-order TSC with helical Majorana hinge modes, a close-to-hinge vortex line will not 
bind robust MZMs at its ends  because of hybridization with the helical Majorana hinge modes. (c) When the superconductor 
is a second-order TSC with chiral Majorana hinge modes, the 
vortex line terminated at the surfaces with Zeeman-field-dominated Dirac mass will not bind MZMs, even when it is far away from the hinges. }
\label{sketch}
\end{figure}

Another remarkable advance in TSCs is the recent birth of the concept named higher-order
TSCs \cite{Benalcazar2017,Schindler2018,Langbehn2017,Shapourian2018SOTSC,Khalaf2018,Geier2018,
Zhu2018hosc,Yan2018hosc,Wang2018hosc,Wangyuxuan2018hosc,Hsu2018hosc,Liu2018hosc,Wuzhigang2019hosc,Zhang2019hinge,
Volpez2019SOTSC,Zhang2019hoscb,Wu2019hoscb,Hsu2019HOSC,Zeng2019mcm,Bultinck2019,Ghorashi2019,Peng2019hinge,Zhu2019mixed,
Laubscher2019hosc,Pan2019SOTSC,Yan2019hosca,Yan2019hoscb,Franca2019SOTSC,Majid2020hosca,
Zhang2020SOTSC,Ahn2020hosc,Suman2020,Bitan2020hosc,Wu2020SOTSC,Majid2020hoscb,ghorashi2019vortex,Chen2019hosc,Niu2020hosc}.
It is interesting that several theoretical works have revealed that a combination of inverted band structure and
$s_{\pm}$-wave pairing can also give rise to second-order topological superconductivity \cite{Yan2018hosc,Wang2018hosc,Zhang2019hinge},
which is manifested by the presence of Majorana corner modes in two dimensions (2D) and  Majorana hinge modes (MHMs)
in three dimensions (3D). Remarkably, experimental evidence for the existence of
MHMs in a 3D FeSC has also been reported \cite{Gray2019helical}. These previous works
motivate us to study the open question about the potential impact of second-order topology
on the vortex lines in 3D FeSCs.

By focusing on the weakly doped regime of most interest and relevance
to the experiments, our main findings can be summarized as follows: (i)
In the weak-Zeeman-field regime, we find that if the second-order topology is absent, then the vortex line will bind one MZM at each end [see
Fig.~\ref{sketch}(a)], 
regardless of its position, resembling the situation with conventional $s$-wave pairing \cite{Hosur2011MZM}. (ii) When the second-order topology is present, a vortex line far away from the hinges will also bind one MZM at each end. However, 
if the topological vortex line is moved close to the hinges, it 
will be trivialized by hybridizing with the helical MHMs [see
Fig.~\ref{sketch}(b)]. Because MZMs are always created  or annihilated in pairs, this $\mathbb{Z}_{2}$ nature  leads the transfer of robust MZMs to the hinges. 
(iii) While the helical MHMs are counterintuitively quite stable against the Zeeman field, 
they will become chiral MHMs when the Zeeman field is larger than a critical value. In this regime, MZMs are absent at the vortex ends, 
no matter how far away they are from the chiral MHMs [see
Fig.~\ref{sketch}(c)]. Nevertheless, MZMs can appear on the hinges 
if the number of vortex lines is odd. 
%======================

{\it Theoretical formalism.---} 
The inverted band structure of FeSCs
near the Fermi level can be described by a four-band minimal model of 3D topological insulators \cite{hasan2010colloquium,qi2011topological}. In combination with the $s_{\pm}$-wave pairing, the underlying physics can be described by
the Bogoliubov-de Gennes Hamiltonian $H = \frac{1}{2} \sum_{\bm{k}} \psi^{\dagger}_{\bm{k}} \mathcal{H}(\bm{k}) \psi_{\bm{k}}$,
where
\begin{align}
\mathcal{H}(\bm{k}) = &~
m(\bm{k}) \sigma_{z} s_{0} \tau_z
+ 2\lambda \hspace{-2mm} \sum_{i=x,y,z} \hspace{-1.5mm} \sin k_{i} \sigma_{x} s_{i} \tau_z
- \mu \sigma_{0}s_{0}\tau_z
\nonumber
\\&
+ \sigma_{0} \bm{h}\cdot \bm{s} \tau_0
- \Delta(\bm{k})\sigma_{0}s_{0} \tau_x,
\label{BdG}
\end{align}
with the basis $\psi^{\text{T}}_{\bm{k}} = (\bm{c}_{k},-i\sigma_0 s_y \bm{c}^{\dagger}_{-k})$ where $\bm{c}_{k} = (c_{\bm{k},a,\uparrow},
c_{\bm{k},b,\uparrow},
c_{\bm{k},a,\downarrow},
c_{\bm{k},b,\downarrow})^{\text{T}}$. For notational simplicity, the lattice constants are set to unity
throughout this work. In Eq.~(\ref{BdG}), the Pauli matrices
$s_i$, $\sigma_i$, and $\tau_i$ act on the spin ($\uparrow,\downarrow$), orbital ($a,b$), and particle-hole degrees of freedom, respectively, and $s_0$, $\sigma_0$, and $\tau_0$ are identity matrices. $\lambda$ denotes the strength of spin-orbit coupling, $\mu$ is the chemical potential, and $\bm{h}=(h_{x},h_{y},h_{z})$ denotes the Zeeman field, which is assumed to be present accompanying the generation of vortex lines.
The $m(\bm{k})$ term and the $\Delta(\bm{k})$ term respectively characterize the band inversion and the $s_{\pm}$-wave pairing,
with their explicit forms given by
\begin{align}
m(\bm{k}) =&~ m_0 - 2t \hspace{-2mm} \sum_{i=x,y,z} \hspace{-1.5mm} \cos k_i,
\\
\Delta(\bm{k}) =&~ \Delta_0 + \Delta_s( \cos k_x +  \cos k_y),
\label{pairing_1}
\end{align}
where $t$ denotes the hopping amplitude, and $\Delta_0$ and $\Delta_s$ represent on-site and extended $s$-wave superconducting pairing magnitudes, respectively. When the equal-energy contour $m(\bm{k})=0$, known as the band inversion surface (BIS), encloses
a single time-reversal invariant (TRI) momentum, the normal state hosts a topological gap and a single Dirac cone on each
surface \cite{fu2007a}. After becoming superconducting below $T_{c}$, the surface Dirac cones are gapped by the  superconductivity. For a conventional $s$-wave pairing, {\it i.e.}, $\Delta(\bm{k})=\Delta_{0}$ in Eq.~(\ref{pairing_1}),
it is known that its induced Dirac mass on the surfaces is uniform if the effect of disorder is negligible.
In other words, there does not exist any kind of
gapless boundary modes. However, for the $s_{\pm}$-wave pairing
given in Eq.~(\ref{pairing_1}), it has been shown in Ref.~\cite{Zhang2019hinge} that even though the first-order topology
is trivial, a nontrivial TRI second-order topology, which is manifested by the presence of helical MHMs, can be achieved in this system.

\begin{figure}[t!]
\centering
\includegraphics[scale=0.31]{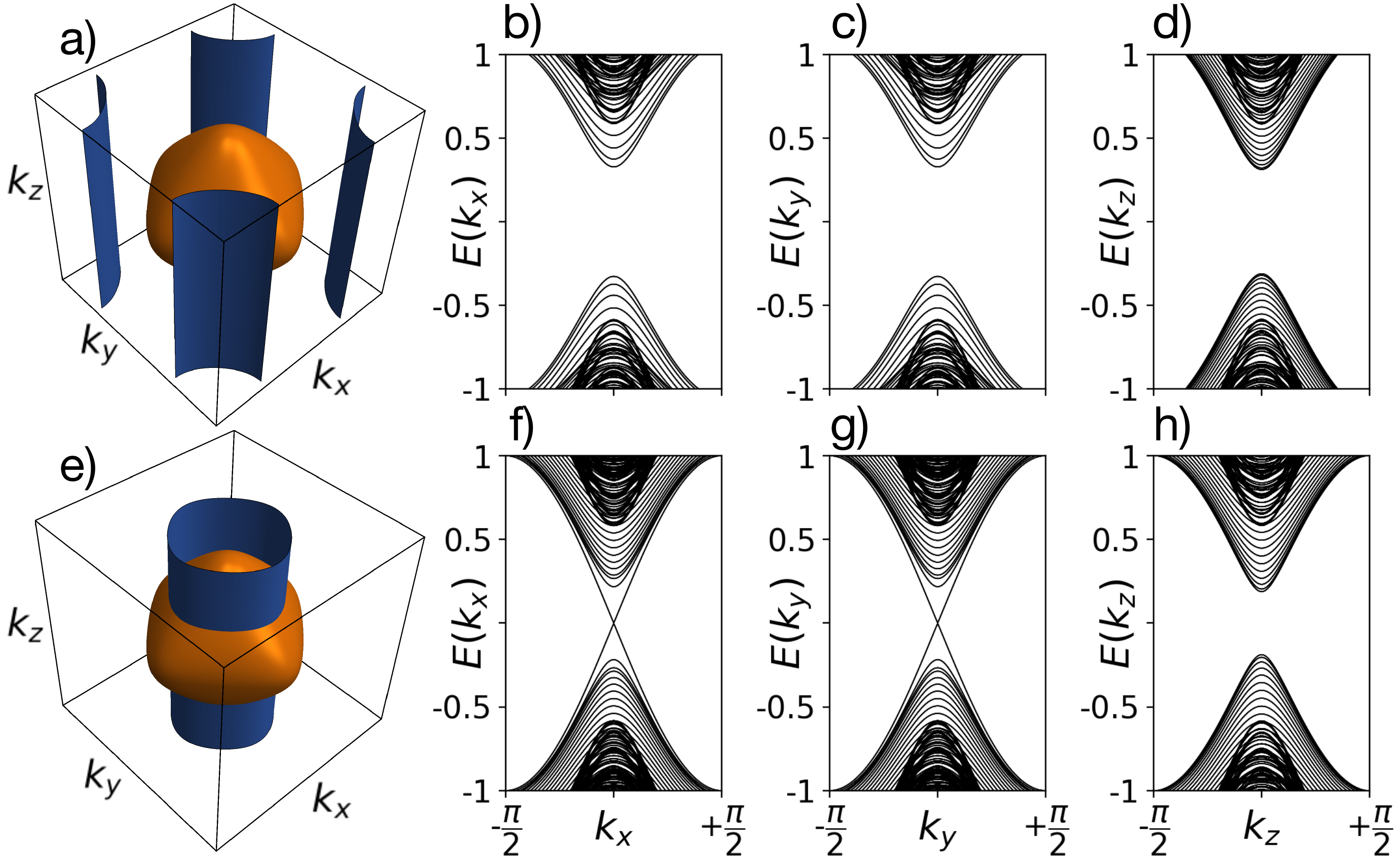}
\caption{(Color online) (a)-(d) No gapless modes appear on the hinges when the band inversion surface (BIS) in orange and the pairing node surface (PNS) in blue do not cross. (e)-(h) Helical Majorana
modes appear on the hinges between $z$-normal surfaces and $x$- and $y$-normal surfaces when the BIS and PNS cross.
Common parameters are: $t=1$, $m_0 =2.5$, $\lambda = 0.5$, $\mu = 0$, $\bm{h}=0$, and $\Delta_{0} = 0.25$. (a)-(d) $\Delta_{s} = 0.25$,
(e)-(h) $\Delta_{s} =-0.25$, without vortex lines. The two directions with open boundary conditions are $y$ and $z$ in (b)(f), $x$ and $z$ in (c)(g), and
$x$ and $y$ in (d)(h), and their lengths contain $24$ lattice sites.  }\label{crossing}
\end{figure}

At $\mu=0$ and $\bm{h}=0$, we can establish a simple geometric criterion for the realization of nontrivial TRI second-order topology.
The geometric picture of the criterion is the crossing of the BIS and the pairing node surface (PNS),
which is the equal-energy contour $\Delta(\bm{k})=0$ (see Supplemental Material \cite{supplemental}).
To show the validity of the geometric criterion explicitly,
we consider a cubic geometry with open boundary condition
in two directions and periodic boundary
condition in the remaining direction. As shown in the first row
of Fig.~\ref{crossing}, when the BIS and PNS do not cross [see the orange and the blue surfaces in Fig.~\ref{crossing}(a)], there is no gapless hinge mode [Figs.~\ref{crossing}(b)-\ref{crossing}(d)],
confirming the trivial topology. On the contrary, when the BIS and PNS
cross, helical Majorana modes are found to appear on the interfacing hinges between $z$-normal surfaces and
$x$- and $y$-normal surfaces [Figs.~\ref{crossing}(f) and \ref{crossing}(g)], with their distributions the same as
in Fig.~\ref{sketch}(b) (here without vortex lines).

\begin{figure*}[t!]
\centering
\includegraphics[scale=0.495]{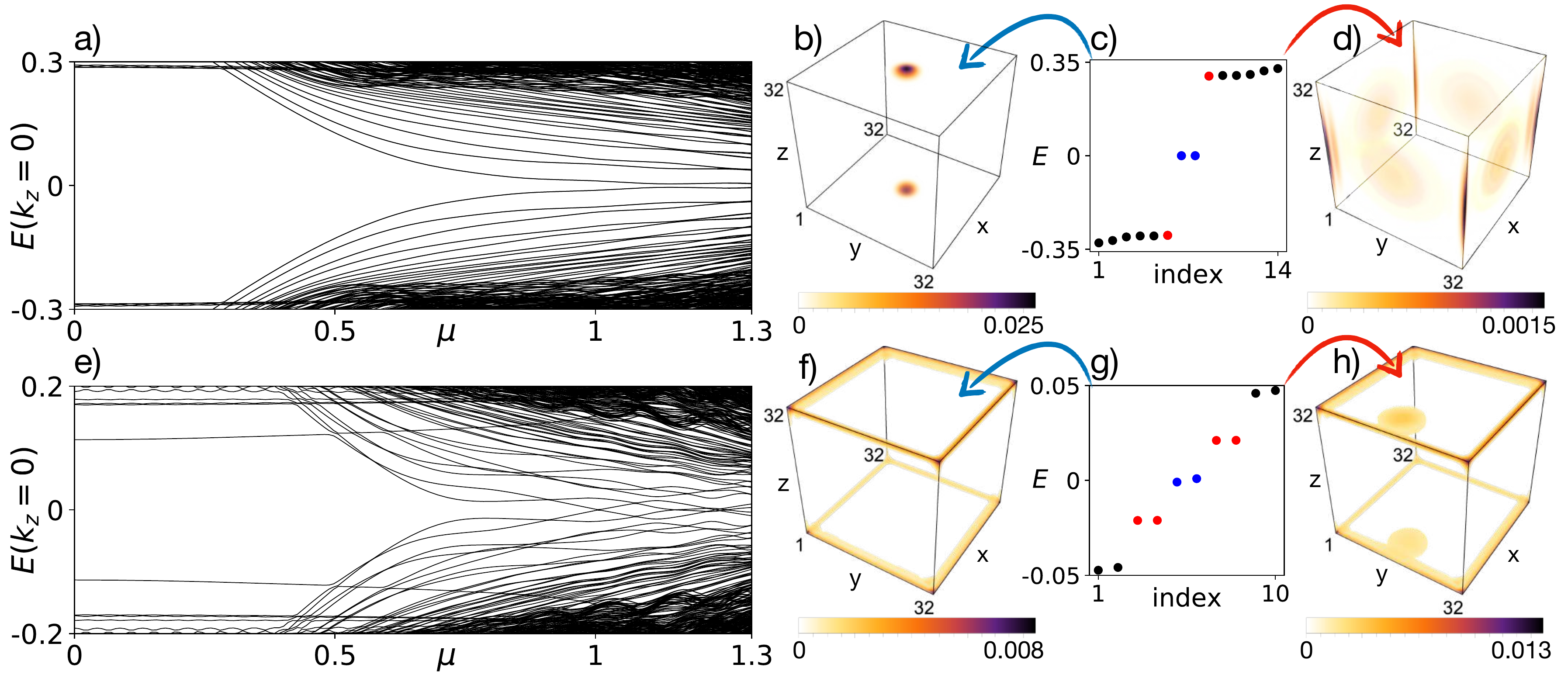}
\caption{(Color online) The evolution of energy dispersions at $k_{z}=0$ with respect to the doping level
and the locations of MZMs. (a)-(d) and (e)-(h) show the results without and with second-order topology, respectively.
For the parameters considered, the normal state becomes
metallic when $\mu>0.5$, and we restrict
$\mu<1.3$ to guarantee the vortex-free bulk spectrum to be gapped. Panel (a) shows that the lowest excitation spectrum  remains gapped and has an accidental double degeneracy (read from data)  when the vortex line is located at the center 
of the system, so $\nu$ keeps the nontrivial value $-1$ for $0<\mu<1.3$.  
In (c), the two blue dots correspond to two robust MZMs, and the two red dots correspond to two high-energy modes. Panel (b) shows
that the probability density of the MZMs is strongly localized at the ends of the vortex line, and (d) shows that the two high-energy modes
correspond to gapped surface states. In (e), the vortex line is centered at $(x_{c},y_{c})=(6.01,16.01)$  so that it is very 
close to the hinges of $x$-normal surface. The lowest excitation spectrum no longer has the accidental double degeneracy  present in (a). One can 
find from (e) that the gap closes at $\mu\approx1$, which signals a vortex  phase transition with $\nu$ changing from 
$-1$ to $1$. In (g), the two blue dots also correspond to two robust MZMs (not at exact zero energy due to finite-size 
effects). Panel (f) shows that the probability density of the robust MZMs is localized on the hinges. (h)
The probability density of the four modes in red suggests that they originate from the hybridization 
of the helical MHMs and vortex-end MZMs. Common parameters are $t=1$, $m_0 = 2.5$, $\lambda =0.5$, $\bm{h}=0$, $\xi = 4$, and $\Delta_0 = 0.25$. In (a)-(d), $\Delta_s = 0.25$, and in (e)-(h), $\Delta_s = -0.25$. In (a)(e), $L_{x}=L_{y}=32$ lattice sites, and in (b)-(d) and (f)-(h), $\mu=0$, $L_{x}=L_{y}=L_{z}=32$ lattice sites.}
\label{MZM}
\end{figure*}

When $\mu$ or $\bm{h}$ becomes nonzero, the above simple geometric criterion generally no longer holds. Nevertheless,
the second-order topology can  survive even when the normal state becomes metallic \cite{supplemental}. 
Generally, it is known that gapless helical modes will become gapped when the time-reversal symmetry 
is lifted by the Zeeman field. However, here we  find that 
the helical MHMs shown in Figs.~\ref{crossing}(f) and \ref{crossing}(g)
are counter-intuitively quite robust against the Zeeman field.
Only when the Zeeman field is strong enough to dominate over 
the superconductivity-induced Dirac mass on some of the surfaces, {\it e.g.}, $|h_{z}|>\sqrt{\mu^{2}+(\Delta_{0}+2\Delta_{s})^{2}}$ \cite{supplemental}, 
will the nature of the domain walls
change and the helical MHMs will become chiral MHMs,
as illustrated in Fig.~\ref{sketch}(c).
We find that the robustness of the helical MHMs is simply because the Zeeman field cannot 
directly act on the subspace of the domain walls \cite{supplemental}.
These results suggest that the simple Hamiltonian in Eq.~(\ref{BdG}) can realize both TRI and chiral second-order TSCs.

%======================
{\it Interplay of second-order topology and vortex lines.---}
We follow the experiments and focus on
the configurations presented in Fig.~\ref{sketch}, where the
vortex lines are generated along the $z$ direction.
To illustrate the key physics in a clear way, we simplify the real
situation with multiple  vortex lines to an ideal situation with
just a single $\pi$-flux vortex line. Such a simplification is
justified in the weak-field regime, where the vortex lines are dilute
and well-separated from each other. Accordingly,
we assume that both on-site and nearest-neighbor pairings follow such a spatial
dependence,
\begin{eqnarray}
\Delta_{0,s}(\bm{r})=\Delta_{0,s}\tanh(\frac{\sqrt{ \bar{x}^{2}+\bar{y}^{2}}}{\xi})\frac{\bar{x}+i\bar{y}}{\sqrt{\bar{x}^{2}+\bar{y}^{2}}},
\label{pairing}
\end{eqnarray}
where $\xi$ denotes the superconducting coherence length, $\bar{x}=x-x_{c}$ and $\bar{y}=y-y_{c}$, with $(x_{c},y_{c})$ denoting the core of the vortex in
the $xy$ plane, and $(x,y)$ representing the coordinates
of the lattice sites for on-site pairing $\Delta_{0}(\bm{r})$ and the coordinates
of the lattice-bond centers for nearest-neighbor pairing $\Delta_{s}(\bm{r})$.

Since the vortex line breaks the time-reversal symmetry but preserves the translational symmetry
in the $z$ direction, the system can be viewed as a quasi-1D superconductor belonging to the class D
of the Atland-Zirnbauer classification \cite{schnyder2008,kitaev2009}. Accordingly, whether MZMs exist or not is characterized by
the $\mathbb{Z}_{2}$ invariant \cite{kitaev2001unpaired,wimmer2012pfaffian,supplemental}
\begin{eqnarray}
\nu=\text{sgn}\{\text{Pf}[H_{\rm M}(0)]\}~\text{sgn}\{\text{Pf}[H_{\rm M}(\pi)]\},\label{pfaffian}
\end{eqnarray}
where $H_{\rm M}(k_{z})$ represents the Hamiltonian in the Majorana representation and ``Pf'' is a shorthand notation
of Pfaffian \cite{supplemental}. $\nu=-1$ and $+1$ indicate the presence and absence of one robust MZM at each end of the quasi-1D superconductor,
respectively. The expression of the $\mathbb{Z}_{2}$ invariant suggests that only the band information at the two TRI momenta are
important.

We first study the configuration in Fig.~\ref{crossing}(a) for which the band
topology of the homogeneous case is trivial. Since the band inversion takes place at the $\Gamma$ point, we only need to focus on the dispersion at $k_{z}=0$ in the weakly doped regime. From Fig.~\ref{MZM}(a),
we find that the vortex line has a finite energy gap and the $\mathbb{Z}_{2}$ invariant is $\nu=-1$
in the weakly doped regime.
When all sides of the sample are  open, we find, as expected, that each end of the vortex line binds one robust MZM, as shown in Figs.~\ref{MZM}(b)-\ref{MZM}(d).

Next we study the configuration in Fig.~\ref{crossing}(e), for which the second-order topology
is nontrivial. We first consider that the Zeeman field is negligible, which can be
a good approximation for materials with small $g$ factor. From a local surface perspective, 
the effects of on-site and extended $s$-wave pairings are equivalent, namely, 
both of them induce a Dirac mass to gap the surface Dirac cones. Since the Dirac mass 
is nearly uniform far away from the hinges, regardless of whether the second-order 
topology is trivial or nontrivial, then a $\pi$-flux vortex far away from the hinges
will also bind one robust MZM at its core in the weakly-doped regime even when 
the second-order topology is nontrivial.
However, this picture will dramatically 
change when the vortex line is moved close to the hinges. To see this, we consider that the
vortex line is very close to the $x$-normal boundary [see Fig.~\ref{sketch}(b)]. The results 
in Fig.~\ref{MZM}(e) indicate that the dispersion of the vortex line remains gapped in the weakly doped regime. 
However, the results in Figs.~\ref{MZM}(f)-\ref{MZM}(h) indicate that the MZMs
are no longer localized at the vortex ends; instead they are localized along the hinges. 

The underlying reason for this transfer of robust MZMs from the vortex ends to the hinges is as follows. 
When the helical Majorana modes go along the hinges of the $z$-normal surfaces, they will pick up 
a $\pi$ phase from the $\pi$-flux vortex line after one circle, which will accordingly change 
their boundary conditions. It is known that the linear dispersion of the helical Majorana modes 
for a finite-size system should follow $E=\pm vq_{m}$, with $v$ a constant and the discrete momentum $q_{m}$ 
determined by the boundary condition $e^{iq_{m}L}=1$ [periodic, $q_{m}=2m\pi/L$] or $-1$ [antiperiodic, $q_{m}=(2m+1)\pi/L$], 
where $m$ is an integer, and $L$ is the total length of the closed path along the hinges. Notably, here we find 
that the $\pi$-flux vortex line will change the boundary condition from an antiperiodic one
to a periodic one \cite{supplemental}. As a result, each pair of helical Majorana modes will contain two MZMs. 
Let us focus on the top $z$-normal surface. When the vortex line is moved 
close to the hinges, the MZM at the top vortex end will hybridize with the two MZMs on 
the hinges and then get split. Nevertheless, the $\mathbb{Z}_{2}$ nature of the MZMs guarantees 
that one of the three will remain to be robust (the energy 
will be exactly zero if the length is infinity in the $z$ direction), 
which turns out to be mostly localized along the hinges. 

\begin{figure}[b!]
\centering
\includegraphics[scale=0.6]{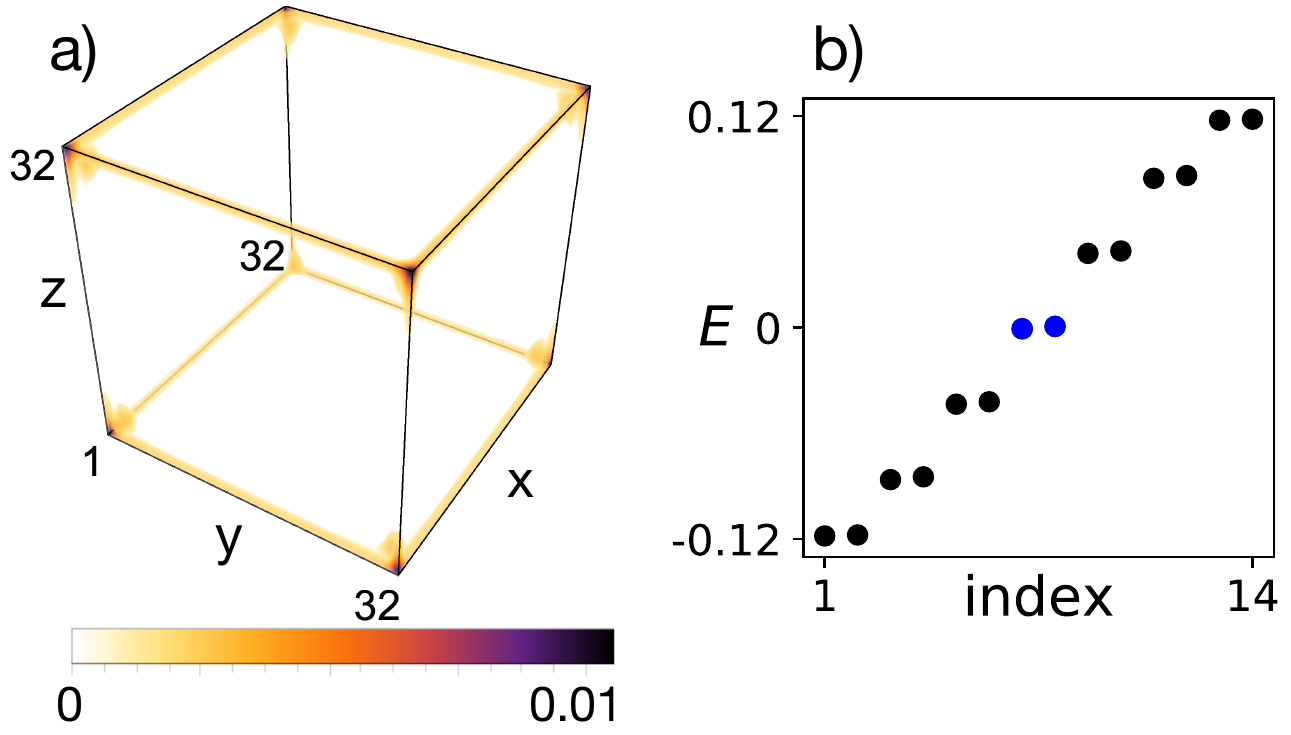}
\caption{(Color online) (a) For a second-order TSC with
chiral MHMs, the probability density of MZMs [the two blue dots in (b)]
are also localized on the hinges rather than at the vortex ends. We take $t=1$, $m_0 = 2.5$, $\lambda =0.5$, $\xi = 4$, $\bm{h}=(h_x,h_y,h_z) = (0, 0, 0.3)$, $\mu=0$, $\Delta_0 =-\Delta_{s}=0.25$, and $L_{x}=L_{y}=L_{z}=32$ lattice sites.}
\label{chiral}
\end{figure}

Finally, we take the Zeeman field into account and study the
situation with chiral MHMs. For the $z$-directional Zeeman field, 
we find that the condition for the realization 
of a chiral second-order TSC with chiral MHMs 
is $|h_{z}|>\sqrt{\mu^{2}+(\Delta_{0}+2\Delta_{s})^{2}}$ while maintaining 
a finite bulk gap~\cite{supplemental}. 
In this regime, we find that the vortex line also
does not bind robust MZMs, and the MZMs also appear on the hinges, 
as shown in Figs.~\ref{chiral}(a)
and \ref{chiral}(b). Notably, the trivialization of the vortex line in 
this regime does not depend on the distance away from the chiral Majorana modes
as well as the existence of other nearby vortex lines. 
This can be intuitively understood by noting that the vortex-end MZMs must 
originate from certain gapless modes in the normal state, so the prerequisite 
is that the chemical potential crosses the surface bands. As the $z$-directional Zeeman field opens a gap of size $2|h_{z}|$ to the Dirac surface states in the normal state, 
the condition for the chemical potential to cross the surface bands is $|\mu|>|h_{z}|$, which is apparently not compatible with the condition for 
the chiral second-order TSC. When $h_{z}$ is finite and the chemical potential 
crosses the surface bands ($|\mu|>|h_{z}|$), 
the scenario is found to be similar to that at the limit $h_{z}=0$. That is, whether the vortex line binds robust MZMs or not depends 
on its position to the hinges, which harbor counterpropagating gapless Majorana modes~\cite{supplemental}.
It is worth noting that, in the regime with chiral MHMs, 
the presence of MZMs on the hinges is
similar to the helical case. That is, an odd number of $\pi$ fluxes
can change the boundary condition of the chiral MHMs,
consequently allowing the presence of MZMs~\cite{read2000}.

%=========================

{\it Discussions and Conclusions.---}
Let us now apply our findings to explain
the coexistence of topological and trivial vortex lines observed in experiments.
On the experimental side, 
strong signatures of helical Majorana modes have been observed in FeTe$_{0.55}$Se$_{0.45}$ not only 
at the hinges \cite{Gray2019helical}, but also at certain crystalline domain walls \cite{Wang2020helical}.
According to our theory, when the positions of the generated vortex lines are close to these 
helical Majorana modes, their hybridizations can trivialize the vortex-line topology, which accordingly
provides an explanation for the existence of trivial vortex lines, even in the weak-Zeeman-field regime. Moreover, some of the topological vortex lines are observed in experiments to be trivialized 
with the increase of magnetic field \cite{chen2019observation}. Since the positions of these vortex lines do not change, according to our theory,
their trivialization is potentially caused by a change of the surface Dirac-mass nature.  
It is noteworthy that the superconducting pairing gap has been observed to
display considerable inhomogeneity on the surface of FeTe$_{0.55}$Se$_{0.45}$
using scanning tunneling microscopy \cite{Wang2020helical,Singh2013}. Meanwhile, the chemical potential also inevitably displays certain fluctuations in real space. Therefore the increase of magnetic field will increase the regions dominated by the Zeeman field, which will consequently increase the ratio of trivial vortex lines, 
agreeing with the experiment observation.

In conclusion, we have developed a mechanism to explain the peculiar coexistence 
and evolution of topological and trivial vortex lines in iron-based superconductors, 
hopefully advancing the understanding of the topological physics in these materials.

%========================
{\it Note added.---} We are grateful to the authors of Ref.~\cite{PhysRevResearch.3.013066} for calling their work to our attention.

%========================
{\it Acknowledgements.---}
We thank the  anonymous referees for the suggestion to consider systems with larger size, which helps us to avoid misinterpretations and establish a simple and natural understanding of the main results. This work was supported in part by the Natural Sciences and Engineering Research Council of Canada (NSERC)
and a Major Innovation Fund (MIF) grant from the Province of Alberta, Canada. Z.Y. is supported by a startup grant from Sun Yat-sen University (Grant No.~74130-18841219), the National Natural Science Foundation of China (Grant No.~11904417), and the Natural Science Foundation of Guangdong Province
(Grant No.~2021B1515020026).

\bibliography{ref.bib}

\begin{thebibliography}{99}%
\makeatletter
\providecommand \@ifxundefined [1]{%
 \@ifx{#1\undefined}
}%
\providecommand \@ifnum [1]{%
 \ifnum #1\expandafter \@firstoftwo
 \else \expandafter \@secondoftwo
 \fi
}%
\providecommand \@ifx [1]{%
 \ifx #1\expandafter \@firstoftwo
 \else \expandafter \@secondoftwo
 \fi
}%
\providecommand \natexlab [1]{#1}%
\providecommand \enquote  [1]{``#1''}%
\providecommand \bibnamefont  [1]{#1}%
\providecommand \bibfnamefont [1]{#1}%
\providecommand \citenamefont [1]{#1}%
\providecommand \href@noop [0]{\@secondoftwo}%
\providecommand \href [0]{\begingroup \@sanitize@url \@href}%
\providecommand \@href[1]{\@@startlink{#1}\@@href}%
\providecommand \@@href[1]{\endgroup#1\@@endlink}%
\providecommand \@sanitize@url [0]{\catcode `\\12\catcode `\$12\catcode
  `\&12\catcode `\#12\catcode `\^12\catcode `\_12\catcode `\%12\relax}%
\providecommand \@@startlink[1]{}%
\providecommand \@@endlink[0]{}%
\providecommand \url  [0]{\begingroup\@sanitize@url \@url }%
\providecommand \@url [1]{\endgroup\@href {#1}{\urlprefix }}%
\providecommand \urlprefix  [0]{URL }%
\providecommand \Eprint [0]{\href }%
\providecommand \doibase [0]{http://dx.doi.org/}%
\providecommand \selectlanguage [0]{\@gobble}%
\providecommand \bibinfo  [0]{\@secondoftwo}%
\providecommand \bibfield  [0]{\@secondoftwo}%
\providecommand \translation [1]{[#1]}%
\providecommand \BibitemOpen [0]{}%
\providecommand \bibitemStop [0]{}%
\providecommand \bibitemNoStop [0]{.\EOS\space}%
\providecommand \EOS [0]{\spacefactor3000\relax}%
\providecommand \BibitemShut  [1]{\csname bibitem#1\endcsname}%
\let\auto@bib@innerbib\@empty
%</preamble>
\bibitem [{\citenamefont {Read}\ and\ \citenamefont {Green}(2000)}]{read2000}%
  \BibitemOpen
  \bibfield  {author} {\bibinfo {author} {\bibfnamefont {N.}~\bibnamefont
  {Read}}\ and\ \bibinfo {author} {\bibfnamefont {Dmitry}\ \bibnamefont
  {Green}},\ }\bibfield  {title} {\enquote {\bibinfo {title} {Paired states of
  fermions in two dimensions with breaking of parity and time-reversal
  symmetries and the fractional quantum {H}all effect},}\ }\href {\doibase
  10.1103/PhysRevB.61.10267} {\bibfield  {journal} {\bibinfo  {journal} {Phys.
  Rev. B}\ }\textbf {\bibinfo {volume} {61}},\ \bibinfo {pages} {10267--10297}
  (\bibinfo {year} {2000})}\BibitemShut {NoStop}%
\bibitem [{\citenamefont {Ivanov}(2001)}]{ivanov2001}%
  \BibitemOpen
  \bibfield  {author} {\bibinfo {author} {\bibfnamefont {D.~A.}\ \bibnamefont
  {Ivanov}},\ }\bibfield  {title} {\enquote {\bibinfo {title} {Non-abelian
  statistics of half-quantum vortices in $\mathit{p}$-wave superconductors},}\
  }\href {\doibase 10.1103/PhysRevLett.86.268} {\bibfield  {journal} {\bibinfo
  {journal} {Phys. Rev. Lett.}\ }\textbf {\bibinfo {volume} {86}},\ \bibinfo
  {pages} {268--271} (\bibinfo {year} {2001})}\BibitemShut {NoStop}%
\bibitem [{\citenamefont {Kitaev}(2001)}]{kitaev2001unpaired}%
  \BibitemOpen
  \bibfield  {author} {\bibinfo {author} {\bibfnamefont {Alexei}\ \bibnamefont
  {Kitaev}},\ }\bibfield  {title} {\enquote {\bibinfo {title} {Unpaired
  {M}ajorana fermions in quantum wires},}\ }\href {\doibase
  10.1070/1063-7869/44/10S/S29} {\bibfield  {journal} {\bibinfo  {journal}
  {Physics-Uspekhi}\ }\textbf {\bibinfo {volume} {44}},\ \bibinfo {pages} {131}
  (\bibinfo {year} {2001})}\BibitemShut {NoStop}%
\bibitem [{\citenamefont {Nayak}\ \emph {et~al.}(2008)\citenamefont {Nayak},
  \citenamefont {Simon}, \citenamefont {Stern}, \citenamefont {Freedman},\ and\
  \citenamefont {Das~Sarma}}]{nayak2008review}%
  \BibitemOpen
  \bibfield  {author} {\bibinfo {author} {\bibfnamefont {Chetan}\ \bibnamefont
  {Nayak}}, \bibinfo {author} {\bibfnamefont {Steven~H.}\ \bibnamefont
  {Simon}}, \bibinfo {author} {\bibfnamefont {Ady}\ \bibnamefont {Stern}},
  \bibinfo {author} {\bibfnamefont {Michael}\ \bibnamefont {Freedman}}, \ and\
  \bibinfo {author} {\bibfnamefont {Sankar}\ \bibnamefont {Das~Sarma}},\
  }\bibfield  {title} {\enquote {\bibinfo {title} {Non-{A}belian anyons and
  topological quantum computation},}\ }\href {\doibase
  10.1103/RevModPhys.80.1083} {\bibfield  {journal} {\bibinfo  {journal} {Rev.
  Mod. Phys.}\ }\textbf {\bibinfo {volume} {80}},\ \bibinfo {pages}
  {1083--1159} (\bibinfo {year} {2008})}\BibitemShut {NoStop}%
\bibitem [{\citenamefont {Fu}\ and\ \citenamefont {Kane}(2008)}]{fu2008}%
  \BibitemOpen
  \bibfield  {author} {\bibinfo {author} {\bibfnamefont {Liang}\ \bibnamefont
  {Fu}}\ and\ \bibinfo {author} {\bibfnamefont {C.~L.}\ \bibnamefont {Kane}},\
  }\bibfield  {title} {\enquote {\bibinfo {title} {Superconducting proximity
  effect and {M}ajorana fermions at the surface of a topological insulator},}\
  }\href {\doibase 10.1103/PhysRevLett.100.096407} {\bibfield  {journal}
  {\bibinfo  {journal} {Phys. Rev. Lett.}\ }\textbf {\bibinfo {volume} {100}},\
  \bibinfo {pages} {096407} (\bibinfo {year} {2008})}\BibitemShut {NoStop}%
\bibitem [{\citenamefont {Lutchyn}\ \emph {et~al.}(2010)\citenamefont
  {Lutchyn}, \citenamefont {Sau},\ and\ \citenamefont
  {Das~Sarma}}]{lutchyn2010}%
  \BibitemOpen
  \bibfield  {author} {\bibinfo {author} {\bibfnamefont {Roman~M.}\
  \bibnamefont {Lutchyn}}, \bibinfo {author} {\bibfnamefont {Jay~D.}\
  \bibnamefont {Sau}}, \ and\ \bibinfo {author} {\bibfnamefont
  {S.}~\bibnamefont {Das~Sarma}},\ }\bibfield  {title} {\enquote {\bibinfo
  {title} {Majorana fermions and a topological phase transition in
  semiconductor-superconductor heterostructures},}\ }\href {\doibase
  10.1103/PhysRevLett.105.077001} {\bibfield  {journal} {\bibinfo  {journal}
  {Phys. Rev. Lett.}\ }\textbf {\bibinfo {volume} {105}},\ \bibinfo {pages}
  {077001} (\bibinfo {year} {2010})}\BibitemShut {NoStop}%
\bibitem [{\citenamefont {Oreg}\ \emph {et~al.}(2010)\citenamefont {Oreg},
  \citenamefont {Refael},\ and\ \citenamefont {von Oppen}}]{oreg2010helical}%
  \BibitemOpen
  \bibfield  {author} {\bibinfo {author} {\bibfnamefont {Yuval}\ \bibnamefont
  {Oreg}}, \bibinfo {author} {\bibfnamefont {Gil}\ \bibnamefont {Refael}}, \
  and\ \bibinfo {author} {\bibfnamefont {Felix}\ \bibnamefont {von Oppen}},\
  }\bibfield  {title} {\enquote {\bibinfo {title} {Helical liquids and
  {M}ajorana bound states in quantum wires},}\ }\href {\doibase
  10.1103/PhysRevLett.105.177002} {\bibfield  {journal} {\bibinfo  {journal}
  {Phys. Rev. Lett.}\ }\textbf {\bibinfo {volume} {105}},\ \bibinfo {pages}
  {177002} (\bibinfo {year} {2010})}\BibitemShut {NoStop}%
\bibitem [{\citenamefont {Alicea}\ \emph {et~al.}(2011)\citenamefont {Alicea},
  \citenamefont {Oreg}, \citenamefont {Refael}, \citenamefont {von Oppen},\
  and\ \citenamefont {Fisher}}]{Alicea2011non}%
  \BibitemOpen
  \bibfield  {author} {\bibinfo {author} {\bibfnamefont {Jason}\ \bibnamefont
  {Alicea}}, \bibinfo {author} {\bibfnamefont {Yuval}\ \bibnamefont {Oreg}},
  \bibinfo {author} {\bibfnamefont {Gil}\ \bibnamefont {Refael}}, \bibinfo
  {author} {\bibfnamefont {Felix}\ \bibnamefont {von Oppen}}, \ and\ \bibinfo
  {author} {\bibfnamefont {Matthew P.~A.}\ \bibnamefont {Fisher}},\ }\bibfield
  {title} {\enquote {\bibinfo {title} {Non-abelian statistics and topological
  quantum information processing in 1{D} wire networks},}\ }\href {\doibase
  10.1038/nphys1915} {\bibfield  {journal} {\bibinfo  {journal} {Nature
  Physics}\ }\textbf {\bibinfo {volume} {7}},\ \bibinfo {pages} {412--417}
  (\bibinfo {year} {2011})}\BibitemShut {NoStop}%
\bibitem [{\citenamefont {Zhang}\ \emph
  {et~al.}(2013{\natexlab{a}})\citenamefont {Zhang}, \citenamefont {Kane},\
  and\ \citenamefont {Mele}}]{zhang2013kramers}%
  \BibitemOpen
  \bibfield  {author} {\bibinfo {author} {\bibfnamefont {Fan}\ \bibnamefont
  {Zhang}}, \bibinfo {author} {\bibfnamefont {C.~L.}\ \bibnamefont {Kane}}, \
  and\ \bibinfo {author} {\bibfnamefont {E.~J.}\ \bibnamefont {Mele}},\
  }\bibfield  {title} {\enquote {\bibinfo {title} {Time-reversal-invariant
  topological superconductivity and majorana {K}ramers pairs},}\ }\href
  {\doibase 10.1103/PhysRevLett.111.056402} {\bibfield  {journal} {\bibinfo
  {journal} {Phys. Rev. Lett.}\ }\textbf {\bibinfo {volume} {111}},\ \bibinfo
  {pages} {056402} (\bibinfo {year} {2013}{\natexlab{a}})}\BibitemShut
  {NoStop}%
\bibitem [{\citenamefont {Zhang}\ \emph
  {et~al.}(2013{\natexlab{b}})\citenamefont {Zhang}, \citenamefont {Kane},\
  and\ \citenamefont {Mele}}]{Zhang2013mirror}%
  \BibitemOpen
  \bibfield  {author} {\bibinfo {author} {\bibfnamefont {Fan}\ \bibnamefont
  {Zhang}}, \bibinfo {author} {\bibfnamefont {C.~L.}\ \bibnamefont {Kane}}, \
  and\ \bibinfo {author} {\bibfnamefont {E.~J.}\ \bibnamefont {Mele}},\
  }\bibfield  {title} {\enquote {\bibinfo {title} {Topological mirror
  superconductivity},}\ }\href {\doibase 10.1103/PhysRevLett.111.056403}
  {\bibfield  {journal} {\bibinfo  {journal} {Phys. Rev. Lett.}\ }\textbf
  {\bibinfo {volume} {111}},\ \bibinfo {pages} {056403} (\bibinfo {year}
  {2013}{\natexlab{b}})}\BibitemShut {NoStop}%
\bibitem [{\citenamefont {Paglione}\ and\ \citenamefont
  {Greene}(2010)}]{Paglione2010}%
  \BibitemOpen
  \bibfield  {author} {\bibinfo {author} {\bibfnamefont {Johnpierre}\
  \bibnamefont {Paglione}}\ and\ \bibinfo {author} {\bibfnamefont {Richard~L.}\
  \bibnamefont {Greene}},\ }\bibfield  {title} {\enquote {\bibinfo {title}
  {High-temperature superconductivity in iron-based materials},}\ }\href
  {\doibase 10.1038/nphys1759} {\bibfield  {journal} {\bibinfo  {journal}
  {Nature Physics}\ }\textbf {\bibinfo {volume} {6}},\ \bibinfo {pages}
  {645--658} (\bibinfo {year} {2010})}\BibitemShut {NoStop}%
\bibitem [{\citenamefont {Hirschfeld}\ \emph {et~al.}(2011)\citenamefont
  {Hirschfeld}, \citenamefont {Korshunov},\ and\ \citenamefont
  {Mazin}}]{Hirschfeld2011}%
  \BibitemOpen
  \bibfield  {author} {\bibinfo {author} {\bibfnamefont {P~J}\ \bibnamefont
  {Hirschfeld}}, \bibinfo {author} {\bibfnamefont {M~M}\ \bibnamefont
  {Korshunov}}, \ and\ \bibinfo {author} {\bibfnamefont {I~I}\ \bibnamefont
  {Mazin}},\ }\bibfield  {title} {\enquote {\bibinfo {title} {Gap symmetry and
  structure of {F}e-based superconductors},}\ }\href {\doibase
  10.1088/0034-4885/74/12/124508} {\bibfield  {journal} {\bibinfo  {journal}
  {Reports on Progress in Physics}\ }\textbf {\bibinfo {volume} {74}},\
  \bibinfo {pages} {124508} (\bibinfo {year} {2011})}\BibitemShut {NoStop}%
\bibitem [{\citenamefont {Stewart}(2011)}]{Stewart2011review}%
  \BibitemOpen
  \bibfield  {author} {\bibinfo {author} {\bibfnamefont {G.~R.}\ \bibnamefont
  {Stewart}},\ }\bibfield  {title} {\enquote {\bibinfo {title}
  {Superconductivity in iron compounds},}\ }\href {\doibase
  10.1103/RevModPhys.83.1589} {\bibfield  {journal} {\bibinfo  {journal} {Rev.
  Mod. Phys.}\ }\textbf {\bibinfo {volume} {83}},\ \bibinfo {pages}
  {1589--1652} (\bibinfo {year} {2011})}\BibitemShut {NoStop}%
\bibitem [{\citenamefont {Chubukov}(2012)}]{Chubukov2012review}%
  \BibitemOpen
  \bibfield  {author} {\bibinfo {author} {\bibfnamefont {Andrey}\ \bibnamefont
  {Chubukov}},\ }\bibfield  {title} {\enquote {\bibinfo {title} {Pairing
  mechanism in {F}e-based superconductors},}\ }\href {\doibase
  10.1146/annurev-conmatphys-020911-125055} {\bibfield  {journal} {\bibinfo
  {journal} {Annual Review of Condensed Matter Physics}\ }\textbf {\bibinfo
  {volume} {3}},\ \bibinfo {pages} {57--92} (\bibinfo {year}
  {2012})}\BibitemShut {NoStop}%
\bibitem [{\citenamefont {Dagotto}(2013)}]{Dagotto2013}%
  \BibitemOpen
  \bibfield  {author} {\bibinfo {author} {\bibfnamefont {Elbio}\ \bibnamefont
  {Dagotto}},\ }\bibfield  {title} {\enquote {\bibinfo {title} {Colloquium: The
  unexpected properties of alkali metal iron selenide superconductors},}\
  }\href {\doibase 10.1103/RevModPhys.85.849} {\bibfield  {journal} {\bibinfo
  {journal} {Rev. Mod. Phys.}\ }\textbf {\bibinfo {volume} {85}},\ \bibinfo
  {pages} {849--867} (\bibinfo {year} {2013})}\BibitemShut {NoStop}%
\bibitem [{\citenamefont {Dai}(2015)}]{Dai2015review}%
  \BibitemOpen
  \bibfield  {author} {\bibinfo {author} {\bibfnamefont {Pengcheng}\
  \bibnamefont {Dai}},\ }\bibfield  {title} {\enquote {\bibinfo {title}
  {Antiferromagnetic order and spin dynamics in iron-based superconductors},}\
  }\href {\doibase 10.1103/RevModPhys.87.855} {\bibfield  {journal} {\bibinfo
  {journal} {Rev. Mod. Phys.}\ }\textbf {\bibinfo {volume} {87}},\ \bibinfo
  {pages} {855--896} (\bibinfo {year} {2015})}\BibitemShut {NoStop}%
\bibitem [{\citenamefont {Mazin}\ \emph {et~al.}(2008)\citenamefont {Mazin},
  \citenamefont {Singh}, \citenamefont {Johannes},\ and\ \citenamefont
  {Du}}]{Mazin2008}%
  \BibitemOpen
  \bibfield  {author} {\bibinfo {author} {\bibfnamefont {I.~I.}\ \bibnamefont
  {Mazin}}, \bibinfo {author} {\bibfnamefont {D.~J.}\ \bibnamefont {Singh}},
  \bibinfo {author} {\bibfnamefont {M.~D.}\ \bibnamefont {Johannes}}, \ and\
  \bibinfo {author} {\bibfnamefont {M.~H.}\ \bibnamefont {Du}},\ }\bibfield
  {title} {\enquote {\bibinfo {title} {Unconventional superconductivity with a
  sign reversal in the order parameter of \ce{LaFeAsO_{1-x}F_x}},}\ }\href
  {\doibase 10.1103/PhysRevLett.101.057003} {\bibfield  {journal} {\bibinfo
  {journal} {Phys. Rev. Lett.}\ }\textbf {\bibinfo {volume} {101}},\ \bibinfo
  {pages} {057003} (\bibinfo {year} {2008})}\BibitemShut {NoStop}%
\bibitem [{\citenamefont {Wang}\ and\ \citenamefont
  {Lee}(2011)}]{Wang2011iron}%
  \BibitemOpen
  \bibfield  {author} {\bibinfo {author} {\bibfnamefont {Fa}~\bibnamefont
  {Wang}}\ and\ \bibinfo {author} {\bibfnamefont {Dung-Hai}\ \bibnamefont
  {Lee}},\ }\bibfield  {title} {\enquote {\bibinfo {title} {The
  electron-pairing mechanism of iron-based superconductors},}\ }\href {\doibase
  10.1126/science.1200182} {\bibfield  {journal} {\bibinfo  {journal}
  {Science}\ }\textbf {\bibinfo {volume} {332}},\ \bibinfo {pages} {200--204}
  (\bibinfo {year} {2011})}\BibitemShut {NoStop}%
\bibitem [{\citenamefont {Wang}\ \emph {et~al.}(2015)\citenamefont {Wang},
  \citenamefont {Zhang}, \citenamefont {Xu}, \citenamefont {Zeng},
  \citenamefont {Miao}, \citenamefont {Xu}, \citenamefont {Qian}, \citenamefont
  {Weng}, \citenamefont {Richard}, \citenamefont {Fedorov}, \citenamefont
  {Ding}, \citenamefont {Dai},\ and\ \citenamefont {Fang}}]{Wang2015iFeSC}%
  \BibitemOpen
  \bibfield  {author} {\bibinfo {author} {\bibfnamefont {Zhijun}\ \bibnamefont
  {Wang}}, \bibinfo {author} {\bibfnamefont {P.}~\bibnamefont {Zhang}},
  \bibinfo {author} {\bibfnamefont {Gang}\ \bibnamefont {Xu}}, \bibinfo
  {author} {\bibfnamefont {L.~K.}\ \bibnamefont {Zeng}}, \bibinfo {author}
  {\bibfnamefont {H.}~\bibnamefont {Miao}}, \bibinfo {author} {\bibfnamefont
  {Xiaoyan}\ \bibnamefont {Xu}}, \bibinfo {author} {\bibfnamefont
  {T.}~\bibnamefont {Qian}}, \bibinfo {author} {\bibfnamefont {Hongming}\
  \bibnamefont {Weng}}, \bibinfo {author} {\bibfnamefont {P.}~\bibnamefont
  {Richard}}, \bibinfo {author} {\bibfnamefont {A.~V.}\ \bibnamefont
  {Fedorov}}, \bibinfo {author} {\bibfnamefont {H.}~\bibnamefont {Ding}},
  \bibinfo {author} {\bibfnamefont {Xi}~\bibnamefont {Dai}}, \ and\ \bibinfo
  {author} {\bibfnamefont {Zhong}\ \bibnamefont {Fang}},\ }\bibfield  {title}
  {\enquote {\bibinfo {title} {Topological nature of the
  \ce{FeSe_{0.5}Te_{0.5}} superconductor},}\ }\href {\doibase
  10.1103/PhysRevB.92.115119} {\bibfield  {journal} {\bibinfo  {journal} {Phys.
  Rev. B}\ }\textbf {\bibinfo {volume} {92}},\ \bibinfo {pages} {115119}
  (\bibinfo {year} {2015})}\BibitemShut {NoStop}%
\bibitem [{\citenamefont {Wu}\ \emph {et~al.}(2016)\citenamefont {Wu},
  \citenamefont {Qin}, \citenamefont {Liang}, \citenamefont {Fan},\ and\
  \citenamefont {Hu}}]{Wu2016iron}%
  \BibitemOpen
  \bibfield  {author} {\bibinfo {author} {\bibfnamefont {Xianxin}\ \bibnamefont
  {Wu}}, \bibinfo {author} {\bibfnamefont {Shengshan}\ \bibnamefont {Qin}},
  \bibinfo {author} {\bibfnamefont {Yi}~\bibnamefont {Liang}}, \bibinfo
  {author} {\bibfnamefont {Heng}\ \bibnamefont {Fan}}, \ and\ \bibinfo {author}
  {\bibfnamefont {Jiangping}\ \bibnamefont {Hu}},\ }\bibfield  {title}
  {\enquote {\bibinfo {title} {Topological characters in
  \ce{Fe(Te_{1-x}Se_{x})} thin films},}\ }\href {\doibase
  10.1103/PhysRevB.93.115129} {\bibfield  {journal} {\bibinfo  {journal} {Phys.
  Rev. B}\ }\textbf {\bibinfo {volume} {93}},\ \bibinfo {pages} {115129}
  (\bibinfo {year} {2016})}\BibitemShut {NoStop}%
\bibitem [{\citenamefont {Xu}\ \emph {et~al.}(2016)\citenamefont {Xu},
  \citenamefont {Lian}, \citenamefont {Tang}, \citenamefont {Qi},\ and\
  \citenamefont {Zhang}}]{Xu2016FeSC}%
  \BibitemOpen
  \bibfield  {author} {\bibinfo {author} {\bibfnamefont {Gang}\ \bibnamefont
  {Xu}}, \bibinfo {author} {\bibfnamefont {Biao}\ \bibnamefont {Lian}},
  \bibinfo {author} {\bibfnamefont {Peizhe}\ \bibnamefont {Tang}}, \bibinfo
  {author} {\bibfnamefont {Xiao-Liang}\ \bibnamefont {Qi}}, \ and\ \bibinfo
  {author} {\bibfnamefont {Shou-Cheng}\ \bibnamefont {Zhang}},\ }\bibfield
  {title} {\enquote {\bibinfo {title} {Topological superconductivity on the
  surface of {F}e-based superconductors},}\ }\href {\doibase
  10.1103/PhysRevLett.117.047001} {\bibfield  {journal} {\bibinfo  {journal}
  {Phys. Rev. Lett.}\ }\textbf {\bibinfo {volume} {117}},\ \bibinfo {pages}
  {047001} (\bibinfo {year} {2016})}\BibitemShut {NoStop}%
\bibitem [{\citenamefont {Hao}\ and\ \citenamefont
  {Hu}(2019)}]{Hao2019topological}%
  \BibitemOpen
  \bibfield  {author} {\bibinfo {author} {\bibfnamefont {Ning}\ \bibnamefont
  {Hao}}\ and\ \bibinfo {author} {\bibfnamefont {Jiangping}\ \bibnamefont
  {Hu}},\ }\bibfield  {title} {\enquote {\bibinfo {title} {Topological quantum
  states of matter in iron-based superconductors: from concept to material
  realization},}\ }\href {https://doi.org/10.1093/nsr/nwy142} {\bibfield
  {journal} {\bibinfo  {journal} {National Science Review}\ }\textbf {\bibinfo
  {volume} {6}},\ \bibinfo {pages} {213--226} (\bibinfo {year}
  {2019})}\BibitemShut {NoStop}%
\bibitem [{\citenamefont {Kreisel}\ \emph {et~al.}(2020)\citenamefont
  {Kreisel}, \citenamefont {Hirschfeld},\ and\ \citenamefont
  {Andersen}}]{Kreise2020review}%
  \BibitemOpen
  \bibfield  {author} {\bibinfo {author} {\bibfnamefont {Andreas}\ \bibnamefont
  {Kreisel}}, \bibinfo {author} {\bibfnamefont {Peter~J}\ \bibnamefont
  {Hirschfeld}}, \ and\ \bibinfo {author} {\bibfnamefont {Brian~M}\
  \bibnamefont {Andersen}},\ }\bibfield  {title} {\enquote {\bibinfo {title}
  {{O}n the {R}emarkable {S}uperconductivity of {F}e{S}e and its {C}lose
  {C}ousins},}\ }\href {https://www.mdpi.com/2073-8994/12/9/1402} {\bibfield
  {journal} {\bibinfo  {journal} {Symmetry}\ }\textbf {\bibinfo {volume}
  {12}},\ \bibinfo {pages} {1402} (\bibinfo {year} {2020})}\BibitemShut
  {NoStop}%
\bibitem [{\citenamefont {Zhang}\ \emph
  {et~al.}(2018{\natexlab{a}})\citenamefont {Zhang}, \citenamefont {Yaji},
  \citenamefont {Hashimoto}, \citenamefont {Ota}, \citenamefont {Kondo},
  \citenamefont {Okazaki}, \citenamefont {Wang}, \citenamefont {Wen},
  \citenamefont {Gu}, \citenamefont {Ding} \emph {et~al.}}]{zhang2018iron}%
  \BibitemOpen
  \bibfield  {author} {\bibinfo {author} {\bibfnamefont {Peng}\ \bibnamefont
  {Zhang}}, \bibinfo {author} {\bibfnamefont {Koichiro}\ \bibnamefont {Yaji}},
  \bibinfo {author} {\bibfnamefont {Takahiro}\ \bibnamefont {Hashimoto}},
  \bibinfo {author} {\bibfnamefont {Yuichi}\ \bibnamefont {Ota}}, \bibinfo
  {author} {\bibfnamefont {Takeshi}\ \bibnamefont {Kondo}}, \bibinfo {author}
  {\bibfnamefont {Kozo}\ \bibnamefont {Okazaki}}, \bibinfo {author}
  {\bibfnamefont {Zhijun}\ \bibnamefont {Wang}}, \bibinfo {author}
  {\bibfnamefont {Jinsheng}\ \bibnamefont {Wen}}, \bibinfo {author}
  {\bibfnamefont {GD}~\bibnamefont {Gu}}, \bibinfo {author} {\bibfnamefont
  {Hong}\ \bibnamefont {Ding}},  \emph {et~al.},\ }\bibfield  {title} {\enquote
  {\bibinfo {title} {Observation of topological superconductivity on the
  surface of an iron-based superconductor},}\ }\href {\doibase
  10.1126/science.aan4596} {\bibfield  {journal} {\bibinfo  {journal}
  {Science}\ }\textbf {\bibinfo {volume} {360}},\ \bibinfo {pages} {182--186}
  (\bibinfo {year} {2018}{\natexlab{a}})}\BibitemShut {NoStop}%
\bibitem [{\citenamefont {Zhang}\ \emph
  {et~al.}(2019{\natexlab{a}})\citenamefont {Zhang}, \citenamefont {Wang},
  \citenamefont {Wu}, \citenamefont {Yaji}, \citenamefont {Ishida},
  \citenamefont {Kohama}, \citenamefont {Dai}, \citenamefont {Sun},
  \citenamefont {Bareille}, \citenamefont {Kuroda} \emph
  {et~al.}}]{zhang2019multiple}%
  \BibitemOpen
  \bibfield  {author} {\bibinfo {author} {\bibfnamefont {Peng}\ \bibnamefont
  {Zhang}}, \bibinfo {author} {\bibfnamefont {Zhijun}\ \bibnamefont {Wang}},
  \bibinfo {author} {\bibfnamefont {Xianxin}\ \bibnamefont {Wu}}, \bibinfo
  {author} {\bibfnamefont {Koichiro}\ \bibnamefont {Yaji}}, \bibinfo {author}
  {\bibfnamefont {Yukiaki}\ \bibnamefont {Ishida}}, \bibinfo {author}
  {\bibfnamefont {Yoshimitsu}\ \bibnamefont {Kohama}}, \bibinfo {author}
  {\bibfnamefont {Guangyang}\ \bibnamefont {Dai}}, \bibinfo {author}
  {\bibfnamefont {Yue}\ \bibnamefont {Sun}}, \bibinfo {author} {\bibfnamefont
  {Cedric}\ \bibnamefont {Bareille}}, \bibinfo {author} {\bibfnamefont {Kenta}\
  \bibnamefont {Kuroda}},  \emph {et~al.},\ }\bibfield  {title} {\enquote
  {\bibinfo {title} {Multiple topological states in iron-based
  superconductors},}\ }\href {\doibase 10.1038/s41567-018-0280-z} {\bibfield
  {journal} {\bibinfo  {journal} {Nature Physics}\ }\textbf {\bibinfo {volume}
  {15}},\ \bibinfo {pages} {41} (\bibinfo {year}
  {2019}{\natexlab{a}})}\BibitemShut {NoStop}%
\bibitem [{\citenamefont {Wang}\ \emph
  {et~al.}(2018{\natexlab{a}})\citenamefont {Wang}, \citenamefont {Kong},
  \citenamefont {Fan}, \citenamefont {Chen}, \citenamefont {Zhu}, \citenamefont
  {Liu}, \citenamefont {Cao}, \citenamefont {Sun}, \citenamefont {Du},
  \citenamefont {Schneeloch} \emph {et~al.}}]{wang2018evidence}%
  \BibitemOpen
  \bibfield  {author} {\bibinfo {author} {\bibfnamefont {Dongfei}\ \bibnamefont
  {Wang}}, \bibinfo {author} {\bibfnamefont {Lingyuan}\ \bibnamefont {Kong}},
  \bibinfo {author} {\bibfnamefont {Peng}\ \bibnamefont {Fan}}, \bibinfo
  {author} {\bibfnamefont {Hui}\ \bibnamefont {Chen}}, \bibinfo {author}
  {\bibfnamefont {Shiyu}\ \bibnamefont {Zhu}}, \bibinfo {author} {\bibfnamefont
  {Wenyao}\ \bibnamefont {Liu}}, \bibinfo {author} {\bibfnamefont
  {Lu}~\bibnamefont {Cao}}, \bibinfo {author} {\bibfnamefont {Yujie}\
  \bibnamefont {Sun}}, \bibinfo {author} {\bibfnamefont {Shixuan}\ \bibnamefont
  {Du}}, \bibinfo {author} {\bibfnamefont {John}\ \bibnamefont {Schneeloch}},
  \emph {et~al.},\ }\bibfield  {title} {\enquote {\bibinfo {title} {Evidence
  for {M}ajorana bound states in an iron-based superconductor},}\ }\href
  {\doibase 10.1126/science.aao1797} {\bibfield  {journal} {\bibinfo  {journal}
  {Science}\ }\textbf {\bibinfo {volume} {362}},\ \bibinfo {pages} {333--335}
  (\bibinfo {year} {2018}{\natexlab{a}})}\BibitemShut {NoStop}%
\bibitem [{\citenamefont {Kong}\ \emph {et~al.}(2019)\citenamefont {Kong},
  \citenamefont {Zhu}, \citenamefont {Papaj}, \citenamefont {Chen},
  \citenamefont {Cao}, \citenamefont {Isobe}, \citenamefont {Xing},
  \citenamefont {Liu}, \citenamefont {Wang}, \citenamefont {Fan} \emph
  {et~al.}}]{kong2019half}%
  \BibitemOpen
  \bibfield  {author} {\bibinfo {author} {\bibfnamefont {Lingyuan}\
  \bibnamefont {Kong}}, \bibinfo {author} {\bibfnamefont {Shiyu}\ \bibnamefont
  {Zhu}}, \bibinfo {author} {\bibfnamefont {Micha{\l}}\ \bibnamefont {Papaj}},
  \bibinfo {author} {\bibfnamefont {Hui}\ \bibnamefont {Chen}}, \bibinfo
  {author} {\bibfnamefont {Lu}~\bibnamefont {Cao}}, \bibinfo {author}
  {\bibfnamefont {Hiroki}\ \bibnamefont {Isobe}}, \bibinfo {author}
  {\bibfnamefont {Yuqing}\ \bibnamefont {Xing}}, \bibinfo {author}
  {\bibfnamefont {Wenyao}\ \bibnamefont {Liu}}, \bibinfo {author}
  {\bibfnamefont {Dongfei}\ \bibnamefont {Wang}}, \bibinfo {author}
  {\bibfnamefont {Peng}\ \bibnamefont {Fan}},  \emph {et~al.},\ }\bibfield
  {title} {\enquote {\bibinfo {title} {Half-integer level shift of vortex bound
  states in an iron-based superconductor},}\ }\href
  {https://doi.org/10.1038/s41567-019-0630-5} {\bibfield  {journal} {\bibinfo
  {journal} {Nature Physics}\ }\textbf {\bibinfo {volume} {15}},\ \bibinfo
  {pages} {1181--1187} (\bibinfo {year} {2019})}\BibitemShut {NoStop}%
\bibitem [{\citenamefont {Machida}\ \emph {et~al.}(2019)\citenamefont
  {Machida}, \citenamefont {Sun}, \citenamefont {Pyon}, \citenamefont {Takeda},
  \citenamefont {Kohsaka}, \citenamefont {Hanaguri}, \citenamefont {Sasagawa},\
  and\ \citenamefont {Tamegai}}]{machida2019zero}%
  \BibitemOpen
  \bibfield  {author} {\bibinfo {author} {\bibfnamefont {T}~\bibnamefont
  {Machida}}, \bibinfo {author} {\bibfnamefont {Y}~\bibnamefont {Sun}},
  \bibinfo {author} {\bibfnamefont {S}~\bibnamefont {Pyon}}, \bibinfo {author}
  {\bibfnamefont {S}~\bibnamefont {Takeda}}, \bibinfo {author} {\bibfnamefont
  {Y}~\bibnamefont {Kohsaka}}, \bibinfo {author} {\bibfnamefont
  {T}~\bibnamefont {Hanaguri}}, \bibinfo {author} {\bibfnamefont
  {T}~\bibnamefont {Sasagawa}}, \ and\ \bibinfo {author} {\bibfnamefont
  {T}~\bibnamefont {Tamegai}},\ }\bibfield  {title} {\enquote {\bibinfo {title}
  {Zero-energy vortex bound state in the superconducting topological surface
  state of {F}e ({S}e, {T}e)},}\ }\href
  {https://www.nature.com/articles/s41563-019-0397-1} {\bibfield  {journal}
  {\bibinfo  {journal} {Nature materials}\ }\textbf {\bibinfo {volume} {18}},\
  \bibinfo {pages} {811--815} (\bibinfo {year} {2019})}\BibitemShut {NoStop}%
\bibitem [{\citenamefont {Liu}\ \emph {et~al.}(2018{\natexlab{a}})\citenamefont
  {Liu}, \citenamefont {Chen}, \citenamefont {Zhang}, \citenamefont {Peng},
  \citenamefont {Yan}, \citenamefont {Wen}, \citenamefont {Lou}, \citenamefont
  {Huang}, \citenamefont {Tian}, \citenamefont {Dong}, \citenamefont {Wang},
  \citenamefont {Bao}, \citenamefont {Wang}, \citenamefont {Yin}, \citenamefont
  {Zhao},\ and\ \citenamefont {Feng}}]{Liu2018MZM}%
  \BibitemOpen
  \bibfield  {author} {\bibinfo {author} {\bibfnamefont {Qin}\ \bibnamefont
  {Liu}}, \bibinfo {author} {\bibfnamefont {Chen}\ \bibnamefont {Chen}},
  \bibinfo {author} {\bibfnamefont {Tong}\ \bibnamefont {Zhang}}, \bibinfo
  {author} {\bibfnamefont {Rui}\ \bibnamefont {Peng}}, \bibinfo {author}
  {\bibfnamefont {Ya-Jun}\ \bibnamefont {Yan}}, \bibinfo {author}
  {\bibfnamefont {Chen-Hao-Ping}\ \bibnamefont {Wen}}, \bibinfo {author}
  {\bibfnamefont {Xia}\ \bibnamefont {Lou}}, \bibinfo {author} {\bibfnamefont
  {Yu-Long}\ \bibnamefont {Huang}}, \bibinfo {author} {\bibfnamefont
  {Jin-Peng}\ \bibnamefont {Tian}}, \bibinfo {author} {\bibfnamefont {Xiao-Li}\
  \bibnamefont {Dong}}, \bibinfo {author} {\bibfnamefont {Guang-Wei}\
  \bibnamefont {Wang}}, \bibinfo {author} {\bibfnamefont {Wei-Cheng}\
  \bibnamefont {Bao}}, \bibinfo {author} {\bibfnamefont {Qiang-Hua}\
  \bibnamefont {Wang}}, \bibinfo {author} {\bibfnamefont {Zhi-Ping}\
  \bibnamefont {Yin}}, \bibinfo {author} {\bibfnamefont {Zhong-Xian}\
  \bibnamefont {Zhao}}, \ and\ \bibinfo {author} {\bibfnamefont {Dong-Lai}\
  \bibnamefont {Feng}},\ }\bibfield  {title} {\enquote {\bibinfo {title}
  {Robust and clean {M}ajorana zero mode in the vortex core of high-temperature
  superconductor \ce{(Li_{0.84}Fe_{0.16})OHFeSe}},}\ }\href {\doibase
  10.1103/PhysRevX.8.041056} {\bibfield  {journal} {\bibinfo  {journal} {Phys.
  Rev. X}\ }\textbf {\bibinfo {volume} {8}},\ \bibinfo {pages} {041056}
  (\bibinfo {year} {2018}{\natexlab{a}})}\BibitemShut {NoStop}%
\bibitem [{\citenamefont {Chen}\ \emph
  {et~al.}(2019{\natexlab{a}})\citenamefont {Chen}, \citenamefont {Liu},
  \citenamefont {Zhang}, \citenamefont {Li}, \citenamefont {Shen},
  \citenamefont {Dong}, \citenamefont {Zhao}, \citenamefont {Zhang},\ and\
  \citenamefont {Feng}}]{chen2019quantized}%
  \BibitemOpen
  \bibfield  {author} {\bibinfo {author} {\bibfnamefont {C}~\bibnamefont
  {Chen}}, \bibinfo {author} {\bibfnamefont {Q}~\bibnamefont {Liu}}, \bibinfo
  {author} {\bibfnamefont {TZ}~\bibnamefont {Zhang}}, \bibinfo {author}
  {\bibfnamefont {D}~\bibnamefont {Li}}, \bibinfo {author} {\bibfnamefont
  {PP}~\bibnamefont {Shen}}, \bibinfo {author} {\bibfnamefont {XL}~\bibnamefont
  {Dong}}, \bibinfo {author} {\bibfnamefont {Z-X}\ \bibnamefont {Zhao}},
  \bibinfo {author} {\bibfnamefont {T}~\bibnamefont {Zhang}}, \ and\ \bibinfo
  {author} {\bibfnamefont {DL}~\bibnamefont {Feng}},\ }\bibfield  {title}
  {\enquote {\bibinfo {title} {Quantized conductance of {M}ajorana zero mode in
  the vortex of the topological superconductor
  \ce{(Li_{0.84}Fe_{0.16})OHFeSe}},}\ }\href {\doibase
  10.1088/0256-307x/36/5/057403} {\bibfield  {journal} {\bibinfo  {journal}
  {Chinese Physics Letters}\ }\textbf {\bibinfo {volume} {36}},\ \bibinfo
  {pages} {057403} (\bibinfo {year} {2019}{\natexlab{a}})}\BibitemShut
  {NoStop}%
\bibitem [{\citenamefont {Zhu}\ \emph {et~al.}(2020)\citenamefont {Zhu},
  \citenamefont {Kong}, \citenamefont {Cao}, \citenamefont {Chen},
  \citenamefont {Papaj}, \citenamefont {Du}, \citenamefont {Xing},
  \citenamefont {Liu}, \citenamefont {Wang}, \citenamefont {Shen},
  \citenamefont {Yang}, \citenamefont {Schneeloch}, \citenamefont {Zhong},
  \citenamefont {Gu}, \citenamefont {Fu}, \citenamefont {Zhang}, \citenamefont
  {Ding},\ and\ \citenamefont {Gao}}]{Zhu2019MZM}%
  \BibitemOpen
  \bibfield  {author} {\bibinfo {author} {\bibfnamefont {Shiyu}\ \bibnamefont
  {Zhu}}, \bibinfo {author} {\bibfnamefont {Lingyuan}\ \bibnamefont {Kong}},
  \bibinfo {author} {\bibfnamefont {Lu}~\bibnamefont {Cao}}, \bibinfo {author}
  {\bibfnamefont {Hui}\ \bibnamefont {Chen}}, \bibinfo {author} {\bibfnamefont
  {Micha{\l}}\ \bibnamefont {Papaj}}, \bibinfo {author} {\bibfnamefont
  {Shixuan}\ \bibnamefont {Du}}, \bibinfo {author} {\bibfnamefont {Yuqing}\
  \bibnamefont {Xing}}, \bibinfo {author} {\bibfnamefont {Wenyao}\ \bibnamefont
  {Liu}}, \bibinfo {author} {\bibfnamefont {Dongfei}\ \bibnamefont {Wang}},
  \bibinfo {author} {\bibfnamefont {Chengmin}\ \bibnamefont {Shen}}, \bibinfo
  {author} {\bibfnamefont {Fazhi}\ \bibnamefont {Yang}}, \bibinfo {author}
  {\bibfnamefont {John}\ \bibnamefont {Schneeloch}}, \bibinfo {author}
  {\bibfnamefont {Ruidan}\ \bibnamefont {Zhong}}, \bibinfo {author}
  {\bibfnamefont {Genda}\ \bibnamefont {Gu}}, \bibinfo {author} {\bibfnamefont
  {Liang}\ \bibnamefont {Fu}}, \bibinfo {author} {\bibfnamefont {Yu-Yang}\
  \bibnamefont {Zhang}}, \bibinfo {author} {\bibfnamefont {Hong}\ \bibnamefont
  {Ding}}, \ and\ \bibinfo {author} {\bibfnamefont {Hong-Jun}\ \bibnamefont
  {Gao}},\ }\bibfield  {title} {\enquote {\bibinfo {title} {Nearly quantized
  conductance plateau of vortex zero mode in an iron-based superconductor},}\
  }\href {\doibase 10.1126/science.aax0274} {\bibfield  {journal} {\bibinfo
  {journal} {Science}\ }\textbf {\bibinfo {volume} {367}},\ \bibinfo {pages}
  {189--192} (\bibinfo {year} {2020})}\BibitemShut {NoStop}%
\bibitem [{\citenamefont {Mourik}\ \emph {et~al.}(2012)\citenamefont {Mourik},
  \citenamefont {Zuo}, \citenamefont {Frolov}, \citenamefont {Plissard},
  \citenamefont {Bakkers},\ and\ \citenamefont {Kouwenhoven}}]{Mourik2012MZM}%
  \BibitemOpen
  \bibfield  {author} {\bibinfo {author} {\bibfnamefont {V.}~\bibnamefont
  {Mourik}}, \bibinfo {author} {\bibfnamefont {K.}~\bibnamefont {Zuo}},
  \bibinfo {author} {\bibfnamefont {S.~M.}\ \bibnamefont {Frolov}}, \bibinfo
  {author} {\bibfnamefont {S.~R.}\ \bibnamefont {Plissard}}, \bibinfo {author}
  {\bibfnamefont {E.~P. A.~M.}\ \bibnamefont {Bakkers}}, \ and\ \bibinfo
  {author} {\bibfnamefont {L.~P.}\ \bibnamefont {Kouwenhoven}},\ }\bibfield
  {title} {\enquote {\bibinfo {title} {Signatures of {M}ajorana fermions in
  hybrid superconductor-semiconductor nanowire devices},}\ }\href {\doibase
  10.1126/science.1222360} {\bibfield  {journal} {\bibinfo  {journal}
  {Science}\ }\textbf {\bibinfo {volume} {336}},\ \bibinfo {pages} {1003--1007}
  (\bibinfo {year} {2012})}\BibitemShut {NoStop}%
\bibitem [{\citenamefont {Nadj-Perge}\ \emph {et~al.}(2014)\citenamefont
  {Nadj-Perge}, \citenamefont {Drozdov}, \citenamefont {Li}, \citenamefont
  {Chen}, \citenamefont {Jeon}, \citenamefont {Seo}, \citenamefont {MacDonald},
  \citenamefont {Bernevig},\ and\ \citenamefont {Yazdani}}]{Nadj2014MZM}%
  \BibitemOpen
  \bibfield  {author} {\bibinfo {author} {\bibfnamefont {Stevan}\ \bibnamefont
  {Nadj-Perge}}, \bibinfo {author} {\bibfnamefont {Ilya~K.}\ \bibnamefont
  {Drozdov}}, \bibinfo {author} {\bibfnamefont {Jian}\ \bibnamefont {Li}},
  \bibinfo {author} {\bibfnamefont {Hua}\ \bibnamefont {Chen}}, \bibinfo
  {author} {\bibfnamefont {Sangjun}\ \bibnamefont {Jeon}}, \bibinfo {author}
  {\bibfnamefont {Jungpil}\ \bibnamefont {Seo}}, \bibinfo {author}
  {\bibfnamefont {Allan~H.}\ \bibnamefont {MacDonald}}, \bibinfo {author}
  {\bibfnamefont {B.~Andrei}\ \bibnamefont {Bernevig}}, \ and\ \bibinfo
  {author} {\bibfnamefont {Ali}\ \bibnamefont {Yazdani}},\ }\bibfield  {title}
  {\enquote {\bibinfo {title} {Observation of {M}ajorana fermions in
  ferromagnetic atomic chains on a superconductor},}\ }\href {\doibase
  10.1126/science.1259327} {\bibfield  {journal} {\bibinfo  {journal}
  {Science}\ }\textbf {\bibinfo {volume} {346}},\ \bibinfo {pages} {602--607}
  (\bibinfo {year} {2014})}\BibitemShut {NoStop}%
\bibitem [{\citenamefont {Sun}\ \emph {et~al.}(2016)\citenamefont {Sun},
  \citenamefont {Zhang}, \citenamefont {Hu}, \citenamefont {Li}, \citenamefont
  {Wang}, \citenamefont {Ma}, \citenamefont {Xu}, \citenamefont {Gao},
  \citenamefont {Guan}, \citenamefont {Li}, \citenamefont {Liu}, \citenamefont
  {Qian}, \citenamefont {Zhou}, \citenamefont {Fu}, \citenamefont {Li},
  \citenamefont {Zhang},\ and\ \citenamefont {Jia}}]{Sun2016Majorana}%
  \BibitemOpen
  \bibfield  {author} {\bibinfo {author} {\bibfnamefont {Hao-Hua}\ \bibnamefont
  {Sun}}, \bibinfo {author} {\bibfnamefont {Kai-Wen}\ \bibnamefont {Zhang}},
  \bibinfo {author} {\bibfnamefont {Lun-Hui}\ \bibnamefont {Hu}}, \bibinfo
  {author} {\bibfnamefont {Chuang}\ \bibnamefont {Li}}, \bibinfo {author}
  {\bibfnamefont {Guan-Yong}\ \bibnamefont {Wang}}, \bibinfo {author}
  {\bibfnamefont {Hai-Yang}\ \bibnamefont {Ma}}, \bibinfo {author}
  {\bibfnamefont {Zhu-An}\ \bibnamefont {Xu}}, \bibinfo {author} {\bibfnamefont
  {Chun-Lei}\ \bibnamefont {Gao}}, \bibinfo {author} {\bibfnamefont {Dan-Dan}\
  \bibnamefont {Guan}}, \bibinfo {author} {\bibfnamefont {Yao-Yi}\ \bibnamefont
  {Li}}, \bibinfo {author} {\bibfnamefont {Canhua}\ \bibnamefont {Liu}},
  \bibinfo {author} {\bibfnamefont {Dong}\ \bibnamefont {Qian}}, \bibinfo
  {author} {\bibfnamefont {Yi}~\bibnamefont {Zhou}}, \bibinfo {author}
  {\bibfnamefont {Liang}\ \bibnamefont {Fu}}, \bibinfo {author} {\bibfnamefont
  {Shao-Chun}\ \bibnamefont {Li}}, \bibinfo {author} {\bibfnamefont {Fu-Chun}\
  \bibnamefont {Zhang}}, \ and\ \bibinfo {author} {\bibfnamefont {Jin-Feng}\
  \bibnamefont {Jia}},\ }\bibfield  {title} {\enquote {\bibinfo {title}
  {Majorana zero mode detected with spin selective {A}ndreev reflection in the
  vortex of a topological superconductor},}\ }\href {\doibase
  10.1103/PhysRevLett.116.257003} {\bibfield  {journal} {\bibinfo  {journal}
  {Phys. Rev. Lett.}\ }\textbf {\bibinfo {volume} {116}},\ \bibinfo {pages}
  {257003} (\bibinfo {year} {2016})}\BibitemShut {NoStop}%
\bibitem [{\citenamefont {Zhang}\ \emph
  {et~al.}(2018{\natexlab{b}})\citenamefont {Zhang}, \citenamefont {Liu},
  \citenamefont {Gazibegovic}, \citenamefont {Xu}, \citenamefont {Logan},
  \citenamefont {Wang}, \citenamefont {van Loo}, \citenamefont {Bommer},
  \citenamefont {de~Moor}, \citenamefont {Car}, \citenamefont {Op~het Veld},
  \citenamefont {van Veldhoven}, \citenamefont {Koelling}, \citenamefont
  {Verheijen}, \citenamefont {Pendharkar}, \citenamefont {Pennachio},
  \citenamefont {Shojaei}, \citenamefont {Lee}, \citenamefont {Palmstrøm},
  \citenamefont {Bakkers}, \citenamefont {Sarma},\ and\ \citenamefont
  {Kouwenhoven}}]{Zhang2018quantized}%
  \BibitemOpen
  \bibfield  {author} {\bibinfo {author} {\bibfnamefont {Hao}\ \bibnamefont
  {Zhang}}, \bibinfo {author} {\bibfnamefont {Chun-Xiao}\ \bibnamefont {Liu}},
  \bibinfo {author} {\bibfnamefont {Sasa}\ \bibnamefont {Gazibegovic}},
  \bibinfo {author} {\bibfnamefont {Di}~\bibnamefont {Xu}}, \bibinfo {author}
  {\bibfnamefont {John~A}\ \bibnamefont {Logan}}, \bibinfo {author}
  {\bibfnamefont {Guanzhong}\ \bibnamefont {Wang}}, \bibinfo {author}
  {\bibfnamefont {Nick}\ \bibnamefont {van Loo}}, \bibinfo {author}
  {\bibfnamefont {Jouri~DS}\ \bibnamefont {Bommer}}, \bibinfo {author}
  {\bibfnamefont {Michiel~WA}\ \bibnamefont {de~Moor}}, \bibinfo {author}
  {\bibfnamefont {Diana}\ \bibnamefont {Car}}, \bibinfo {author} {\bibfnamefont
  {Roy L.~M.}\ \bibnamefont {Op~het Veld}}, \bibinfo {author} {\bibfnamefont
  {Petrus~J.}\ \bibnamefont {van Veldhoven}}, \bibinfo {author} {\bibfnamefont
  {Sebastian}\ \bibnamefont {Koelling}}, \bibinfo {author} {\bibfnamefont
  {Marcel~A.}\ \bibnamefont {Verheijen}}, \bibinfo {author} {\bibfnamefont
  {Mihir}\ \bibnamefont {Pendharkar}}, \bibinfo {author} {\bibfnamefont
  {Daniel~J.}\ \bibnamefont {Pennachio}}, \bibinfo {author} {\bibfnamefont
  {Borzoyeh}\ \bibnamefont {Shojaei}}, \bibinfo {author} {\bibfnamefont
  {Joon~Sue}\ \bibnamefont {Lee}}, \bibinfo {author} {\bibfnamefont {Chris~J.}\
  \bibnamefont {Palmstrøm}}, \bibinfo {author} {\bibfnamefont {Erik P. A.~M.}\
  \bibnamefont {Bakkers}}, \bibinfo {author} {\bibfnamefont {S.~Das}\
  \bibnamefont {Sarma}}, \ and\ \bibinfo {author} {\bibfnamefont {Leo~P.}\
  \bibnamefont {Kouwenhoven}},\ }\bibfield  {title} {\enquote {\bibinfo {title}
  {Quantized {M}ajorana conductance},}\ }\href {\doibase 10.1038/nature26142}
  {\bibfield  {journal} {\bibinfo  {journal} {Nature}\ }\textbf {\bibinfo
  {volume} {556}},\ \bibinfo {pages} {74} (\bibinfo {year}
  {2018}{\natexlab{b}})}\BibitemShut {NoStop}%
\bibitem [{\citenamefont {Fornieri}\ \emph {et~al.}(2019)\citenamefont
  {Fornieri}, \citenamefont {Whiticar}, \citenamefont {Setiawan}, \citenamefont
  {Portol{\'e}s}, \citenamefont {Drachmann}, \citenamefont {Keselman},
  \citenamefont {Gronin}, \citenamefont {Thomas}, \citenamefont {Wang},
  \citenamefont {Kallaher}, \citenamefont {Gardner}, \citenamefont {Berg},
  \citenamefont {Manfra}, \citenamefont {Stern}, \citenamefont {Marcus},\ and\
  \citenamefont {Nichele}}]{Fornieri2019}%
  \BibitemOpen
  \bibfield  {author} {\bibinfo {author} {\bibfnamefont {Antonio}\ \bibnamefont
  {Fornieri}}, \bibinfo {author} {\bibfnamefont {Alexander~M.}\ \bibnamefont
  {Whiticar}}, \bibinfo {author} {\bibfnamefont {F.}~\bibnamefont {Setiawan}},
  \bibinfo {author} {\bibfnamefont {El{\'i}as}\ \bibnamefont {Portol{\'e}s}},
  \bibinfo {author} {\bibfnamefont {Asbj{\o}rn C.~C.}\ \bibnamefont
  {Drachmann}}, \bibinfo {author} {\bibfnamefont {Anna}\ \bibnamefont
  {Keselman}}, \bibinfo {author} {\bibfnamefont {Sergei}\ \bibnamefont
  {Gronin}}, \bibinfo {author} {\bibfnamefont {Candice}\ \bibnamefont
  {Thomas}}, \bibinfo {author} {\bibfnamefont {Tian}\ \bibnamefont {Wang}},
  \bibinfo {author} {\bibfnamefont {Ray}\ \bibnamefont {Kallaher}}, \bibinfo
  {author} {\bibfnamefont {Geoffrey~C.}\ \bibnamefont {Gardner}}, \bibinfo
  {author} {\bibfnamefont {Erez}\ \bibnamefont {Berg}}, \bibinfo {author}
  {\bibfnamefont {Michael~J.}\ \bibnamefont {Manfra}}, \bibinfo {author}
  {\bibfnamefont {Ady}\ \bibnamefont {Stern}}, \bibinfo {author} {\bibfnamefont
  {Charles~M.}\ \bibnamefont {Marcus}}, \ and\ \bibinfo {author} {\bibfnamefont
  {Fabrizio}\ \bibnamefont {Nichele}},\ }\bibfield  {title} {\enquote {\bibinfo
  {title} {Evidence of topological superconductivity in planar {J}osephson
  junctions},}\ }\href {\doibase 10.1038/s41586-019-1068-8} {\bibfield
  {journal} {\bibinfo  {journal} {Nature}\ }\textbf {\bibinfo {volume} {569}},\
  \bibinfo {pages} {89--92} (\bibinfo {year} {2019})}\BibitemShut {NoStop}%
\bibitem [{\citenamefont {Jiang}\ \emph {et~al.}(2019)\citenamefont {Jiang},
  \citenamefont {Dai},\ and\ \citenamefont {Wang}}]{Jiang2019vortex}%
  \BibitemOpen
  \bibfield  {author} {\bibinfo {author} {\bibfnamefont {Kun}\ \bibnamefont
  {Jiang}}, \bibinfo {author} {\bibfnamefont {Xi}~\bibnamefont {Dai}}, \ and\
  \bibinfo {author} {\bibfnamefont {Ziqiang}\ \bibnamefont {Wang}},\ }\bibfield
   {title} {\enquote {\bibinfo {title} {Quantum anomalous vortex and {M}ajorana
  zero mode in iron-based superconductor {F}e({T}e,{S}e)},}\ }\href {\doibase
  10.1103/PhysRevX.9.011033} {\bibfield  {journal} {\bibinfo  {journal} {Phys.
  Rev. X}\ }\textbf {\bibinfo {volume} {9}},\ \bibinfo {pages} {011033}
  (\bibinfo {year} {2019})}\BibitemShut {NoStop}%
\bibitem [{\citenamefont {Peng}\ \emph {et~al.}(2019)\citenamefont {Peng},
  \citenamefont {Li}, \citenamefont {Wu}, \citenamefont {Deng}, \citenamefont
  {Shi}, \citenamefont {Fan}, \citenamefont {Li}, \citenamefont {Huang},
  \citenamefont {Qian}, \citenamefont {Richard}, \citenamefont {Hu},
  \citenamefont {Pan}, \citenamefont {Mao}, \citenamefont {Sun},\ and\
  \citenamefont {Ding}}]{Peng2019iron}%
  \BibitemOpen
  \bibfield  {author} {\bibinfo {author} {\bibfnamefont {X.-L.}\ \bibnamefont
  {Peng}}, \bibinfo {author} {\bibfnamefont {Y.}~\bibnamefont {Li}}, \bibinfo
  {author} {\bibfnamefont {X.-X.}\ \bibnamefont {Wu}}, \bibinfo {author}
  {\bibfnamefont {H.-B.}\ \bibnamefont {Deng}}, \bibinfo {author}
  {\bibfnamefont {X.}~\bibnamefont {Shi}}, \bibinfo {author} {\bibfnamefont
  {W.-H.}\ \bibnamefont {Fan}}, \bibinfo {author} {\bibfnamefont
  {M.}~\bibnamefont {Li}}, \bibinfo {author} {\bibfnamefont {Y.-B.}\
  \bibnamefont {Huang}}, \bibinfo {author} {\bibfnamefont {T.}~\bibnamefont
  {Qian}}, \bibinfo {author} {\bibfnamefont {P.}~\bibnamefont {Richard}},
  \bibinfo {author} {\bibfnamefont {J.-P.}\ \bibnamefont {Hu}}, \bibinfo
  {author} {\bibfnamefont {S.-H.}\ \bibnamefont {Pan}}, \bibinfo {author}
  {\bibfnamefont {H.-Q.}\ \bibnamefont {Mao}}, \bibinfo {author} {\bibfnamefont
  {Y.-J.}\ \bibnamefont {Sun}}, \ and\ \bibinfo {author} {\bibfnamefont
  {H.}~\bibnamefont {Ding}},\ }\bibfield  {title} {\enquote {\bibinfo {title}
  {Observation of topological transition in high-${T}_{c}$ superconducting
  monolayer \ce{FeTe_{1-x}Se_{x}} films on \ce{SrTiO_{3}(001)}},}\ }\href
  {\doibase 10.1103/PhysRevB.100.155134} {\bibfield  {journal} {\bibinfo
  {journal} {Phys. Rev. B}\ }\textbf {\bibinfo {volume} {100}},\ \bibinfo
  {pages} {155134} (\bibinfo {year} {2019})}\BibitemShut {NoStop}%
\bibitem [{\citenamefont {Chiu}\ \emph {et~al.}(2020)\citenamefont {Chiu},
  \citenamefont {Machida}, \citenamefont {Huang}, \citenamefont {Hanaguri},\
  and\ \citenamefont {Zhang}}]{Chiu2020vortex}%
  \BibitemOpen
  \bibfield  {author} {\bibinfo {author} {\bibfnamefont {Ching-Kai}\
  \bibnamefont {Chiu}}, \bibinfo {author} {\bibfnamefont {T}~\bibnamefont
  {Machida}}, \bibinfo {author} {\bibfnamefont {Yingyi}\ \bibnamefont {Huang}},
  \bibinfo {author} {\bibfnamefont {T}~\bibnamefont {Hanaguri}}, \ and\
  \bibinfo {author} {\bibfnamefont {Fu-Chun}\ \bibnamefont {Zhang}},\
  }\bibfield  {title} {\enquote {\bibinfo {title} {Scalable {M}ajorana vortex
  modes in iron-based superconductors},}\ }\href
  {https://advances.sciencemag.org/content/6/9/eaay0443} {\bibfield  {journal}
  {\bibinfo  {journal} {Science Advances}\ }\textbf {\bibinfo {volume} {6}},\
  \bibinfo {pages} {eaay0443} (\bibinfo {year} {2020})}\BibitemShut {NoStop}%
\bibitem [{\citenamefont {Wang}\ \emph {et~al.}(2020)\citenamefont {Wang},
  \citenamefont {Rodriguez}, \citenamefont {Jiao}, \citenamefont {Howard},
  \citenamefont {Graham}, \citenamefont {Gu}, \citenamefont {Hughes},
  \citenamefont {Morr},\ and\ \citenamefont {Madhavan}}]{Wang2020helical}%
  \BibitemOpen
  \bibfield  {author} {\bibinfo {author} {\bibfnamefont {Zhenyu}\ \bibnamefont
  {Wang}}, \bibinfo {author} {\bibfnamefont {Jorge~Olivares}\ \bibnamefont
  {Rodriguez}}, \bibinfo {author} {\bibfnamefont {Lin}\ \bibnamefont {Jiao}},
  \bibinfo {author} {\bibfnamefont {Sean}\ \bibnamefont {Howard}}, \bibinfo
  {author} {\bibfnamefont {Martin}\ \bibnamefont {Graham}}, \bibinfo {author}
  {\bibfnamefont {G.~D.}\ \bibnamefont {Gu}}, \bibinfo {author} {\bibfnamefont
  {Taylor~L.}\ \bibnamefont {Hughes}}, \bibinfo {author} {\bibfnamefont
  {Dirk~K.}\ \bibnamefont {Morr}}, \ and\ \bibinfo {author} {\bibfnamefont
  {Vidya}\ \bibnamefont {Madhavan}},\ }\bibfield  {title} {\enquote {\bibinfo
  {title} {Evidence for dispersing 1{D} {M}ajorana channels in an iron-based
  superconductor},}\ }\href {\doibase 10.1126/science.aaw8419} {\bibfield
  {journal} {\bibinfo  {journal} {Science}\ }\textbf {\bibinfo {volume}
  {367}},\ \bibinfo {pages} {104--108} (\bibinfo {year} {2020})}\BibitemShut
  {NoStop}%
\bibitem [{\citenamefont {Chen}\ \emph {et~al.}(2020)\citenamefont {Chen},
  \citenamefont {Jiang}, \citenamefont {Zhang}, \citenamefont {Liu},
  \citenamefont {Liu}, \citenamefont {Wang},\ and\ \citenamefont
  {Wang}}]{Chen2020MZM}%
  \BibitemOpen
  \bibfield  {author} {\bibinfo {author} {\bibfnamefont {Cheng}\ \bibnamefont
  {Chen}}, \bibinfo {author} {\bibfnamefont {Kun}\ \bibnamefont {Jiang}},
  \bibinfo {author} {\bibfnamefont {Yi}~\bibnamefont {Zhang}}, \bibinfo
  {author} {\bibfnamefont {Chaofei}\ \bibnamefont {Liu}}, \bibinfo {author}
  {\bibfnamefont {Yi}~\bibnamefont {Liu}}, \bibinfo {author} {\bibfnamefont
  {Ziqiang}\ \bibnamefont {Wang}}, \ and\ \bibinfo {author} {\bibfnamefont
  {Jian}\ \bibnamefont {Wang}},\ }\bibfield  {title} {\enquote {\bibinfo
  {title} {Atomic line defects and zero-energy end states in monolayer
  {F}e({T}e,{S}e) high-temperature superconductors},}\ }\href {\doibase
  10.1038/s41567-020-0813-0} {\bibfield  {journal} {\bibinfo  {journal} {Nature
  Physics}\ }\textbf {\bibinfo {volume} {16}},\ \bibinfo {pages} {536--540}
  (\bibinfo {year} {2020})}\BibitemShut {NoStop}%
\bibitem [{\citenamefont {Zhang}\ \emph {et~al.}(2021)\citenamefont {Zhang},
  \citenamefont {Jiang}, \citenamefont {Zhang}, \citenamefont {Wang},\ and\
  \citenamefont {Wang}}]{PhysRevX.11.011041}%
  \BibitemOpen
  \bibfield  {author} {\bibinfo {author} {\bibfnamefont {Yi}~\bibnamefont
  {Zhang}}, \bibinfo {author} {\bibfnamefont {Kun}\ \bibnamefont {Jiang}},
  \bibinfo {author} {\bibfnamefont {Fuchun}\ \bibnamefont {Zhang}}, \bibinfo
  {author} {\bibfnamefont {Jian}\ \bibnamefont {Wang}}, \ and\ \bibinfo
  {author} {\bibfnamefont {Ziqiang}\ \bibnamefont {Wang}},\ }\bibfield  {title}
  {\enquote {\bibinfo {title} {Atomic line defects and topological
  superconductivity in unconventional superconductors},}\ }\href {\doibase
  10.1103/PhysRevX.11.011041} {\bibfield  {journal} {\bibinfo  {journal} {Phys.
  Rev. X}\ }\textbf {\bibinfo {volume} {11}},\ \bibinfo {pages} {011041}
  (\bibinfo {year} {2021})}\BibitemShut {NoStop}%
\bibitem [{\citenamefont {Wu}\ \emph {et~al.}(2020{\natexlab{a}})\citenamefont
  {Wu}, \citenamefont {Yin}, \citenamefont {Liu},\ and\ \citenamefont
  {Hu}}]{Wu2020ironline}%
  \BibitemOpen
  \bibfield  {author} {\bibinfo {author} {\bibfnamefont {Xianxin}\ \bibnamefont
  {Wu}}, \bibinfo {author} {\bibfnamefont {Jia-Xin}\ \bibnamefont {Yin}},
  \bibinfo {author} {\bibfnamefont {Chao-Xing}\ \bibnamefont {Liu}}, \ and\
  \bibinfo {author} {\bibfnamefont {Jiangping}\ \bibnamefont {Hu}},\ }\bibfield
   {title} {\enquote {\bibinfo {title} {Topological magnetic line defects in
  {F}e({T}e, {S}e) high-temperature superconductors},}\ }\href
  {https://arxiv.org/abs/2004.05848} {\bibfield  {journal} {\bibinfo  {journal}
  {arXiv preprint arXiv:2004.05848}\ } (\bibinfo {year}
  {2020}{\natexlab{a}})}\BibitemShut {NoStop}%
\bibitem [{\citenamefont {Ghazaryan}\ \emph {et~al.}(2020)\citenamefont
  {Ghazaryan}, \citenamefont {Lopes}, \citenamefont {Hosur}, \citenamefont
  {Gilbert},\ and\ \citenamefont {Ghaemi}}]{Ghazaryan2020vortex}%
  \BibitemOpen
  \bibfield  {author} {\bibinfo {author} {\bibfnamefont {Areg}\ \bibnamefont
  {Ghazaryan}}, \bibinfo {author} {\bibfnamefont {P.~L.~S.}\ \bibnamefont
  {Lopes}}, \bibinfo {author} {\bibfnamefont {Pavan}\ \bibnamefont {Hosur}},
  \bibinfo {author} {\bibfnamefont {Matthew~J.}\ \bibnamefont {Gilbert}}, \
  and\ \bibinfo {author} {\bibfnamefont {Pouyan}\ \bibnamefont {Ghaemi}},\
  }\bibfield  {title} {\enquote {\bibinfo {title} {Effect of zeeman coupling on
  the {M}ajorana vortex modes in iron-based topological superconductors},}\
  }\href {\doibase 10.1103/PhysRevB.101.020504} {\bibfield  {journal} {\bibinfo
   {journal} {Phys. Rev. B}\ }\textbf {\bibinfo {volume} {101}},\ \bibinfo
  {pages} {020504} (\bibinfo {year} {2020})}\BibitemShut {NoStop}%
\bibitem [{\citenamefont {Qin}\ \emph {et~al.}(2019{\natexlab{a}})\citenamefont
  {Qin}, \citenamefont {Hu}, \citenamefont {Le}, \citenamefont {Zeng},
  \citenamefont {Zhang}, \citenamefont {Fang},\ and\ \citenamefont
  {Hu}}]{Qin2019vortex}%
  \BibitemOpen
  \bibfield  {author} {\bibinfo {author} {\bibfnamefont {Shengshan}\
  \bibnamefont {Qin}}, \bibinfo {author} {\bibfnamefont {Lunhui}\ \bibnamefont
  {Hu}}, \bibinfo {author} {\bibfnamefont {Congcong}\ \bibnamefont {Le}},
  \bibinfo {author} {\bibfnamefont {Jinfeng}\ \bibnamefont {Zeng}}, \bibinfo
  {author} {\bibfnamefont {Fu-chun}\ \bibnamefont {Zhang}}, \bibinfo {author}
  {\bibfnamefont {Chen}\ \bibnamefont {Fang}}, \ and\ \bibinfo {author}
  {\bibfnamefont {Jiangping}\ \bibnamefont {Hu}},\ }\bibfield  {title}
  {\enquote {\bibinfo {title} {Quasi-1{D} topological nodal vortex line phase
  in doped superconducting 3{D} {D}irac semimetals},}\ }\href {\doibase
  10.1103/PhysRevLett.123.027003} {\bibfield  {journal} {\bibinfo  {journal}
  {Phys. Rev. Lett.}\ }\textbf {\bibinfo {volume} {123}},\ \bibinfo {pages}
  {027003} (\bibinfo {year} {2019}{\natexlab{a}})}\BibitemShut {NoStop}%
\bibitem [{\citenamefont {Qin}\ \emph {et~al.}(2019{\natexlab{b}})\citenamefont
  {Qin}, \citenamefont {Hu}, \citenamefont {Wu}, \citenamefont {Dai},
  \citenamefont {Fang}, \citenamefont {Zhang},\ and\ \citenamefont
  {Hu}}]{Qin2019vortex2}%
  \BibitemOpen
  \bibfield  {author} {\bibinfo {author} {\bibfnamefont {Shengshan}\
  \bibnamefont {Qin}}, \bibinfo {author} {\bibfnamefont {Lunhui}\ \bibnamefont
  {Hu}}, \bibinfo {author} {\bibfnamefont {Xianxin}\ \bibnamefont {Wu}},
  \bibinfo {author} {\bibfnamefont {Xia}\ \bibnamefont {Dai}}, \bibinfo
  {author} {\bibfnamefont {Chen}\ \bibnamefont {Fang}}, \bibinfo {author}
  {\bibfnamefont {Fu-Chun}\ \bibnamefont {Zhang}}, \ and\ \bibinfo {author}
  {\bibfnamefont {Jiangping}\ \bibnamefont {Hu}},\ }\bibfield  {title}
  {\enquote {\bibinfo {title} {Topological vortex phase transitions in
  iron-based superconductors},}\ }\href {\doibase
  https://doi.org/10.1016/j.scib.2019.07.011} {\bibfield  {journal} {\bibinfo
  {journal} {Science Bulletin}\ }\textbf {\bibinfo {volume} {64}},\ \bibinfo
  {pages} {1207 -- 1214} (\bibinfo {year} {2019}{\natexlab{b}})}\BibitemShut
  {NoStop}%
\bibitem [{\citenamefont {K\"onig}\ and\ \citenamefont
  {Coleman}(2019)}]{Konig2019votex}%
  \BibitemOpen
  \bibfield  {author} {\bibinfo {author} {\bibfnamefont {Elio~J.}\ \bibnamefont
  {K\"onig}}\ and\ \bibinfo {author} {\bibfnamefont {Piers}\ \bibnamefont
  {Coleman}},\ }\bibfield  {title} {\enquote {\bibinfo {title}
  {Crystalline-symmetry-protected helical {M}ajorana modes in the {I}ron
  {P}nictides},}\ }\href {\doibase 10.1103/PhysRevLett.122.207001} {\bibfield
  {journal} {\bibinfo  {journal} {Phys. Rev. Lett.}\ }\textbf {\bibinfo
  {volume} {122}},\ \bibinfo {pages} {207001} (\bibinfo {year}
  {2019})}\BibitemShut {NoStop}%
\bibitem [{\citenamefont {Yan}\ \emph {et~al.}(2020)\citenamefont {Yan},
  \citenamefont {Wu},\ and\ \citenamefont {Huang}}]{Yan2020vortex}%
  \BibitemOpen
  \bibfield  {author} {\bibinfo {author} {\bibfnamefont {Zhongbo}\ \bibnamefont
  {Yan}}, \bibinfo {author} {\bibfnamefont {Zhigang}\ \bibnamefont {Wu}}, \
  and\ \bibinfo {author} {\bibfnamefont {Wen}\ \bibnamefont {Huang}},\
  }\bibfield  {title} {\enquote {\bibinfo {title} {Vortex end {M}ajorana zero
  modes in superconducting {D}irac and {W}eyl semimetals},}\ }\href {\doibase
  10.1103/PhysRevLett.124.257001} {\bibfield  {journal} {\bibinfo  {journal}
  {Phys. Rev. Lett.}\ }\textbf {\bibinfo {volume} {124}},\ \bibinfo {pages}
  {257001} (\bibinfo {year} {2020})}\BibitemShut {NoStop}%
\bibitem [{\citenamefont {Chen}\ \emph {et~al.}(2018)\citenamefont {Chen},
  \citenamefont {Chen}, \citenamefont {Yang}, \citenamefont {Du}, \citenamefont
  {Zhu}, \citenamefont {Wang},\ and\ \citenamefont {Wen}}]{Chen2018iron}%
  \BibitemOpen
  \bibfield  {author} {\bibinfo {author} {\bibfnamefont {Mingyang}\
  \bibnamefont {Chen}}, \bibinfo {author} {\bibfnamefont {Xiaoyu}\ \bibnamefont
  {Chen}}, \bibinfo {author} {\bibfnamefont {Huan}\ \bibnamefont {Yang}},
  \bibinfo {author} {\bibfnamefont {Zengyi}\ \bibnamefont {Du}}, \bibinfo
  {author} {\bibfnamefont {Xiyu}\ \bibnamefont {Zhu}}, \bibinfo {author}
  {\bibfnamefont {Enyu}\ \bibnamefont {Wang}}, \ and\ \bibinfo {author}
  {\bibfnamefont {Hai-Hu}\ \bibnamefont {Wen}},\ }\bibfield  {title} {\enquote
  {\bibinfo {title} {Discrete energy levels of {C}aroli-de {G}ennes-{M}atricon
  states in quantum limit in \ce{FeTe_{0.55}Se_{0.45}}},}\ }\href {\doibase
  10.1038/s41467-018-03404-8} {\bibfield  {journal} {\bibinfo  {journal}
  {Nature Communications}\ }\textbf {\bibinfo {volume} {9}},\ \bibinfo {pages}
  {970} (\bibinfo {year} {2018})}\BibitemShut {NoStop}%
\bibitem [{\citenamefont {Benalcazar}\ \emph {et~al.}(2017)\citenamefont
  {Benalcazar}, \citenamefont {Bernevig},\ and\ \citenamefont
  {Hughes}}]{Benalcazar2017}%
  \BibitemOpen
  \bibfield  {author} {\bibinfo {author} {\bibfnamefont {Wladimir~A.}\
  \bibnamefont {Benalcazar}}, \bibinfo {author} {\bibfnamefont {B.~Andrei}\
  \bibnamefont {Bernevig}}, \ and\ \bibinfo {author} {\bibfnamefont
  {Taylor~L.}\ \bibnamefont {Hughes}},\ }\bibfield  {title} {\enquote {\bibinfo
  {title} {Quantized electric multipole insulators},}\ }\href {\doibase
  10.1126/science.aah6442} {\bibfield  {journal} {\bibinfo  {journal}
  {Science}\ }\textbf {\bibinfo {volume} {357}},\ \bibinfo {pages} {61--66}
  (\bibinfo {year} {2017})}\BibitemShut {NoStop}%
\bibitem [{\citenamefont {Schindler}\ \emph {et~al.}(2018)\citenamefont
  {Schindler}, \citenamefont {Cook}, \citenamefont {Vergniory}, \citenamefont
  {Wang}, \citenamefont {Parkin}, \citenamefont {Bernevig},\ and\ \citenamefont
  {Neupert}}]{Schindler2018}%
  \BibitemOpen
  \bibfield  {author} {\bibinfo {author} {\bibfnamefont {Frank}\ \bibnamefont
  {Schindler}}, \bibinfo {author} {\bibfnamefont {Ashley~M.}\ \bibnamefont
  {Cook}}, \bibinfo {author} {\bibfnamefont {Maia~G.}\ \bibnamefont
  {Vergniory}}, \bibinfo {author} {\bibfnamefont {Zhijun}\ \bibnamefont
  {Wang}}, \bibinfo {author} {\bibfnamefont {Stuart S.~P.}\ \bibnamefont
  {Parkin}}, \bibinfo {author} {\bibfnamefont {B.~Andrei}\ \bibnamefont
  {Bernevig}}, \ and\ \bibinfo {author} {\bibfnamefont {Titus}\ \bibnamefont
  {Neupert}},\ }\bibfield  {title} {\enquote {\bibinfo {title} {Higher-order
  topological insulators},}\ }\href {\doibase 10.1126/sciadv.aat0346}
  {\bibfield  {journal} {\bibinfo  {journal} {Science Advances}\ }\textbf
  {\bibinfo {volume} {4}} (\bibinfo {year} {2018}),\
  10.1126/sciadv.aat0346}\BibitemShut {NoStop}%
\bibitem [{\citenamefont {Langbehn}\ \emph {et~al.}(2017)\citenamefont
  {Langbehn}, \citenamefont {Peng}, \citenamefont {Trifunovic}, \citenamefont
  {von Oppen},\ and\ \citenamefont {Brouwer}}]{Langbehn2017}%
  \BibitemOpen
  \bibfield  {author} {\bibinfo {author} {\bibfnamefont {Josias}\ \bibnamefont
  {Langbehn}}, \bibinfo {author} {\bibfnamefont {Yang}\ \bibnamefont {Peng}},
  \bibinfo {author} {\bibfnamefont {Luka}\ \bibnamefont {Trifunovic}}, \bibinfo
  {author} {\bibfnamefont {Felix}\ \bibnamefont {von Oppen}}, \ and\ \bibinfo
  {author} {\bibfnamefont {Piet~W.}\ \bibnamefont {Brouwer}},\ }\bibfield
  {title} {\enquote {\bibinfo {title} {Reflection-symmetric second-order
  topological insulators and superconductors},}\ }\href {\doibase
  10.1103/PhysRevLett.119.246401} {\bibfield  {journal} {\bibinfo  {journal}
  {Phys. Rev. Lett.}\ }\textbf {\bibinfo {volume} {119}},\ \bibinfo {pages}
  {246401} (\bibinfo {year} {2017})}\BibitemShut {NoStop}%
\bibitem [{\citenamefont {Shapourian}\ \emph {et~al.}(2018)\citenamefont
  {Shapourian}, \citenamefont {Wang},\ and\ \citenamefont
  {Ryu}}]{Shapourian2018SOTSC}%
  \BibitemOpen
  \bibfield  {author} {\bibinfo {author} {\bibfnamefont {Hassan}\ \bibnamefont
  {Shapourian}}, \bibinfo {author} {\bibfnamefont {Yuxuan}\ \bibnamefont
  {Wang}}, \ and\ \bibinfo {author} {\bibfnamefont {Shinsei}\ \bibnamefont
  {Ryu}},\ }\bibfield  {title} {\enquote {\bibinfo {title} {Topological
  crystalline superconductivity and second-order topological superconductivity
  in nodal-loop materials},}\ }\href {\doibase 10.1103/PhysRevB.97.094508}
  {\bibfield  {journal} {\bibinfo  {journal} {Phys. Rev. B}\ }\textbf {\bibinfo
  {volume} {97}},\ \bibinfo {pages} {094508} (\bibinfo {year}
  {2018})}\BibitemShut {NoStop}%
\bibitem [{\citenamefont {Khalaf}(2018)}]{Khalaf2018}%
  \BibitemOpen
  \bibfield  {author} {\bibinfo {author} {\bibfnamefont {Eslam}\ \bibnamefont
  {Khalaf}},\ }\bibfield  {title} {\enquote {\bibinfo {title} {Higher-order
  topological insulators and superconductors protected by inversion
  symmetry},}\ }\href {\doibase 10.1103/PhysRevB.97.205136} {\bibfield
  {journal} {\bibinfo  {journal} {Phys. Rev. B}\ }\textbf {\bibinfo {volume}
  {97}},\ \bibinfo {pages} {205136} (\bibinfo {year} {2018})}\BibitemShut
  {NoStop}%
\bibitem [{\citenamefont {Geier}\ \emph {et~al.}(2018)\citenamefont {Geier},
  \citenamefont {Trifunovic}, \citenamefont {Hoskam},\ and\ \citenamefont
  {Brouwer}}]{Geier2018}%
  \BibitemOpen
  \bibfield  {author} {\bibinfo {author} {\bibfnamefont {Max}\ \bibnamefont
  {Geier}}, \bibinfo {author} {\bibfnamefont {Luka}\ \bibnamefont
  {Trifunovic}}, \bibinfo {author} {\bibfnamefont {Max}\ \bibnamefont
  {Hoskam}}, \ and\ \bibinfo {author} {\bibfnamefont {Piet~W.}\ \bibnamefont
  {Brouwer}},\ }\bibfield  {title} {\enquote {\bibinfo {title} {Second-order
  topological insulators and superconductors with an order-two crystalline
  symmetry},}\ }\href {\doibase 10.1103/PhysRevB.97.205135} {\bibfield
  {journal} {\bibinfo  {journal} {Phys. Rev. B}\ }\textbf {\bibinfo {volume}
  {97}},\ \bibinfo {pages} {205135} (\bibinfo {year} {2018})}\BibitemShut
  {NoStop}%
\bibitem [{\citenamefont {Zhu}(2018)}]{Zhu2018hosc}%
  \BibitemOpen
  \bibfield  {author} {\bibinfo {author} {\bibfnamefont {Xiaoyu}\ \bibnamefont
  {Zhu}},\ }\bibfield  {title} {\enquote {\bibinfo {title} {Tunable {M}ajorana
  corner states in a two-dimensional second-order topological superconductor
  induced by magnetic fields},}\ }\href {\doibase 10.1103/PhysRevB.97.205134}
  {\bibfield  {journal} {\bibinfo  {journal} {Phys. Rev. B}\ }\textbf {\bibinfo
  {volume} {97}},\ \bibinfo {pages} {205134} (\bibinfo {year}
  {2018})}\BibitemShut {NoStop}%
\bibitem [{\citenamefont {Yan}\ \emph {et~al.}(2018)\citenamefont {Yan},
  \citenamefont {Song},\ and\ \citenamefont {Wang}}]{Yan2018hosc}%
  \BibitemOpen
  \bibfield  {author} {\bibinfo {author} {\bibfnamefont {Zhongbo}\ \bibnamefont
  {Yan}}, \bibinfo {author} {\bibfnamefont {Fei}\ \bibnamefont {Song}}, \ and\
  \bibinfo {author} {\bibfnamefont {Zhong}\ \bibnamefont {Wang}},\ }\bibfield
  {title} {\enquote {\bibinfo {title} {Majorana corner modes in a
  high-temperature platform},}\ }\href {\doibase
  10.1103/PhysRevLett.121.096803} {\bibfield  {journal} {\bibinfo  {journal}
  {Phys. Rev. Lett.}\ }\textbf {\bibinfo {volume} {121}},\ \bibinfo {pages}
  {096803} (\bibinfo {year} {2018})}\BibitemShut {NoStop}%
\bibitem [{\citenamefont {Wang}\ \emph
  {et~al.}(2018{\natexlab{b}})\citenamefont {Wang}, \citenamefont {Liu},
  \citenamefont {Lu},\ and\ \citenamefont {Zhang}}]{Wang2018hosc}%
  \BibitemOpen
  \bibfield  {author} {\bibinfo {author} {\bibfnamefont {Qiyue}\ \bibnamefont
  {Wang}}, \bibinfo {author} {\bibfnamefont {Cheng-Cheng}\ \bibnamefont {Liu}},
  \bibinfo {author} {\bibfnamefont {Yuan-Ming}\ \bibnamefont {Lu}}, \ and\
  \bibinfo {author} {\bibfnamefont {Fan}\ \bibnamefont {Zhang}},\ }\bibfield
  {title} {\enquote {\bibinfo {title} {High-temperature {M}ajorana corner
  states},}\ }\href {\doibase 10.1103/PhysRevLett.121.186801} {\bibfield
  {journal} {\bibinfo  {journal} {Phys. Rev. Lett.}\ }\textbf {\bibinfo
  {volume} {121}},\ \bibinfo {pages} {186801} (\bibinfo {year}
  {2018}{\natexlab{b}})}\BibitemShut {NoStop}%
\bibitem [{\citenamefont {Wang}\ \emph
  {et~al.}(2018{\natexlab{c}})\citenamefont {Wang}, \citenamefont {Lin},\ and\
  \citenamefont {Hughes}}]{Wangyuxuan2018hosc}%
  \BibitemOpen
  \bibfield  {author} {\bibinfo {author} {\bibfnamefont {Yuxuan}\ \bibnamefont
  {Wang}}, \bibinfo {author} {\bibfnamefont {Mao}\ \bibnamefont {Lin}}, \ and\
  \bibinfo {author} {\bibfnamefont {Taylor~L.}\ \bibnamefont {Hughes}},\
  }\bibfield  {title} {\enquote {\bibinfo {title} {Weak-pairing higher order
  topological superconductors},}\ }\href {\doibase 10.1103/PhysRevB.98.165144}
  {\bibfield  {journal} {\bibinfo  {journal} {Phys. Rev. B}\ }\textbf {\bibinfo
  {volume} {98}},\ \bibinfo {pages} {165144} (\bibinfo {year}
  {2018}{\natexlab{c}})}\BibitemShut {NoStop}%
\bibitem [{\citenamefont {Hsu}\ \emph {et~al.}(2018)\citenamefont {Hsu},
  \citenamefont {Stano}, \citenamefont {Klinovaja},\ and\ \citenamefont
  {Loss}}]{Hsu2018hosc}%
  \BibitemOpen
  \bibfield  {author} {\bibinfo {author} {\bibfnamefont {Chen-Hsuan}\
  \bibnamefont {Hsu}}, \bibinfo {author} {\bibfnamefont {Peter}\ \bibnamefont
  {Stano}}, \bibinfo {author} {\bibfnamefont {Jelena}\ \bibnamefont
  {Klinovaja}}, \ and\ \bibinfo {author} {\bibfnamefont {Daniel}\ \bibnamefont
  {Loss}},\ }\bibfield  {title} {\enquote {\bibinfo {title} {Majorana {K}ramers
  pairs in higher-order topological insulators},}\ }\href {\doibase
  10.1103/PhysRevLett.121.196801} {\bibfield  {journal} {\bibinfo  {journal}
  {Phys. Rev. Lett.}\ }\textbf {\bibinfo {volume} {121}},\ \bibinfo {pages}
  {196801} (\bibinfo {year} {2018})}\BibitemShut {NoStop}%
\bibitem [{\citenamefont {Liu}\ \emph {et~al.}(2018{\natexlab{b}})\citenamefont
  {Liu}, \citenamefont {He},\ and\ \citenamefont {Nori}}]{Liu2018hosc}%
  \BibitemOpen
  \bibfield  {author} {\bibinfo {author} {\bibfnamefont {Tao}\ \bibnamefont
  {Liu}}, \bibinfo {author} {\bibfnamefont {James~Jun}\ \bibnamefont {He}}, \
  and\ \bibinfo {author} {\bibfnamefont {Franco}\ \bibnamefont {Nori}},\
  }\bibfield  {title} {\enquote {\bibinfo {title} {Majorana corner states in a
  two-dimensional magnetic topological insulator on a high-temperature
  superconductor},}\ }\href {\doibase 10.1103/PhysRevB.98.245413} {\bibfield
  {journal} {\bibinfo  {journal} {Phys. Rev. B}\ }\textbf {\bibinfo {volume}
  {98}},\ \bibinfo {pages} {245413} (\bibinfo {year}
  {2018}{\natexlab{b}})}\BibitemShut {NoStop}%
\bibitem [{\citenamefont {Wu}\ \emph {et~al.}(2019{\natexlab{a}})\citenamefont
  {Wu}, \citenamefont {Yan},\ and\ \citenamefont {Huang}}]{Wuzhigang2019hosc}%
  \BibitemOpen
  \bibfield  {author} {\bibinfo {author} {\bibfnamefont {Zhigang}\ \bibnamefont
  {Wu}}, \bibinfo {author} {\bibfnamefont {Zhongbo}\ \bibnamefont {Yan}}, \
  and\ \bibinfo {author} {\bibfnamefont {Wen}\ \bibnamefont {Huang}},\
  }\bibfield  {title} {\enquote {\bibinfo {title} {Higher-order topological
  superconductivity: Possible realization in {F}ermi gases and \ce{Sr2RuO4}},}\
  }\href {\doibase 10.1103/PhysRevB.99.020508} {\bibfield  {journal} {\bibinfo
  {journal} {Phys. Rev. B}\ }\textbf {\bibinfo {volume} {99}},\ \bibinfo
  {pages} {020508} (\bibinfo {year} {2019}{\natexlab{a}})}\BibitemShut
  {NoStop}%
\bibitem [{\citenamefont {Zhang}\ \emph
  {et~al.}(2019{\natexlab{b}})\citenamefont {Zhang}, \citenamefont {Cole},\
  and\ \citenamefont {Das~Sarma}}]{Zhang2019hinge}%
  \BibitemOpen
  \bibfield  {author} {\bibinfo {author} {\bibfnamefont {Rui-Xing}\
  \bibnamefont {Zhang}}, \bibinfo {author} {\bibfnamefont {William~S.}\
  \bibnamefont {Cole}}, \ and\ \bibinfo {author} {\bibfnamefont
  {S.}~\bibnamefont {Das~Sarma}},\ }\bibfield  {title} {\enquote {\bibinfo
  {title} {Helical hinge {M}ajorana modes in iron-based superconductors},}\
  }\href {\doibase 10.1103/PhysRevLett.122.187001} {\bibfield  {journal}
  {\bibinfo  {journal} {Phys. Rev. Lett.}\ }\textbf {\bibinfo {volume} {122}},\
  \bibinfo {pages} {187001} (\bibinfo {year} {2019}{\natexlab{b}})}\BibitemShut
  {NoStop}%
\bibitem [{\citenamefont {Volpez}\ \emph {et~al.}(2019)\citenamefont {Volpez},
  \citenamefont {Loss},\ and\ \citenamefont {Klinovaja}}]{Volpez2019SOTSC}%
  \BibitemOpen
  \bibfield  {author} {\bibinfo {author} {\bibfnamefont {Yanick}\ \bibnamefont
  {Volpez}}, \bibinfo {author} {\bibfnamefont {Daniel}\ \bibnamefont {Loss}}, \
  and\ \bibinfo {author} {\bibfnamefont {Jelena}\ \bibnamefont {Klinovaja}},\
  }\bibfield  {title} {\enquote {\bibinfo {title} {Second-order topological
  superconductivity in $\ensuremath{\pi}$-junction {R}ashba layers},}\ }\href
  {\doibase 10.1103/PhysRevLett.122.126402} {\bibfield  {journal} {\bibinfo
  {journal} {Phys. Rev. Lett.}\ }\textbf {\bibinfo {volume} {122}},\ \bibinfo
  {pages} {126402} (\bibinfo {year} {2019})}\BibitemShut {NoStop}%
\bibitem [{\citenamefont {Zhang}\ \emph
  {et~al.}(2019{\natexlab{c}})\citenamefont {Zhang}, \citenamefont {Cole},
  \citenamefont {Wu},\ and\ \citenamefont {Das~Sarma}}]{Zhang2019hoscb}%
  \BibitemOpen
  \bibfield  {author} {\bibinfo {author} {\bibfnamefont {Rui-Xing}\
  \bibnamefont {Zhang}}, \bibinfo {author} {\bibfnamefont {William~S.}\
  \bibnamefont {Cole}}, \bibinfo {author} {\bibfnamefont {Xianxin}\
  \bibnamefont {Wu}}, \ and\ \bibinfo {author} {\bibfnamefont {S.}~\bibnamefont
  {Das~Sarma}},\ }\bibfield  {title} {\enquote {\bibinfo {title} {Higher-order
  topology and nodal topological superconductivity in {F}e({S}e,{T}e)
  heterostructures},}\ }\href {\doibase 10.1103/PhysRevLett.123.167001}
  {\bibfield  {journal} {\bibinfo  {journal} {Phys. Rev. Lett.}\ }\textbf
  {\bibinfo {volume} {123}},\ \bibinfo {pages} {167001} (\bibinfo {year}
  {2019}{\natexlab{c}})}\BibitemShut {NoStop}%
\bibitem [{\citenamefont {Wu}\ \emph {et~al.}(2019{\natexlab{b}})\citenamefont
  {Wu}, \citenamefont {Liu}, \citenamefont {Thomale},\ and\ \citenamefont
  {Liu}}]{Wu2019hoscb}%
  \BibitemOpen
  \bibfield  {author} {\bibinfo {author} {\bibfnamefont {Xianxin}\ \bibnamefont
  {Wu}}, \bibinfo {author} {\bibfnamefont {Xin}\ \bibnamefont {Liu}}, \bibinfo
  {author} {\bibfnamefont {Ronny}\ \bibnamefont {Thomale}}, \ and\ \bibinfo
  {author} {\bibfnamefont {Chao-Xing}\ \bibnamefont {Liu}},\ }\bibfield
  {title} {\enquote {\bibinfo {title} {{High-$T_c$ superconductor Fe(Se,Te)
  monolayer: An intrinsic, scalable and electrically-tunable Majorana
  platform}},}\ }\href {https://arxiv.org/abs/1905.10648} {\bibfield  {journal}
  {\bibinfo  {journal} {arXiv preprint arXiv:1905.10648}\ } (\bibinfo {year}
  {2019}{\natexlab{b}})}\BibitemShut {NoStop}%
\bibitem [{\citenamefont {Hsu}\ \emph {et~al.}(2020)\citenamefont {Hsu},
  \citenamefont {Cole}, \citenamefont {Zhang},\ and\ \citenamefont
  {Sau}}]{Hsu2019HOSC}%
  \BibitemOpen
  \bibfield  {author} {\bibinfo {author} {\bibfnamefont {Yi-Ting}\ \bibnamefont
  {Hsu}}, \bibinfo {author} {\bibfnamefont {William~S.}\ \bibnamefont {Cole}},
  \bibinfo {author} {\bibfnamefont {Rui-Xing}\ \bibnamefont {Zhang}}, \ and\
  \bibinfo {author} {\bibfnamefont {Jay~D.}\ \bibnamefont {Sau}},\ }\bibfield
  {title} {\enquote {\bibinfo {title} {Inversion-protected higher-order
  topological superconductivity in monolayer \ce{WTe2}},}\ }\href {\doibase
  10.1103/PhysRevLett.125.097001} {\bibfield  {journal} {\bibinfo  {journal}
  {Phys. Rev. Lett.}\ }\textbf {\bibinfo {volume} {125}},\ \bibinfo {pages}
  {097001} (\bibinfo {year} {2020})}\BibitemShut {NoStop}%
\bibitem [{\citenamefont {Zeng}\ \emph {et~al.}(2019)\citenamefont {Zeng},
  \citenamefont {Stanescu}, \citenamefont {Zhang}, \citenamefont {Scarola},\
  and\ \citenamefont {Tewari}}]{Zeng2019mcm}%
  \BibitemOpen
  \bibfield  {author} {\bibinfo {author} {\bibfnamefont {Chuanchang}\
  \bibnamefont {Zeng}}, \bibinfo {author} {\bibfnamefont {T.~D.}\ \bibnamefont
  {Stanescu}}, \bibinfo {author} {\bibfnamefont {Chuanwei}\ \bibnamefont
  {Zhang}}, \bibinfo {author} {\bibfnamefont {V.~W.}\ \bibnamefont {Scarola}},
  \ and\ \bibinfo {author} {\bibfnamefont {Sumanta}\ \bibnamefont {Tewari}},\
  }\bibfield  {title} {\enquote {\bibinfo {title} {Majorana corner modes with
  solitons in an attractive {H}ubbard-{H}ofstadter model of cold atom optical
  lattices},}\ }\href {\doibase 10.1103/PhysRevLett.123.060402} {\bibfield
  {journal} {\bibinfo  {journal} {Phys. Rev. Lett.}\ }\textbf {\bibinfo
  {volume} {123}},\ \bibinfo {pages} {060402} (\bibinfo {year}
  {2019})}\BibitemShut {NoStop}%
\bibitem [{\citenamefont {Bultinck}\ \emph {et~al.}(2019)\citenamefont
  {Bultinck}, \citenamefont {Bernevig},\ and\ \citenamefont
  {Zaletel}}]{Bultinck2019}%
  \BibitemOpen
  \bibfield  {author} {\bibinfo {author} {\bibfnamefont {Nick}\ \bibnamefont
  {Bultinck}}, \bibinfo {author} {\bibfnamefont {B.~Andrei}\ \bibnamefont
  {Bernevig}}, \ and\ \bibinfo {author} {\bibfnamefont {Michael~P.}\
  \bibnamefont {Zaletel}},\ }\bibfield  {title} {\enquote {\bibinfo {title}
  {Three-dimensional superconductors with hybrid higher-order topology},}\
  }\href {\doibase 10.1103/PhysRevB.99.125149} {\bibfield  {journal} {\bibinfo
  {journal} {Phys. Rev. B}\ }\textbf {\bibinfo {volume} {99}},\ \bibinfo
  {pages} {125149} (\bibinfo {year} {2019})}\BibitemShut {NoStop}%
\bibitem [{\citenamefont {Ghorashi}\ \emph {et~al.}(2019)\citenamefont
  {Ghorashi}, \citenamefont {Hu}, \citenamefont {Hughes},\ and\ \citenamefont
  {Rossi}}]{Ghorashi2019}%
  \BibitemOpen
  \bibfield  {author} {\bibinfo {author} {\bibfnamefont {Sayed Ali~Akbar}\
  \bibnamefont {Ghorashi}}, \bibinfo {author} {\bibfnamefont {Xiang}\
  \bibnamefont {Hu}}, \bibinfo {author} {\bibfnamefont {Taylor~L.}\
  \bibnamefont {Hughes}}, \ and\ \bibinfo {author} {\bibfnamefont {Enrico}\
  \bibnamefont {Rossi}},\ }\bibfield  {title} {\enquote {\bibinfo {title}
  {Second-order {D}irac superconductors and magnetic field induced {M}ajorana
  hinge modes},}\ }\href {\doibase 10.1103/PhysRevB.100.020509} {\bibfield
  {journal} {\bibinfo  {journal} {Phys. Rev. B}\ }\textbf {\bibinfo {volume}
  {100}},\ \bibinfo {pages} {020509} (\bibinfo {year} {2019})}\BibitemShut
  {NoStop}%
\bibitem [{\citenamefont {Peng}\ and\ \citenamefont
  {Xu}(2019)}]{Peng2019hinge}%
  \BibitemOpen
  \bibfield  {author} {\bibinfo {author} {\bibfnamefont {Yang}\ \bibnamefont
  {Peng}}\ and\ \bibinfo {author} {\bibfnamefont {Yong}\ \bibnamefont {Xu}},\
  }\bibfield  {title} {\enquote {\bibinfo {title} {Proximity-induced {M}ajorana
  hinge modes in antiferromagnetic topological insulators},}\ }\href {\doibase
  10.1103/PhysRevB.99.195431} {\bibfield  {journal} {\bibinfo  {journal} {Phys.
  Rev. B}\ }\textbf {\bibinfo {volume} {99}},\ \bibinfo {pages} {195431}
  (\bibinfo {year} {2019})}\BibitemShut {NoStop}%
\bibitem [{\citenamefont {Zhu}(2019)}]{Zhu2019mixed}%
  \BibitemOpen
  \bibfield  {author} {\bibinfo {author} {\bibfnamefont {Xiaoyu}\ \bibnamefont
  {Zhu}},\ }\bibfield  {title} {\enquote {\bibinfo {title} {Second-order
  topological superconductors with mixed pairing},}\ }\href {\doibase
  10.1103/PhysRevLett.122.236401} {\bibfield  {journal} {\bibinfo  {journal}
  {Phys. Rev. Lett.}\ }\textbf {\bibinfo {volume} {122}},\ \bibinfo {pages}
  {236401} (\bibinfo {year} {2019})}\BibitemShut {NoStop}%
\bibitem [{\citenamefont {Laubscher}\ \emph {et~al.}(2019)\citenamefont
  {Laubscher}, \citenamefont {Loss},\ and\ \citenamefont
  {Klinovaja}}]{Laubscher2019hosc}%
  \BibitemOpen
  \bibfield  {author} {\bibinfo {author} {\bibfnamefont {Katharina}\
  \bibnamefont {Laubscher}}, \bibinfo {author} {\bibfnamefont {Daniel}\
  \bibnamefont {Loss}}, \ and\ \bibinfo {author} {\bibfnamefont {Jelena}\
  \bibnamefont {Klinovaja}},\ }\bibfield  {title} {\enquote {\bibinfo {title}
  {Fractional topological superconductivity and parafermion corner states},}\
  }\href {\doibase 10.1103/PhysRevResearch.1.032017} {\bibfield  {journal}
  {\bibinfo  {journal} {Phys. Rev. Research}\ }\textbf {\bibinfo {volume}
  {1}},\ \bibinfo {pages} {032017} (\bibinfo {year} {2019})}\BibitemShut
  {NoStop}%
\bibitem [{\citenamefont {Pan}\ \emph {et~al.}(2019)\citenamefont {Pan},
  \citenamefont {Yang}, \citenamefont {Chen}, \citenamefont {Xu}, \citenamefont
  {Liu},\ and\ \citenamefont {Liu}}]{Pan2019SOTSC}%
  \BibitemOpen
  \bibfield  {author} {\bibinfo {author} {\bibfnamefont {Xiao-Hong}\
  \bibnamefont {Pan}}, \bibinfo {author} {\bibfnamefont {Kai-Jie}\ \bibnamefont
  {Yang}}, \bibinfo {author} {\bibfnamefont {Li}~\bibnamefont {Chen}}, \bibinfo
  {author} {\bibfnamefont {Gang}\ \bibnamefont {Xu}}, \bibinfo {author}
  {\bibfnamefont {Chao-Xing}\ \bibnamefont {Liu}}, \ and\ \bibinfo {author}
  {\bibfnamefont {Xin}\ \bibnamefont {Liu}},\ }\bibfield  {title} {\enquote
  {\bibinfo {title} {Lattice-symmetry-assisted second-order topological
  superconductors and {M}ajorana patterns},}\ }\href {\doibase
  10.1103/PhysRevLett.123.156801} {\bibfield  {journal} {\bibinfo  {journal}
  {Phys. Rev. Lett.}\ }\textbf {\bibinfo {volume} {123}},\ \bibinfo {pages}
  {156801} (\bibinfo {year} {2019})}\BibitemShut {NoStop}%
\bibitem [{\citenamefont {Yan}(2019{\natexlab{a}})}]{Yan2019hosca}%
  \BibitemOpen
  \bibfield  {author} {\bibinfo {author} {\bibfnamefont {Zhongbo}\ \bibnamefont
  {Yan}},\ }\bibfield  {title} {\enquote {\bibinfo {title} {Higher-order
  topological odd-parity superconductors},}\ }\href {\doibase
  10.1103/PhysRevLett.123.177001} {\bibfield  {journal} {\bibinfo  {journal}
  {Phys. Rev. Lett.}\ }\textbf {\bibinfo {volume} {123}},\ \bibinfo {pages}
  {177001} (\bibinfo {year} {2019}{\natexlab{a}})}\BibitemShut {NoStop}%
\bibitem [{\citenamefont {Yan}(2019{\natexlab{b}})}]{Yan2019hoscb}%
  \BibitemOpen
  \bibfield  {author} {\bibinfo {author} {\bibfnamefont {Zhongbo}\ \bibnamefont
  {Yan}},\ }\bibfield  {title} {\enquote {\bibinfo {title} {Majorana corner and
  hinge modes in second-order topological insulator/superconductor
  heterostructures},}\ }\href {\doibase 10.1103/PhysRevB.100.205406} {\bibfield
   {journal} {\bibinfo  {journal} {Phys. Rev. B}\ }\textbf {\bibinfo {volume}
  {100}},\ \bibinfo {pages} {205406} (\bibinfo {year}
  {2019}{\natexlab{b}})}\BibitemShut {NoStop}%
\bibitem [{\citenamefont {Franca}\ \emph {et~al.}(2019)\citenamefont {Franca},
  \citenamefont {Efremov},\ and\ \citenamefont {Fulga}}]{Franca2019SOTSC}%
  \BibitemOpen
  \bibfield  {author} {\bibinfo {author} {\bibfnamefont {S.}~\bibnamefont
  {Franca}}, \bibinfo {author} {\bibfnamefont {D.~V.}\ \bibnamefont {Efremov}},
  \ and\ \bibinfo {author} {\bibfnamefont {I.~C.}\ \bibnamefont {Fulga}},\
  }\bibfield  {title} {\enquote {\bibinfo {title} {Phase-tunable second-order
  topological superconductor},}\ }\href {\doibase 10.1103/PhysRevB.100.075415}
  {\bibfield  {journal} {\bibinfo  {journal} {Phys. Rev. B}\ }\textbf {\bibinfo
  {volume} {100}},\ \bibinfo {pages} {075415} (\bibinfo {year}
  {2019})}\BibitemShut {NoStop}%
\bibitem [{\citenamefont {Kheirkhah}\ \emph
  {et~al.}(2020{\natexlab{a}})\citenamefont {Kheirkhah}, \citenamefont {Nagai},
  \citenamefont {Chen},\ and\ \citenamefont {Marsiglio}}]{Majid2020hosca}%
  \BibitemOpen
  \bibfield  {author} {\bibinfo {author} {\bibfnamefont {Majid}\ \bibnamefont
  {Kheirkhah}}, \bibinfo {author} {\bibfnamefont {Yuki}\ \bibnamefont {Nagai}},
  \bibinfo {author} {\bibfnamefont {Chun}\ \bibnamefont {Chen}}, \ and\
  \bibinfo {author} {\bibfnamefont {Frank}\ \bibnamefont {Marsiglio}},\
  }\bibfield  {title} {\enquote {\bibinfo {title} {Majorana corner flat bands
  in two-dimensional second-order topological superconductors},}\ }\href
  {\doibase 10.1103/PhysRevB.101.104502} {\bibfield  {journal} {\bibinfo
  {journal} {Phys. Rev. B}\ }\textbf {\bibinfo {volume} {101}},\ \bibinfo
  {pages} {104502} (\bibinfo {year} {2020}{\natexlab{a}})}\BibitemShut
  {NoStop}%
\bibitem [{\citenamefont {Zhang}\ and\ \citenamefont
  {Trauzettel}(2020)}]{Zhang2020SOTSC}%
  \BibitemOpen
  \bibfield  {author} {\bibinfo {author} {\bibfnamefont {Song-Bo}\ \bibnamefont
  {Zhang}}\ and\ \bibinfo {author} {\bibfnamefont {Bj\"orn}\ \bibnamefont
  {Trauzettel}},\ }\bibfield  {title} {\enquote {\bibinfo {title} {Detection of
  second-order topological superconductors by {J}osephson junctions},}\ }\href
  {\doibase 10.1103/PhysRevResearch.2.012018} {\bibfield  {journal} {\bibinfo
  {journal} {Phys. Rev. Research}\ }\textbf {\bibinfo {volume} {2}},\ \bibinfo
  {pages} {012018} (\bibinfo {year} {2020})}\BibitemShut {NoStop}%
\bibitem [{\citenamefont {Ahn}\ and\ \citenamefont {Yang}(2020)}]{Ahn2020hosc}%
  \BibitemOpen
  \bibfield  {author} {\bibinfo {author} {\bibfnamefont {Junyeong}\
  \bibnamefont {Ahn}}\ and\ \bibinfo {author} {\bibfnamefont {Bohm-Jung}\
  \bibnamefont {Yang}},\ }\bibfield  {title} {\enquote {\bibinfo {title}
  {Higher-order topological superconductivity of spin-polarized fermions},}\
  }\href {\doibase 10.1103/PhysRevResearch.2.012060} {\bibfield  {journal}
  {\bibinfo  {journal} {Phys. Rev. Research}\ }\textbf {\bibinfo {volume}
  {2}},\ \bibinfo {pages} {012060} (\bibinfo {year} {2020})}\BibitemShut
  {NoStop}%
\bibitem [{\citenamefont {De}\ \emph {et~al.}(2020)\citenamefont {De},
  \citenamefont {Khanna},\ and\ \citenamefont {Rao}}]{Suman2020}%
  \BibitemOpen
  \bibfield  {author} {\bibinfo {author} {\bibfnamefont {Suman~Jyoti}\
  \bibnamefont {De}}, \bibinfo {author} {\bibfnamefont {Udit}\ \bibnamefont
  {Khanna}}, \ and\ \bibinfo {author} {\bibfnamefont {Sumathi}\ \bibnamefont
  {Rao}},\ }\bibfield  {title} {\enquote {\bibinfo {title} {Magnetic flux
  periodicity in second order topological superconductors},}\ }\href {\doibase
  10.1103/PhysRevB.101.125429} {\bibfield  {journal} {\bibinfo  {journal}
  {Phys. Rev. B}\ }\textbf {\bibinfo {volume} {101}},\ \bibinfo {pages}
  {125429} (\bibinfo {year} {2020})}\BibitemShut {NoStop}%
\bibitem [{\citenamefont {Roy}(2020)}]{Bitan2020hosc}%
  \BibitemOpen
  \bibfield  {author} {\bibinfo {author} {\bibfnamefont {Bitan}\ \bibnamefont
  {Roy}},\ }\bibfield  {title} {\enquote {\bibinfo {title} {Higher-order
  topological superconductors in $\mathcal{P}$-, $\mathcal{T}$-odd quadrupolar
  {D}irac materials},}\ }\href {\doibase 10.1103/PhysRevB.101.220506}
  {\bibfield  {journal} {\bibinfo  {journal} {Phys. Rev. B}\ }\textbf {\bibinfo
  {volume} {101}},\ \bibinfo {pages} {220506} (\bibinfo {year}
  {2020})}\BibitemShut {NoStop}%
\bibitem [{\citenamefont {Wu}\ \emph {et~al.}(2020{\natexlab{b}})\citenamefont
  {Wu}, \citenamefont {Hou}, \citenamefont {Li}, \citenamefont {Luo},
  \citenamefont {Shi},\ and\ \citenamefont {Zhang}}]{Wu2020SOTSC}%
  \BibitemOpen
  \bibfield  {author} {\bibinfo {author} {\bibfnamefont {Ya-Jie}\ \bibnamefont
  {Wu}}, \bibinfo {author} {\bibfnamefont {Junpeng}\ \bibnamefont {Hou}},
  \bibinfo {author} {\bibfnamefont {Yun-Mei}\ \bibnamefont {Li}}, \bibinfo
  {author} {\bibfnamefont {Xi-Wang}\ \bibnamefont {Luo}}, \bibinfo {author}
  {\bibfnamefont {Xiaoyan}\ \bibnamefont {Shi}}, \ and\ \bibinfo {author}
  {\bibfnamefont {Chuanwei}\ \bibnamefont {Zhang}},\ }\bibfield  {title}
  {\enquote {\bibinfo {title} {In-plane {Z}eeman-field-induced {M}ajorana
  corner and hinge modes in an $s$-wave superconductor heterostructure},}\
  }\href {\doibase 10.1103/PhysRevLett.124.227001} {\bibfield  {journal}
  {\bibinfo  {journal} {Phys. Rev. Lett.}\ }\textbf {\bibinfo {volume} {124}},\
  \bibinfo {pages} {227001} (\bibinfo {year} {2020}{\natexlab{b}})}\BibitemShut
  {NoStop}%
\bibitem [{\citenamefont {Kheirkhah}\ \emph
  {et~al.}(2020{\natexlab{b}})\citenamefont {Kheirkhah}, \citenamefont {Yan},
  \citenamefont {Nagai},\ and\ \citenamefont {Marsiglio}}]{Majid2020hoscb}%
  \BibitemOpen
  \bibfield  {author} {\bibinfo {author} {\bibfnamefont {Majid}\ \bibnamefont
  {Kheirkhah}}, \bibinfo {author} {\bibfnamefont {Zhongbo}\ \bibnamefont
  {Yan}}, \bibinfo {author} {\bibfnamefont {Yuki}\ \bibnamefont {Nagai}}, \
  and\ \bibinfo {author} {\bibfnamefont {Frank}\ \bibnamefont {Marsiglio}},\
  }\bibfield  {title} {\enquote {\bibinfo {title} {First- and second-order
  topological superconductivity and temperature-driven topological phase
  transitions in the extended {H}ubbard model with spin-orbit coupling},}\
  }\href {\doibase 10.1103/PhysRevLett.125.017001} {\bibfield  {journal}
  {\bibinfo  {journal} {Phys. Rev. Lett.}\ }\textbf {\bibinfo {volume} {125}},\
  \bibinfo {pages} {017001} (\bibinfo {year} {2020}{\natexlab{b}})}\BibitemShut
  {NoStop}%
\bibitem [{\citenamefont {Ghorashi}\ \emph {et~al.}(2020)\citenamefont
  {Ghorashi}, \citenamefont {Hughes},\ and\ \citenamefont
  {Rossi}}]{ghorashi2019vortex}%
  \BibitemOpen
  \bibfield  {author} {\bibinfo {author} {\bibfnamefont {Sayed Ali~Akbar}\
  \bibnamefont {Ghorashi}}, \bibinfo {author} {\bibfnamefont {Taylor~L.}\
  \bibnamefont {Hughes}}, \ and\ \bibinfo {author} {\bibfnamefont {Enrico}\
  \bibnamefont {Rossi}},\ }\bibfield  {title} {\enquote {\bibinfo {title}
  {Vortex and surface phase transitions in superconducting higher-order
  topological insulators},}\ }\href {\doibase 10.1103/PhysRevLett.125.037001}
  {\bibfield  {journal} {\bibinfo  {journal} {Phys. Rev. Lett.}\ }\textbf
  {\bibinfo {volume} {125}},\ \bibinfo {pages} {037001} (\bibinfo {year}
  {2020})}\BibitemShut {NoStop}%
\bibitem [{\citenamefont {Chen}\ \emph
  {et~al.}(2019{\natexlab{b}})\citenamefont {Chen}, \citenamefont {Liu},
  \citenamefont {Xu},\ and\ \citenamefont {Liu}}]{Chen2019hosc}%
  \BibitemOpen
  \bibfield  {author} {\bibinfo {author} {\bibfnamefont {Li}~\bibnamefont
  {Chen}}, \bibinfo {author} {\bibfnamefont {Bin}\ \bibnamefont {Liu}},
  \bibinfo {author} {\bibfnamefont {Gang}\ \bibnamefont {Xu}}, \ and\ \bibinfo
  {author} {\bibfnamefont {Xin}\ \bibnamefont {Liu}},\ }\bibfield  {title}
  {\enquote {\bibinfo {title} {{Lattice distortion induced first and second
  order topological phase transition in rectangular high-T$_{\rm c}$
  superconducting monolayer}},}\ }\href {https://arxiv.org/abs/1909.10402}
  {\bibfield  {journal} {\bibinfo  {journal} {arXiv preprint arXiv:1909.10402}\
  } (\bibinfo {year} {2019}{\natexlab{b}})}\BibitemShut {NoStop}%
\bibitem [{\citenamefont {Niu}\ \emph {et~al.}(2021)\citenamefont {Niu},
  \citenamefont {Yan}, \citenamefont {Zhou}, \citenamefont {Tao}, \citenamefont
  {Li}, \citenamefont {Liu}, \citenamefont {Zhang}, \citenamefont {Jia},
  \citenamefont {Liu}, \citenamefont {Yan}, \citenamefont {Chen},\ and\
  \citenamefont {Yu}}]{Niu2020hosc}%
  \BibitemOpen
  \bibfield  {author} {\bibinfo {author} {\bibfnamefont {Jingjing}\
  \bibnamefont {Niu}}, \bibinfo {author} {\bibfnamefont {Tongxing}\
  \bibnamefont {Yan}}, \bibinfo {author} {\bibfnamefont {Yuxuan}\ \bibnamefont
  {Zhou}}, \bibinfo {author} {\bibfnamefont {Ziyu}\ \bibnamefont {Tao}},
  \bibinfo {author} {\bibfnamefont {Xiaole}\ \bibnamefont {Li}}, \bibinfo
  {author} {\bibfnamefont {Weiyang}\ \bibnamefont {Liu}}, \bibinfo {author}
  {\bibfnamefont {Libo}\ \bibnamefont {Zhang}}, \bibinfo {author}
  {\bibfnamefont {Hao}\ \bibnamefont {Jia}}, \bibinfo {author} {\bibfnamefont
  {Song}\ \bibnamefont {Liu}}, \bibinfo {author} {\bibfnamefont {Zhongbo}\
  \bibnamefont {Yan}}, \bibinfo {author} {\bibfnamefont {Yuanzhen}\
  \bibnamefont {Chen}}, \ and\ \bibinfo {author} {\bibfnamefont {Dapeng}\
  \bibnamefont {Yu}},\ }\bibfield  {title} {\enquote {\bibinfo {title}
  {Simulation of higher-order topological phases and related topological phase
  transitions in a superconducting qubit},}\ }\href {\doibase
  10.1016/j.scib.2021.02.035} {\bibfield  {journal} {\bibinfo  {journal}
  {Science Bulletin}\ } (\bibinfo {year} {2021}),\
  10.1016/j.scib.2021.02.035}\BibitemShut {NoStop}%
\bibitem [{\citenamefont {Gray}\ \emph {et~al.}(2019)\citenamefont {Gray},
  \citenamefont {Freudenstein}, \citenamefont {Zhao}, \citenamefont
  {O’Connor}, \citenamefont {Jenkins}, \citenamefont {Kumar}, \citenamefont
  {Hoek}, \citenamefont {Kopec}, \citenamefont {Huh}, \citenamefont
  {Taniguchi}, \citenamefont {Watanabe}, \citenamefont {Zhong}, \citenamefont
  {Kim}, \citenamefont {Gu},\ and\ \citenamefont {Burch}}]{Gray2019helical}%
  \BibitemOpen
  \bibfield  {author} {\bibinfo {author} {\bibfnamefont {Mason~J.}\
  \bibnamefont {Gray}}, \bibinfo {author} {\bibfnamefont {Josef}\ \bibnamefont
  {Freudenstein}}, \bibinfo {author} {\bibfnamefont {Shu Yang~F.}\ \bibnamefont
  {Zhao}}, \bibinfo {author} {\bibfnamefont {Ryan}\ \bibnamefont {O’Connor}},
  \bibinfo {author} {\bibfnamefont {Samuel}\ \bibnamefont {Jenkins}}, \bibinfo
  {author} {\bibfnamefont {Narendra}\ \bibnamefont {Kumar}}, \bibinfo {author}
  {\bibfnamefont {Marcel}\ \bibnamefont {Hoek}}, \bibinfo {author}
  {\bibfnamefont {Abigail}\ \bibnamefont {Kopec}}, \bibinfo {author}
  {\bibfnamefont {Soonsang}\ \bibnamefont {Huh}}, \bibinfo {author}
  {\bibfnamefont {Takashi}\ \bibnamefont {Taniguchi}}, \bibinfo {author}
  {\bibfnamefont {Kenji}\ \bibnamefont {Watanabe}}, \bibinfo {author}
  {\bibfnamefont {Ruidan}\ \bibnamefont {Zhong}}, \bibinfo {author}
  {\bibfnamefont {Changyoung}\ \bibnamefont {Kim}}, \bibinfo {author}
  {\bibfnamefont {G.~D.}\ \bibnamefont {Gu}}, \ and\ \bibinfo {author}
  {\bibfnamefont {K.~S.}\ \bibnamefont {Burch}},\ }\bibfield  {title} {\enquote
  {\bibinfo {title} {Evidence for helical hinge zero modes in an {F}e-based
  superconductor},}\ }\href {\doibase 10.1021/acs.nanolett.9b00844} {\bibfield
  {journal} {\bibinfo  {journal} {Nano Letters}\ }\textbf {\bibinfo {volume}
  {19}},\ \bibinfo {pages} {4890--4896} (\bibinfo {year} {2019})}\BibitemShut
  {NoStop}%
\bibitem [{\citenamefont {Hosur}\ \emph {et~al.}(2011)\citenamefont {Hosur},
  \citenamefont {Ghaemi}, \citenamefont {Mong},\ and\ \citenamefont
  {Vishwanath}}]{Hosur2011MZM}%
  \BibitemOpen
  \bibfield  {author} {\bibinfo {author} {\bibfnamefont {Pavan}\ \bibnamefont
  {Hosur}}, \bibinfo {author} {\bibfnamefont {Pouyan}\ \bibnamefont {Ghaemi}},
  \bibinfo {author} {\bibfnamefont {Roger S.~K.}\ \bibnamefont {Mong}}, \ and\
  \bibinfo {author} {\bibfnamefont {Ashvin}\ \bibnamefont {Vishwanath}},\
  }\bibfield  {title} {\enquote {\bibinfo {title} {Majorana modes at the ends
  of superconductor vortices in doped topological insulators},}\ }\href
  {\doibase 10.1103/PhysRevLett.107.097001} {\bibfield  {journal} {\bibinfo
  {journal} {Phys. Rev. Lett.}\ }\textbf {\bibinfo {volume} {107}},\ \bibinfo
  {pages} {097001} (\bibinfo {year} {2011})}\BibitemShut {NoStop}%
\bibitem [{\citenamefont {Hasan}\ and\ \citenamefont
  {Kane}(2010)}]{hasan2010colloquium}%
  \BibitemOpen
  \bibfield  {author} {\bibinfo {author} {\bibfnamefont {M~Zahid}\ \bibnamefont
  {Hasan}}\ and\ \bibinfo {author} {\bibfnamefont {Charles~L}\ \bibnamefont
  {Kane}},\ }\bibfield  {title} {\enquote {\bibinfo {title} {Colloquium:
  topological insulators},}\ }\href
  {https://journals.aps.org/rmp/abstract/10.1103/RevModPhys.82.3045} {\bibfield
   {journal} {\bibinfo  {journal} {Reviews of Modern Physics}\ }\textbf
  {\bibinfo {volume} {82}},\ \bibinfo {pages} {3045} (\bibinfo {year}
  {2010})}\BibitemShut {NoStop}%
\bibitem [{\citenamefont {Qi}\ and\ \citenamefont
  {Zhang}(2011)}]{qi2011topological}%
  \BibitemOpen
  \bibfield  {author} {\bibinfo {author} {\bibfnamefont {Xiao-Liang}\
  \bibnamefont {Qi}}\ and\ \bibinfo {author} {\bibfnamefont {Shou-Cheng}\
  \bibnamefont {Zhang}},\ }\bibfield  {title} {\enquote {\bibinfo {title}
  {Topological insulators and superconductors},}\ }\href
  {https://journals.aps.org/rmp/abstract/10.1103/RevModPhys.83.1057} {\bibfield
   {journal} {\bibinfo  {journal} {Reviews of Modern Physics}\ }\textbf
  {\bibinfo {volume} {83}},\ \bibinfo {pages} {1057} (\bibinfo {year}
  {2011})}\BibitemShut {NoStop}%
\bibitem [{\citenamefont {Fu}\ and\ \citenamefont {Kane}(2007)}]{fu2007a}%
  \BibitemOpen
  \bibfield  {author} {\bibinfo {author} {\bibfnamefont {Liang}\ \bibnamefont
  {Fu}}\ and\ \bibinfo {author} {\bibfnamefont {C.~L.}\ \bibnamefont {Kane}},\
  }\bibfield  {title} {\enquote {\bibinfo {title} {Topological insulators with
  inversion symmetry},}\ }\href {\doibase 10.1103/PhysRevB.76.045302}
  {\bibfield  {journal} {\bibinfo  {journal} {Phys. Rev. B}\ }\textbf {\bibinfo
  {volume} {76}},\ \bibinfo {eid} {045302} (\bibinfo {year}
  {2007})}\BibitemShut {NoStop}%
\bibitem [{sup()}]{supplemental}%
  \BibitemOpen
  \href@noop {} {}\bibinfo {howpublished} {See Supplemental Material for
  details of calculations.}\BibitemShut {Stop}%
\bibitem [{\citenamefont {Schnyder}\ \emph {et~al.}(2008)\citenamefont
  {Schnyder}, \citenamefont {Ryu}, \citenamefont {Furusaki},\ and\
  \citenamefont {Ludwig}}]{schnyder2008}%
  \BibitemOpen
  \bibfield  {author} {\bibinfo {author} {\bibfnamefont {Andreas~P.}\
  \bibnamefont {Schnyder}}, \bibinfo {author} {\bibfnamefont {Shinsei}\
  \bibnamefont {Ryu}}, \bibinfo {author} {\bibfnamefont {Akira}\ \bibnamefont
  {Furusaki}}, \ and\ \bibinfo {author} {\bibfnamefont {Andreas W.~W.}\
  \bibnamefont {Ludwig}},\ }\bibfield  {title} {\enquote {\bibinfo {title}
  {Classification of topological insulators and superconductors in three
  spatial dimensions},}\ }\href {\doibase 10.1103/PhysRevB.78.195125}
  {\bibfield  {journal} {\bibinfo  {journal} {Phys. Rev. B}\ }\textbf {\bibinfo
  {volume} {78}},\ \bibinfo {pages} {195125} (\bibinfo {year}
  {2008})}\BibitemShut {NoStop}%
\bibitem [{\citenamefont {Kitaev}(2009)}]{kitaev2009}%
  \BibitemOpen
  \bibfield  {author} {\bibinfo {author} {\bibfnamefont {Alexei}\ \bibnamefont
  {Kitaev}},\ }\bibfield  {title} {\enquote {\bibinfo {title} {Periodic table
  for topological insulators and superconductors},}\ }in\ \href {\doibase
  10.1063/1.3149495} {\emph {\bibinfo {booktitle} {AIP Conference
  Proceedings}}},\ Vol.\ \bibinfo {volume} {1134}\ (\bibinfo {organization}
  {AIP},\ \bibinfo {year} {2009})\ pp.\ \bibinfo {pages} {22--30}\BibitemShut
  {NoStop}%
\bibitem [{\citenamefont {Wimmer}(2012)}]{wimmer2012pfaffian}%
  \BibitemOpen
  \bibfield  {author} {\bibinfo {author} {\bibfnamefont {M.}~\bibnamefont
  {Wimmer}},\ }\bibfield  {title} {\enquote {\bibinfo {title} {Algorithm 923:
  Efficient numerical computation of the {P}faffian for dense and banded
  skew-symmetric matrices},}\ }\href {\doibase 10.1145/2331130.2331138}
  {\bibfield  {journal} {\bibinfo  {journal} {ACM Trans. Math. Softw.}\
  }\textbf {\bibinfo {volume} {38}},\ \bibinfo {pages} {30:1--30:17} (\bibinfo
  {year} {2012})}\BibitemShut {NoStop}%
\bibitem [{\citenamefont {Chen}\ \emph
  {et~al.}(2019{\natexlab{c}})\citenamefont {Chen}, \citenamefont {Chen},
  \citenamefont {Duan}, \citenamefont {Zhu}, \citenamefont {Yang},\ and\
  \citenamefont {Wen}}]{chen2019observation}%
  \BibitemOpen
  \bibfield  {author} {\bibinfo {author} {\bibfnamefont {Xiaoyu}\ \bibnamefont
  {Chen}}, \bibinfo {author} {\bibfnamefont {Mingyang}\ \bibnamefont {Chen}},
  \bibinfo {author} {\bibfnamefont {Wen}\ \bibnamefont {Duan}}, \bibinfo
  {author} {\bibfnamefont {Xiyu}\ \bibnamefont {Zhu}}, \bibinfo {author}
  {\bibfnamefont {Huan}\ \bibnamefont {Yang}}, \ and\ \bibinfo {author}
  {\bibfnamefont {Hai-Hu}\ \bibnamefont {Wen}},\ }\bibfield  {title} {\enquote
  {\bibinfo {title} {Observation and characterization of the zero energy
  conductance peak in the vortex core state of \ce{FeTe_{0.55}Se_{0.45}}},}\
  }\href {https://arxiv.org/abs/1909.01686} {\bibfield  {journal} {\bibinfo
  {journal} {arXiv preprint arXiv:1909.01686}\ } (\bibinfo {year}
  {2019}{\natexlab{c}})}\BibitemShut {NoStop}%
\bibitem [{\citenamefont {Singh}\ \emph {et~al.}(2013)\citenamefont {Singh},
  \citenamefont {White}, \citenamefont {Schmaus}, \citenamefont {Tsurkan},
  \citenamefont {Loidl}, \citenamefont {Deisenhofer},\ and\ \citenamefont
  {Wahl}}]{Singh2013}%
  \BibitemOpen
  \bibfield  {author} {\bibinfo {author} {\bibfnamefont {U.~R.}\ \bibnamefont
  {Singh}}, \bibinfo {author} {\bibfnamefont {S.~C.}\ \bibnamefont {White}},
  \bibinfo {author} {\bibfnamefont {S.}~\bibnamefont {Schmaus}}, \bibinfo
  {author} {\bibfnamefont {V.}~\bibnamefont {Tsurkan}}, \bibinfo {author}
  {\bibfnamefont {A.}~\bibnamefont {Loidl}}, \bibinfo {author} {\bibfnamefont
  {J.}~\bibnamefont {Deisenhofer}}, \ and\ \bibinfo {author} {\bibfnamefont
  {P.}~\bibnamefont {Wahl}},\ }\bibfield  {title} {\enquote {\bibinfo {title}
  {Spatial inhomogeneity of the superconducting gap and order parameter in
  \ce{FeSe_{0.4}Te_{0.6}}},}\ }\href {\doibase 10.1103/PhysRevB.88.155124}
  {\bibfield  {journal} {\bibinfo  {journal} {Phys. Rev. B}\ }\textbf {\bibinfo
  {volume} {88}},\ \bibinfo {pages} {155124} (\bibinfo {year}
  {2013})}\BibitemShut {NoStop}%
\bibitem [{\citenamefont {Wu}\ \emph {et~al.}(2021)\citenamefont {Wu},
  \citenamefont {Chung}, \citenamefont {Liu},\ and\ \citenamefont
  {Kim}}]{PhysRevResearch.3.013066}%
  \BibitemOpen
  \bibfield  {author} {\bibinfo {author} {\bibfnamefont {Xianxin}\ \bibnamefont
  {Wu}}, \bibinfo {author} {\bibfnamefont {Suk~Bum}\ \bibnamefont {Chung}},
  \bibinfo {author} {\bibfnamefont {Chaoxing}\ \bibnamefont {Liu}}, \ and\
  \bibinfo {author} {\bibfnamefont {Eun-Ah}\ \bibnamefont {Kim}},\ }\bibfield
  {title} {\enquote {\bibinfo {title} {Topological orders competing for the
  {D}irac surface state in \ce{FeSeTe} surfaces},}\ }\href {\doibase
  10.1103/PhysRevResearch.3.013066} {\bibfield  {journal} {\bibinfo  {journal}
  {Phys. Rev. Research}\ }\textbf {\bibinfo {volume} {3}},\ \bibinfo {pages}
  {013066} (\bibinfo {year} {2021})}\BibitemShut {NoStop}%
\end{thebibliography}%

\clearpage


\section*{Supplementary material}

# Supplemental Material for "Vortex-line topology in iron-based superconductors with and without second-order topology"

Majid Kheirkhah,[1] Zhongbo Yan,[2, *] and Frank Marsiglio[1]

[1]*Department of Physics, University of Alberta, Edmonton, Alberta T6G 2E1, Canada*
[2]*School of Physics, Sun Yat-Sen University, Guangzhou 510275, China*

## CONTENTS

## I. PHASE DIAGRAM

In momentum space, the minimal BdG Hamiltonian for iron-based superconductors with band inversion can be written as

$$\mathcal{H} = \frac{1}{2} \sum_{\boldsymbol{k}} \begin{pmatrix} \boldsymbol{c}_{\boldsymbol{k}}^\dagger, & \boldsymbol{c}_{-\boldsymbol{k}} \end{pmatrix} \begin{pmatrix} H_0(\boldsymbol{k}) & \Delta(\boldsymbol{k}) \\ \Delta^\dagger(\boldsymbol{k}) & -H_0^*(-\boldsymbol{k}) \end{pmatrix} \begin{pmatrix} \boldsymbol{c}_{\boldsymbol{k}} \\ \boldsymbol{c}_{-\boldsymbol{k}}^\dagger \end{pmatrix}, \quad (S1)$$

where $\boldsymbol{c}_{\boldsymbol{k}} = (c_{\boldsymbol{k},a,\uparrow}, c_{\boldsymbol{k},b,\uparrow}, c_{\boldsymbol{k},a,\downarrow}, c_{\boldsymbol{k},b,\downarrow})^{\mathrm{T}}$ and

$$H_0(\boldsymbol{k}) = (m_0 - 2t \sum_{i=x,y,z} \cos k_i)\sigma_z s_0 + 2\lambda \sum_{i=x,y,z} \sin k_i \sigma_x s_i - \mu \sigma_0 s_0 + \sigma_0 \boldsymbol{h} \cdot \boldsymbol{s}, \quad (S2)$$

$$\Delta(\boldsymbol{k}) = i\{\Delta_0 + \Delta_s(\cos k_x + \cos k_y)\}\sigma_0 s_y. \quad (S3)$$

Here, $\boldsymbol{h} = (h_x, h_y, h_z)$, $\boldsymbol{s} = (s_x, s_y, s_z)$, and the Pauli matrices $\sigma$ and $s$ act in orbital $(a, b)$ and spin $(\uparrow, \downarrow)$ degrees of freedom, respectively, and $\sigma_0$ and $s_0$ are the unit matrices. $H_0(\boldsymbol{k})$ describes the normal state, and $\Delta(\boldsymbol{k})$ represents the $s_\pm$-wave pairing. In this section, we focus on the uniform case without vortex lines and Zeeman field. Throughout the main text, we have chosen $t = 1$ as the unit of energy, $m_0 = 2.5$, $\lambda = 0.5$, so that a band inversion takes place at the $\Gamma = (0, 0, 0)$. It is worth noting that for this set of parameters, the band edges are located at $(0, \pi, 0)$ and $(\pi, 0, 0)$, with the energy gap $E_g = 1.0$. The normal state becomes metallic when $\mu > 0.5$ (we only consider positive $\mu$ because the results are symmetric about $\mu = 0$). After becoming superconducting, the doubly degenerate eigenvalues of the BdG Hamiltonian for $\boldsymbol{h} = 0$ are given by,

$$E_{(\pm,\pm)}(\boldsymbol{k}) = \pm\sqrt{\varepsilon_\pm^2(\boldsymbol{k}) + \Delta^2(\boldsymbol{k})}, \quad (S4)$$

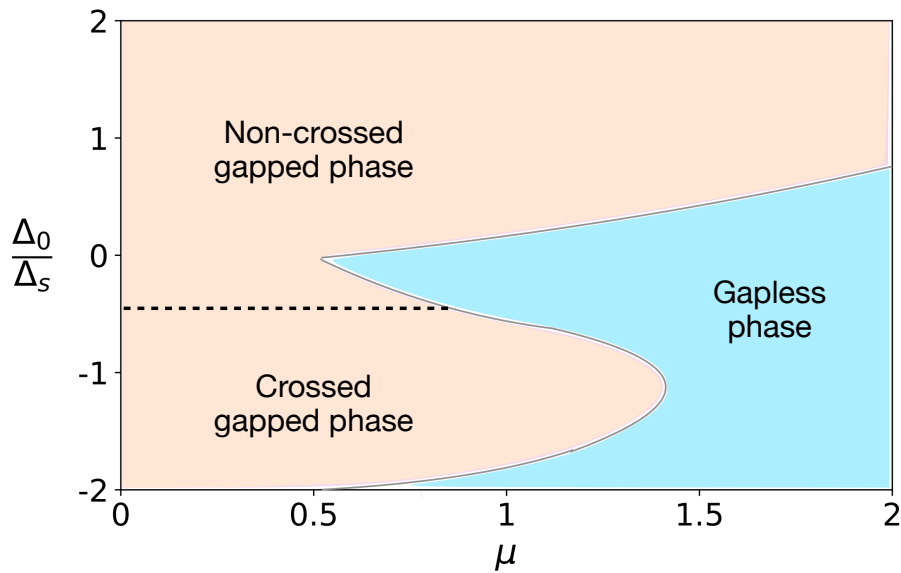

FIG. S1. (Color online) The phase diagram consists of gapped (orange region) and gapless (blue region) phases. We take $t = 1$, $m_0 = 2.5$, $\lambda = 0.5$, and $\boldsymbol{h} = 0$ in the absence of vortex lines. The horizontal dashed line divides the gapped phase into two distinct topological phases. The crossed gapped phase corresponds to a gapped second-order topological superconducting phase with the underlying band inversion surface and pairing node surface crossing with each other. The noncrossed gapped phase corresponds to a gapped trivial superconducting phase with the underlying band inversion surface and pairing node surface separated in the Brillouin zone.

where

$$\varepsilon_\pm(\boldsymbol{k}) = \mu \pm \sqrt{(m_0 - 2t\sum_{i=x,y,z} \cos k_i)^2 + 4\lambda^2 \sum_{i=x,y,z} \sin^2 k_i}. \tag{S5}$$

Because $s_\pm$-wave pairing has nodes when $|\Delta_0| < 2|\Delta_s|$, the superconductor can be gapless when the Fermi surface intersects with the pairing nodes. As the location of pairing nodes is determined by $\Delta(\boldsymbol{k}) = 0$, and the location of Fermi surface is determined by $\varepsilon_-(\boldsymbol{k}) = 0$, we can numerically determine the phase boundary which separates the gapped phase from the gapless phase by checking when the contour for $\Delta(\boldsymbol{k}) = 0$ intersects with the contour for $\varepsilon_-(\boldsymbol{k}) = 0$. The phase boundary is shown in Fig. S1.

As the energy bands of the concerned iron-based superconductors are gapped, therefore, in this work we are interested in the gapped regime. Within the gapped regime, the phase can be further divided as second-order topological superconductor and trivial superconductor. At $\mu = 0$, we have pointed out in the main text that when the band inversion surface and pairing node surface cross, the system is a second-order topological superconductor with helical Majorana hinge modes. The logic behind this statement is as follows. The Hamiltonian (1) of the main text at $\mu = \boldsymbol{h} = 0$ can be divided into two parts, i.e., $H = H_e + H_o$, where $H_e$ and $H_o$ represent respectively the even-parity and odd-parity parts. Because $\{H_e, H_o\} = 0$, the even-parity part $H_e$ plays the role of Dirac mass. When the energy spectra of $H_e$ are gapless, $H$ will be topologically nontrivial since the existence of nodes in $H_e$ implies the existence of domain walls of Dirac mass on the boundary. Such kinds of geometric criteria are universally applicable to higher-order topological models as long as the condition that all terms of the Hamiltonian anticommute with each other is fulfilled. Accordingly, if we start with the crossing case, it is expected that a transition from the second-order topological superconducting phase to the trivial superconducting phase should happen. It is known that a topological phase transition from a first-order topological phase to a trivial phase is always associated with the closure of the bulk energy gap as long as the symmetry class does not change. However, a topological phase transition from a second-order topological phase to a trivial phase in general does not require the closure of the bulk energy gap; what it requires is the closure of the energy gap of the surface states. Interestingly, for the set of isotropic parameters considered in this work, we found that the closure of surface energy gap coincides with the closure of the bulk energy gap, indicating that the phase boundary of the second-order topological superconducting phase with a finite bulk gap coincides with the gapped-gapless phase boundary. These findings establish the phase diagram in Fig. S1.

## II. REAL SPACE BDG HAMILTONIAN

According to the BdG Hamiltonian in momentum space, the real space Hamiltonian is obtained as

$$H = \frac{1}{2}\sum_{i,j} \begin{pmatrix} \boldsymbol{c}_i^\dagger, & \boldsymbol{c}_i \end{pmatrix} \begin{pmatrix} H_{0;ij} & \Delta_{ij} \\ -\Delta_{ij}^* & -H_{0;ij}^* \end{pmatrix} \begin{pmatrix} \boldsymbol{c}_j \\ \boldsymbol{c}_j^\dagger \end{pmatrix}, \tag{S6}$$

with the basis $\boldsymbol{c}_i = (c_{i,a,\uparrow}, c_{i,b,\uparrow}, c_{i,a,\downarrow}, c_{i,b,\downarrow})^{\mathrm{T}}$ and

$$H_{0;ij} = \sum_{\nu=0,x,y,z} A_{ij}^\nu s_\nu, \tag{S7}$$

$$\Delta_{ij} = i\{\Delta_0 \delta_{ij} + \frac{\Delta_s}{2}(\delta_{j,i+\hat{x}} + \delta_{j,i-\hat{x}} + \delta_{j,i+\hat{y}} + \delta_{j,i-\hat{y}})\}\sigma_0 s_y, \tag{S8}$$

where

$$A_{ij}^0 = \{m_0 \delta_{ij} - t\sum_{\nu=x,y,z}(\delta_{j,i+\hat{\nu}} + \delta_{j,i-\hat{\nu}})\}\sigma_z - \mu \delta_{ij}\sigma_0, \tag{S9}$$

$$A_{ij}^\alpha = -i\lambda(\delta_{j,i+\hat{\alpha}} - \delta_{j,i-\hat{\alpha}})\sigma_x + h_\alpha \delta_{ij}\sigma_0, \tag{S10}$$

are four $2N \times 2N$ matrices for $N = L_x L_y L_z$ and $\alpha \in \{x,y,z\}$. In the presence of a $\pi$-flux vortex line, let us say along $z$ direction, we need only modify the $\Delta_{ij}$ matrix. In this case, we should replace the site-independent real numbers $\Delta_0$ and $\Delta_s$ with the site-dependent complex numbers as given by

$$\Delta_{0,s} \longrightarrow \Delta_{0,s}(\boldsymbol{r}) = \Delta_{0,s} \tanh(\frac{\sqrt{\bar{x}^2 + \bar{y}^2}}{\xi})\frac{\bar{x} + i\bar{y}}{\sqrt{\bar{x}^2 + \bar{y}^2}}, \tag{S11}$$

where $\bar{x} = x - x_c$ and $\bar{y} = y - y_c$, with $(x_c, y_c)$ denoting the core of the vortex in the $xy$ plane, and $(x,y)$ representing the coordinates of the lattice sites for on-site pairing $\Delta_0(\boldsymbol{r})$ and the coordinates of the lattice-bond centers for nearest-neighbor pairing $\Delta_s(\boldsymbol{r})$; $\xi$ denotes the superconducting coherence length. In this work, we will not determine the value of $\xi$ through complicated self-consistent calculations, instead we will set its value by hand since the topological property of the vortex line in the weakly doped regime is insensitive to its value. It is noteworthy that similar forms can be adopted if one wants to study vortex lines along other directions.

## III. THE $\mathbb{Z}_2$ INVARIANT CHARACTERIZES THE SYSTEM

While the presence of a straight $z$-directional vortex line breaks the translational symmetry in the $x$ and $y$ directions, the $z$ direction still preserves translational symmetry and $k_z$ is still a good quantum number. Therefore, the whole system can be taken as a quasi-1D superconductor. Since the presence of a vortex line also breaks the time-reversal symmetry, this quasi-1D superconductor thus falls into the D class of the Atland-Zirnbauer classification [S1, S2]. As is known, when a superconductor in the D class is fully gapped, its band topology is characterized by a $\mathbb{Z}_2$ invariant of the form [S3]

$$\nu = \mathrm{sgn}\{\mathrm{Pf}[H_{\mathrm{M}}(0)]\} \cdot \mathrm{sgn}\{\mathrm{Pf}[H_{\mathrm{M}}(\pi)]\}, \tag{S12}$$

where $H_{\mathrm{M}}(k_z)$ represents the Hamiltonian in the Majorana representation, which is an antisymmetric matrix forced by the intrinsic self-adjoint property of Majorana operators; "Pf" is a shorthand notation of Pfaffian. The formula (S12) implies that a change of the $\mathbb{Z}_2$ invariant, or say a topological phase transition, can only occur at $k_z = 0$ or $\pi$, a consequence originating from the particle-hole symmetry. To obtain the explicit form of $H_{\mathrm{M}}(k_z)$, we introduce the following Majorana operators,

$$\gamma_{\alpha,s,\boldsymbol{r};1} = \frac{1}{\sqrt{2}}(c_{\alpha,s,\boldsymbol{r}}^\dagger + c_{\alpha,s,\boldsymbol{r}}) = \gamma_{\alpha,s,\boldsymbol{r};1}^\dagger, \tag{S13}$$

$$\gamma_{\alpha,s,\boldsymbol{r};2} = \frac{i}{\sqrt{2}}(c_{\alpha,s,\boldsymbol{r}}^\dagger - c_{\alpha,s,\boldsymbol{r}}) = \gamma_{\alpha,s,\boldsymbol{r};2}^\dagger, \tag{S14}$$

at orbital $\alpha$, with spin $s$, and at site $\boldsymbol{r}$. Since only the two time-reversal invariant momenta are relevant for the determination of the underlying band topology, in the following we restrict ourselves to these two points. At $k_z = 0$, it is straightforward to find that the real space Hamiltonian is reduced as

$$
\begin{aligned}
H(k_z = 0) = \sum_{\bar{r};\alpha\alpha';s,s'} \Big\{ & -\mu c^\dagger_{\alpha,s,\bar{r}}\delta_{\alpha\alpha'}\delta_{ss'}c_{\alpha',s',\bar{r}} + (m_0 - 2t)c^\dagger_{\alpha,s,\bar{r}}[\sigma_z]_{\alpha\alpha'}\delta_{ss'}c_{\alpha',s',\bar{r}} + h_z c^\dagger_{\alpha,s,\bar{r}}\delta_{\alpha\alpha'}[s_z]_{ss'}c_{\alpha',s',\bar{r}} \\
& -(tc^\dagger_{\alpha,s,\bar{r}}[\sigma_z]_{\alpha\alpha'}\delta_{ss'}c_{\alpha',s',\bar{r}+\hat{x}} + tc^\dagger_{\alpha,s,\bar{r}}[\sigma_z]_{\alpha\alpha'}\delta_{ss'}c_{\alpha',s',\bar{r}+\hat{y}} + \text{h.c.}) \\
& +(-i\lambda c^\dagger_{\alpha,s,\bar{r}}[\sigma_x]_{\alpha\alpha'}[s_x]_{ss'}c_{\alpha',s',\bar{r}+\hat{x}} - i\lambda c^\dagger_{\alpha,s,\bar{r}}[\sigma_x]_{\alpha\alpha'}[s_y]_{ss'}c_{\alpha',s',\bar{r}+\hat{y}} + \text{h.c.}) \\
& +(i\frac{\Delta_0(\bar{r})}{2}c^\dagger_{\alpha,s,\bar{r}}\delta_{\alpha\alpha'}[s_y]_{ss'}c^\dagger_{\alpha',s',\bar{r}} + i(\frac{\Delta_s(\bar{r}+\frac{\hat{x}}{2})}{2}c^\dagger_{\alpha,s,\bar{r}}\delta_{\alpha\alpha'}[s_y]_{ss'}c^\dagger_{\alpha',s',\bar{r}+\hat{x}} \\
& +\frac{\Delta_s(\bar{r}+\frac{\hat{y}}{2})}{2}c^\dagger_{\alpha,s,\bar{r}}\delta_{\alpha\alpha'}[s_y]_{ss'}c^\dagger_{\alpha',s',\bar{r}+\hat{y}}) + \text{h.c.}) \Big\},
\end{aligned}
\tag{S15}
$$

and at $k_z = \pi$, we have

$$
\begin{aligned}
H(k_z = \pi) = \sum_{\bar{r};\alpha\alpha';s,s'} \Big\{ & -\mu c^\dagger_{\alpha,s,\bar{r}}\delta_{\alpha\alpha'}\delta_{ss'}c_{\alpha',s',\bar{r}} + (m_0 + 2t)c^\dagger_{\alpha,s,\bar{r}}[\sigma_z]_{\alpha\alpha'}\delta_{ss'}c_{\alpha',s',\bar{r}} + h_z c^\dagger_{\alpha,s,\bar{r}}\delta_{\alpha\alpha'}[s_z]_{ss'}c_{\alpha',s',\bar{r}} \\
& -(tc^\dagger_{\alpha,s,\bar{r}}[\sigma_z]_{\alpha\alpha'}\delta_{ss'}c_{\alpha',s',\bar{r}+\hat{x}} + tc^\dagger_{\alpha,s,\bar{r}}[\sigma_z]_{\alpha\alpha'}\delta_{ss'}c_{\alpha',s',\bar{r}+\hat{y}} + \text{h.c.}) \\
& +(-i\lambda c^\dagger_{\alpha,s,\bar{r}}[\sigma_x]_{\alpha\alpha'}[s_x]_{ss'}c_{\alpha',s',\bar{r}+\hat{x}} - i\lambda c^\dagger_{\alpha,s,\bar{r}}[\sigma_x]_{\alpha\alpha'}[s_y]_{ss'}c_{\alpha',s',\bar{r}+\hat{y}} + \text{h.c.}) \\
& +(i\frac{\Delta_0(\bar{r})}{2}c^\dagger_{\alpha,s,\bar{r}}\delta_{\alpha\alpha'}[s_y]_{ss'}c^\dagger_{\alpha',s',\bar{r}} + i(\frac{\Delta_s(\bar{r}+\frac{\hat{x}}{2})}{2}c^\dagger_{\alpha,s,\bar{r}}\delta_{\alpha\alpha'}[s_y]_{ss'}c^\dagger_{\alpha',s',\bar{r}+\hat{x}} \\
& +\frac{\Delta_s(\bar{r}+\frac{\hat{y}}{2})}{2}c^\dagger_{\alpha,s,\bar{r}}\delta_{\alpha\alpha'}[s_y]_{ss'}c^\dagger_{\alpha',s',\bar{r}+\hat{y}}) + \text{h.c.}) \Big\},
\end{aligned}
\tag{S16}
$$

where $\bar{r} = (x, y)$ only involves the lattice sites because the kinetic terms in the $z$ direction have been rewritten in the momentum space due to the preservation of translational symmetry in this direction. It is worth noting that because the vortex line is generated along the $z$ direction, above we have accordingly taken $\boldsymbol{h} = (0, 0, h_z)$. Substituting the Dirac fermion operators by Majorana operators, we get

$$
\begin{aligned}
H(k_z = 0) = \sum_{\bar{r};\alpha\alpha';s,s'} \Big\{ & -i\mu\gamma_{\alpha,s,\bar{r};1}\delta_{\alpha\alpha'}\delta_{ss'}\gamma_{\alpha',s',\bar{r};2} + i(m-2t)\gamma_{\alpha,s,\bar{r};1}[\sigma_z]_{\alpha\alpha'}\delta_{ss'}\gamma_{\alpha',s',\bar{r};2} \\
& +ih_z\gamma_{\alpha,s,\bar{r};1}\delta_{\alpha\alpha'}[s_z]_{ss'}\gamma_{\alpha',s',\bar{r};2} - it(\gamma_{\alpha,s,\bar{r};1}[\sigma_z]_{\alpha\alpha'}\delta_{ss'}\gamma_{\alpha',s',\bar{r}+\hat{x};2} - \gamma_{\alpha,s,\bar{r};2}[\sigma_z]_{\alpha\alpha'}\delta_{ss'}\gamma_{\alpha',s',\bar{r}+\hat{x};1}) \\
& -it(\gamma_{\alpha,s,\bar{r};1}[\sigma_z]_{\alpha\alpha'}\delta_{ss'}\gamma_{\alpha',s',\bar{r}+\hat{y};2} - \gamma_{\alpha,s,\bar{r};2}[\sigma_z]_{\alpha\alpha'}\delta_{ss'}\gamma_{\alpha',s',\bar{r}+\hat{y};1}) \\
& -i\lambda(\gamma_{\alpha,s,\bar{r};1}[\sigma_x]_{\alpha\alpha'}[s_x]_{ss'}\gamma_{\alpha',s',\bar{r}+\hat{x};1} + \gamma_{\alpha,s,\bar{r};2}[\sigma_x]_{\alpha\alpha'}[s_x]_{ss'}\gamma_{\alpha',s',\bar{r}+\hat{x};2}) \\
& +\lambda(\gamma_{\alpha,s,\bar{r};1}[\sigma_y]_{\alpha\alpha'}[s_y]_{ss'}\gamma_{\alpha',s',\bar{r}+\hat{y};2} - \gamma_{\alpha,s,\bar{r};2}[\sigma_x]_{\alpha\alpha'}[s_x]_{ss'}\gamma_{\alpha',s',\bar{r}+\hat{y};1}) \\
& +\frac{\text{Re}\Delta_0(\bar{r})}{2}(\gamma_{\alpha,s,\bar{r};2}\delta_{\alpha\alpha'}[s_y]_{ss'}\gamma_{\alpha',s',\bar{r};1} + \gamma_{\alpha,s,r;1}\delta_{\alpha\alpha'}[s_y]_{ss'}\gamma_{\alpha',s',r;2}) \\
& -\frac{\text{Im}\Delta_0(\bar{r})}{2}(\gamma_{\alpha,s,\bar{r};1}\delta_{\alpha\alpha'}[s_y]_{ss'}\gamma_{\alpha',s',\bar{r};1} - \gamma_{\alpha,s,\bar{r};2}\delta_{\alpha\alpha'}[s_y]_{ss'}\gamma_{\alpha',s',\bar{r};2}) \\
& +\frac{\text{Re}\Delta_s(\bar{r}+\frac{\hat{x}}{2})}{2}(\gamma_{\alpha,s,\bar{r};2}\delta_{\alpha\alpha'}[s_y]_{ss'}\gamma_{\alpha',s',\bar{r}+\hat{x};1} + \gamma_{\alpha,s,\bar{r};1}\delta_{\alpha\alpha'}[s_y]_{ss'}\gamma_{\alpha',s',\bar{r}+\hat{x};2}) \\
& -\frac{\text{Im}\Delta_s(\bar{r}+\frac{\hat{x}}{2})}{2}(\gamma_{\alpha,s,\bar{r};1}\delta_{\alpha\alpha'}[s_y]_{ss'}\gamma_{\alpha',s',\bar{r}+\hat{x};1} - \gamma_{\alpha,s,\bar{r};2}\delta_{\alpha\alpha'}[s_y]_{ss'}\gamma_{\alpha',s',\bar{r}+\hat{x};2}) \\
& +\frac{\text{Re}\Delta_s(\bar{r}+\frac{\hat{y}}{2})}{2}(\gamma_{\alpha,s,\bar{r};2}\delta_{\alpha\alpha'}[s_y]_{ss'}\gamma_{\alpha',s',\bar{r}+\hat{y};1} + \gamma_{\alpha,s,r;1}\delta_{\alpha\alpha'}[s_y]_{ss'}\gamma_{\alpha',s',r+\hat{y};2}) \\
& -\frac{\text{Im}\Delta_s(\bar{r}+\frac{\hat{y}}{2})}{2}(\gamma_{\alpha,s,\bar{r};1}\delta_{\alpha\alpha'}[s_y]_{ss'}\gamma_{\alpha',s',\bar{r}+\hat{y};1} - \gamma_{\alpha,s,\bar{r};2}\delta_{\alpha\alpha'}[s_y]_{ss'}\gamma_{\alpha',s',\bar{r}+\hat{y};2}) \Big\},
\end{aligned}
\tag{S17}
$$

and

$$
\begin{aligned}
H(k_z = \pi) = \sum_{\bar{r};\alpha\alpha';s,s'} \Big\{ & -i\mu\gamma_{\alpha,s,\bar{r};1}\delta_{\alpha\alpha'}\delta_{ss'}\gamma_{\alpha',s',\bar{r};2} + i(m+2t)\gamma_{\alpha,s,\bar{r};1}[\sigma_z]_{\alpha\alpha'}\delta_{ss'}\gamma_{\alpha',s',\bar{r};2} \\
& +ih_z\gamma_{\alpha,s,\bar{r};1}\delta_{\alpha\alpha'}[s_z]_{ss'}\gamma_{\alpha',s',\bar{r};2} - it(\gamma_{\alpha,s,\bar{r};1}[\sigma_z]_{\alpha\alpha'}\delta_{ss'}\gamma_{\alpha',s',\bar{r}+\hat{x};2} - \gamma_{\alpha,s,\bar{r};2}[\sigma_z]_{\alpha\alpha'}\delta_{ss'}\gamma_{\alpha',s',\bar{r}+\hat{x};1})
\end{aligned}
$$

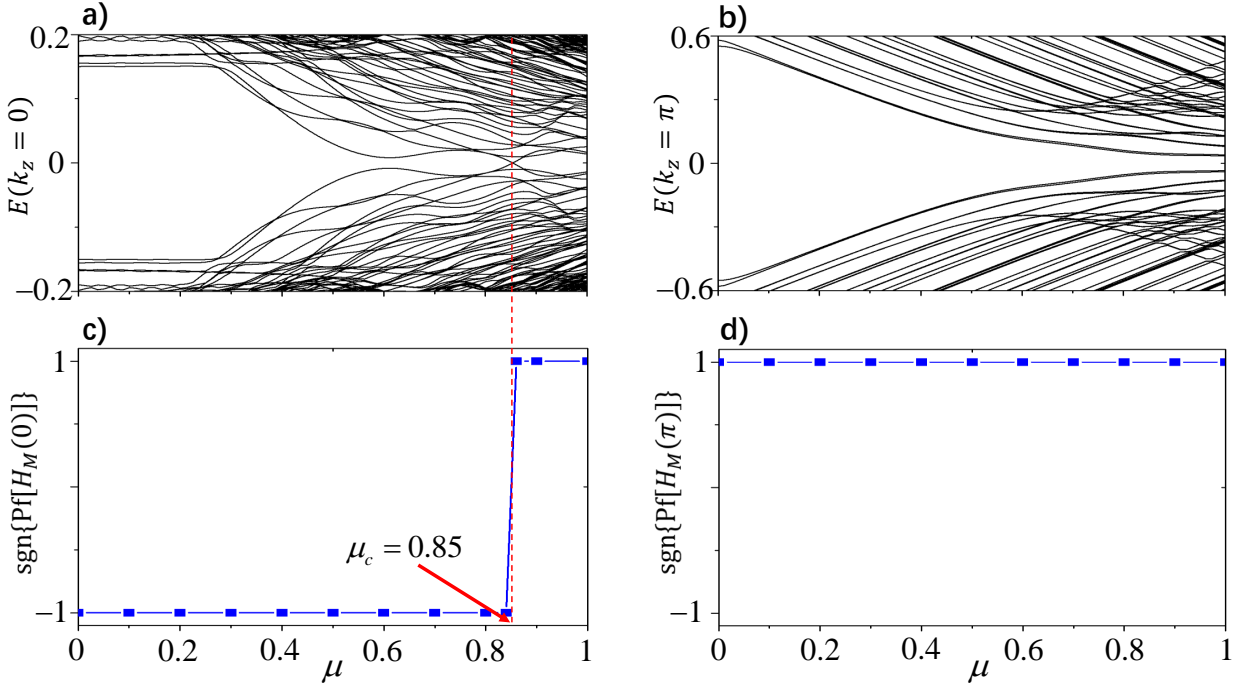

FIG. S2. (Color online) Vortex phase transitions characterized by the $\mathbb{Z}_2$ invariant. (a) The evolution of the energy dispersions at $k_z = 0$ with respect to the doping level $\mu$. A gap closure occurs at $\mu_c = 0.85$ for the parameters considered. (b) The evolution of the energy dispersions at $k_z = \pi$ with respect to the doping level $\mu$. The energy dispersion remains gapped for the range of $\mu$ that we consider here. (c) $\mathrm{sgn}\{\mathrm{Pf}[H_M(0)]\}$ jumps from $-1$ to $1$ when $\mu$ goes across $\mu_c$. (d) $\mathrm{sgn}\{\mathrm{Pf}[H_M(\pi)]\}$ retains the value 1. A combination of (c) and (d) indicates that the $\mathbb{Z}_2$ invariant $\nu$ changes from $-1$ for $\mu < \mu_c$ to $1$ for $\mu > \mu_c$. Common parameters are: $t = 1$, $m_0 = 2.5$, $\lambda_x = 0.25$, $\lambda_y = \lambda_z = 0.5$, $\Delta_0 = -\Delta_s = 0.25$, $\boldsymbol{h} = 0$, $\xi = 4$, and the vortex line is located at the center of the $xy$ plane. The lengths along the $x$ and $y$ directions contain $L_x = L_y = 20$ lattice sites. (a) and (c) share the same horizontal coordinates, and (b) and (d) share the same horizontal coordinates.

$$\begin{aligned}
&-it(\gamma_{\alpha,s,\bar{r};1}[\sigma_z]_{\alpha\alpha'}\delta_{ss'}\gamma_{\alpha',s',\bar{r}+\hat{y};2} - \gamma_{\alpha,s,\bar{r};2}[\sigma_z]_{\alpha\alpha'}\delta_{ss'}\gamma_{\alpha',s',\bar{r}+\hat{y};1}) \\
&-i\lambda(\gamma_{\alpha,s,\bar{r};1}[\sigma_x]_{\alpha\alpha'}[s_x]_{ss'}\gamma_{\alpha',s',\bar{r}+\hat{x};1} + \gamma_{\alpha,s,\bar{r};2}[\sigma_x]_{\alpha\alpha'}[s_x]_{ss'}\gamma_{\alpha',s',\bar{r}+\hat{x};2}) \\
&+\lambda(\gamma_{\alpha,s,\bar{r};1}[\sigma_x]_{\alpha\alpha'}[s_y]_{ss'}\gamma_{\alpha',s',\bar{r}+\hat{y};2} - \gamma_{\alpha,s,\bar{r};2}[\sigma_x]_{\alpha\alpha'}[s_y]_{ss'}\gamma_{\alpha',s',\bar{r}+\hat{y};1}) \\
&+\frac{\mathrm{Re}\Delta_0(\bar{r})}{2}(\gamma_{\alpha,s,\bar{r};2}\delta_{\alpha\alpha'}[s_y]_{ss'}\gamma_{\alpha',s',\bar{r};1} + \gamma_{\alpha,s,r;1}\delta_{\alpha\alpha'}[s_y]_{ss'}\gamma_{\alpha',s',r;2}) \\
&-\frac{\mathrm{Im}\Delta_0(\bar{r})}{2}(\gamma_{\alpha,s,\bar{r};1}\delta_{\alpha\alpha'}[s_y]_{ss'}\gamma_{\alpha',s',\bar{r};1} - \gamma_{\alpha,s,\bar{r};2}\delta_{\alpha\alpha'}[s_y]_{ss'}\gamma_{\alpha',s',\bar{r};2}) \\
&+\frac{\mathrm{Re}\Delta_s(\bar{r}+\frac{\hat{x}}{2})}{2}(\gamma_{\alpha,s,\bar{r};2}\delta_{\alpha\alpha'}[s_y]_{ss'}\gamma_{\alpha',s',\bar{r}+\hat{x};1} + \gamma_{\alpha,s,\bar{r};1}\delta_{\alpha\alpha'}[s_y]_{ss'}\gamma_{\alpha',s',\bar{r}+\hat{x};2}) \\
&-\frac{\mathrm{Im}\Delta_s(\bar{r}+\frac{\hat{x}}{2})}{2}(\gamma_{\alpha,s,\bar{r};1}\delta_{\alpha\alpha'}[s_y]_{ss'}\gamma_{\alpha',s',\bar{r}+\hat{x};1} - \gamma_{\alpha,s,\bar{r};2}\delta_{\alpha\alpha'}[s_y]_{ss'}\gamma_{\alpha',s',\bar{r}+\hat{x};2}) \\
&+\frac{\mathrm{Re}\Delta_s(\bar{r}+\frac{\hat{y}}{2})}{2}(\gamma_{\alpha,s,\bar{r};2}\delta_{\alpha\alpha'}[s_y]_{ss'}\gamma_{\alpha',s',\bar{r}+\hat{y};1} + \gamma_{\alpha,s,r;1}\delta_{\alpha\alpha'}[s_y]_{ss'}\gamma_{\alpha',s',r+\hat{y};2}) \\
&-\frac{\mathrm{Im}\Delta_s(\bar{r}+\frac{\hat{y}}{2})}{2}(\gamma_{\alpha,s,\bar{r};1}\delta_{\alpha\alpha'}[s_y]_{ss'}\gamma_{\alpha',s',\bar{r}+\hat{y};1} - \gamma_{\alpha,s,\bar{r};2}\delta_{\alpha\alpha'}[s_y]_{ss'}\gamma_{\alpha',s',\bar{r}+\hat{y};2})\}.
\end{aligned} \tag{S18}$$

Let us consider a system with open boundary conditions in the both $x$ and $y$ directions where the number of lattice sites is $N_x$ in the $x$ direction and $N_y$ in the $y$ direction. In terms of the Majorana representation

$$\gamma_{k_z=0/\pi} = (\Gamma_{1,1}, \Gamma_{2,1}, ..., \Gamma_{n,m}, ..., \Gamma_{N_x,N_y})^T \tag{S19}$$

with $\Gamma_{n,m} = (\gamma_{a,\uparrow,n\hat{x}+m\hat{y};1}, \gamma_{a,\uparrow,n\hat{x}+m\hat{y};2}, \gamma_{b,\uparrow,n\hat{x}+m\hat{y};1}, \gamma_{b,\uparrow,n\hat{x}+m\hat{y};2}, \gamma_{a,\downarrow,n\hat{x}+m\hat{y};1}, \gamma_{a,\downarrow,n\hat{x}+m\hat{y};2}, \gamma_{b,\downarrow,n\hat{x}+m\hat{y};1}, \gamma_{b,\downarrow,n\hat{x}+m\hat{y};2})$, the Hamiltonian can be written as

$$H(k_z = 0/\pi) = i\gamma_{0/\pi}^\dagger H_\mathrm{M}(k_z = 0/\pi)\gamma_{0/\pi}, \tag{S20}$$

where $H_M$ is an $8N_xN_y \times 8N_xN_y$ real antisymmetric matrix for which the Pfaffian is well defined. In this work, the Pfaffian has been calculated by using the code developed in Ref. [S4].

To show that the formula in Eq. (S12) indeed faithfully characterizes the topological property, here we provide a concrete example with a vortex phase transition and show that the $\mathbb{Z}_2$ invariant $\nu$ predicts this topological phase transition precisely. As is known, a change of the topological invariant is associated with the closure of energy gap. Because here the topological invariant $\nu$ is of $\mathbb{Z}_2$ nature, if the gap closure occurs between two bands with double degeneracy (which is the situation shown in Fig. 3(a) of the main text), then such gap closures do not change $\nu$. As the degeneracy of bands requires some symmetry protection (it is a $C_{4z}$ rotation symmetry in Fig. 3(a) of the main text), in order to avoid such issues, here we consider to lift the $C_{4z}$ rotation symmetry of the Hamiltonian in Eq. (S2) by setting the strengths of the spin-orbit coupling terms to be anisotropic, i.e., $2\lambda \sum_{i=x,y,z} \sin k_i \sigma_x s_i \to 2\sum_{i=x,y,z} \lambda_i \sin k_i \sigma_x s_i$, with $\lambda_x \neq \lambda_y$. To be specific, we also consider that the band inversion occurs at the $k_z = 0$ plane, so that the vortex phase transition is associated with the gap closure at $k_z = 0$. As shown in Fig. S2(a), the energy dispersions at $k_z = 0$ show a gap closure at $\mu_c = 0.85$ for the set of parameters we consider. By contrast, the energy dispersions at $k_z = \pi$ are found to keep gapped for the range of doping level we consider, as shown in Fig. S2(b). A combination of Figs. S2(a) and S2(b) indicates that a vortex phase transition should occur at $\mu_c$. Figs. S2(c) and S2(d) show respectively the corresponding evolution of $\text{sgn}\{\text{Pf}[H_M(0)]\}$ and $\text{sgn}\{\text{Pf}[H_M(\pi)]\}$. One can see that $\text{sgn}\{\text{Pf}[H_M(0)]\}$ jumps from $-1$ to $1$ exactly at $\mu_c$ where the energy dispersions at $k_z = 0$ get closed. Meanwhile, one can see that $\text{sgn}\{\text{Pf}[H_M(\pi)]\}$ keeps its value 1 for the range of $\mu$ considered, agreeing with the fact that the band inversion does not occur at the $k_z = \pi$ plane and the energy dispersions keep gapped as shown in Fig. S2(b). A combination of Figs. S2(c) and S2(d) indicates that the $\mathbb{Z}_2$ invariant $\nu$ jumps from $-1$ for $\mu < \mu_c$ to $1$ for $\mu > \mu_c$, demonstrating that the $\mathbb{Z}_2$ invariant given in Eq. (S12) can precisely predict the vortex phase transition.

## IV. THE EFFECTS OF ZEEMAN FIELD ON THE HELICAL MAJORANA HINGE MODES

In the main article, we showed that in the absence of a Zeeman field, the Majorana hinge modes are helical due to the preservation of time-reversal symmetry. In this section, we show the impact of the Zeeman field on the Majorana hinge modes in the absence of vortex lines in detail. Naively, one may expect that the introduction of Zeeman field would immediately gap the helical Majorana hinge modes due to the breaking of time-reversal symmetry. However, the actual results are in contradiction with this expectation.

While the magnetic field will generate vortices due to the orbital effect, in this section we neglect the generation of vortex lines and focus on the effect from the Zeeman field to the helical Majorana hinge modes. In Fig. S3 and Fig. S4, we show the evolution of the energy spectra with respect to Zeeman field. The underlying cubic geometry takes an

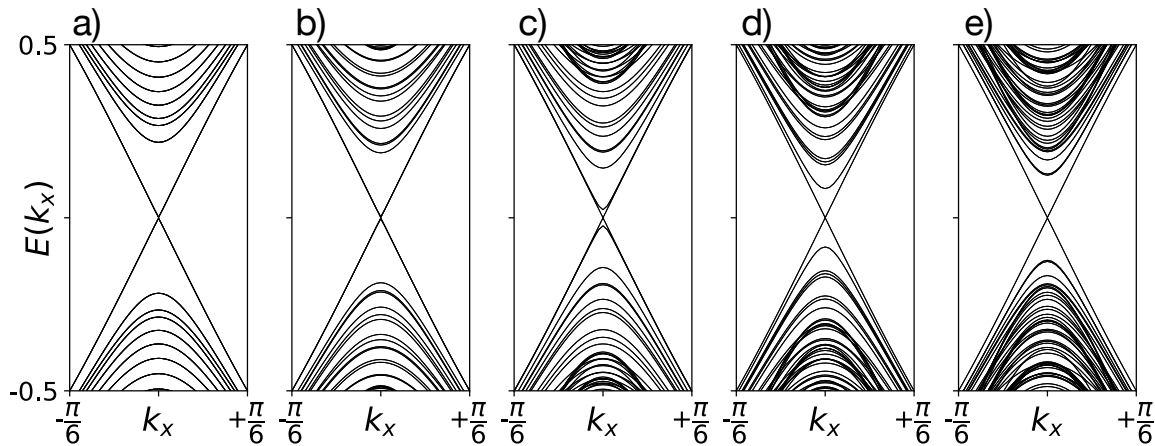

FIG. S3. The effects of Zeeman field (no vortex lines) to the helical Majorana hinge modes. Common parameters are $t = 1$, $m_0 = 2.5$, $\lambda = 0.5$, $\mu = 0$, $\Delta_0 = -\Delta_s = 0.25$, $h_x = h_y = 0$. The lengths along the $y$ and $z$ directions contain $L_y = L_z = 24$ lattice sites. (a) $h_z = 0$, (b) $h_z = 0.1$, (c) $h_z = 0.2$, (d) $h_z = 0.3$, (e) $h_z = 0.4$. In (a) (b), the gapless modes traversing the gap are of four-fold degeneracy (read from data), corresponding to four-pairs of helical Majorana hinge modes. (c) shows the spectra just crossing the transition from a second-order topological superconductor with helical Majorana hinge modes to a second-order topological superconductor with chiral Majorana hinge modes. In (d) (e), the gapless modes traversing the gap are of two-fold degeneracy (read from data), corresponding to four-branches of chiral Majorana hinge modes.

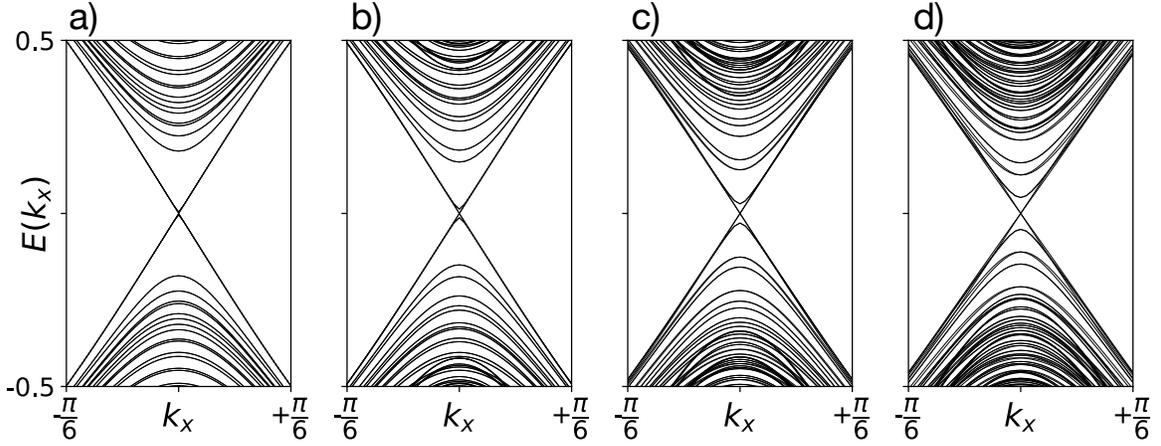

FIG. S4. The effects of Zeeman field (no vortex lines) to the helical Majorana hinge modes. Common parameters are $t = 1$, $m_0 = 2.5$, $\lambda = 0.5$, $\mu = 0$, $\Delta_0 = -\Delta_s = 0.25$. The lengths along the $y$ and $z$ directions contain $L_y = L_z = 24$ lattice sites. $\boldsymbol{h} = (h, h, h)$, with (a) $h = 0.05$, (b) $h = 0.1$, (c) $h = 0.15$, (d) $h = 0.2$. In (a) (b), the gapless modes traversing the gap are of four-fold degeneracy (read from data), corresponding to four-pairs of helical Majorana hinge modes. (c) shows the spectra just crossing the transition from a second-order topological superconductor with helical Majorana hinge modes to a second-order topological superconductor with chiral Majorana hinge modes. In (d), the gapless modes traversing the gap are of two-fold degeneracy (read from data), corresponding to four-branches of chiral Majorana hinge modes.

open boundary conditions in the $y$ and $z$ directions and periodic boundary condition in the $x$ direction. In Fig. S3, the magnetic field is applied in the $z$ direction i.e. $\boldsymbol{h} = (0, 0, h_z)$, and in Fig. S4, the magnetic field is applied in the [111] direction i.e. $\boldsymbol{h} = (h, h, h)$. For comparison, the result for the time-reversal invariant case ($\boldsymbol{h} = 0$) is presented in Fig. S3(a). In Fig. S3(b) and Fig. S4(a), one can see that the Majorana hinge modes remain gapless in a small applied field. Moreover, the degeneracy of the Majorana hinge modes does not change even though the degeneracy of other higher-energy spectra is lifted due to the breaking of time-reversal symmetry. The helical Majorana hinge modes are found to remain intact until the Zeeman field becomes larger than an orientation-dependent critical value. In Fig. S3(c) and Fig. S4(b), the results show that when the Zeeman field becomes a little larger than the critical value, half of the gapless Majorana hinge modes become gapped. We find that this corresponds to a transition from a second-order topological superconductor with helical Majorana hinge modes to a second-order topological superconductor with chiral Majorana hinge modes. With a further increase of Zeeman field, these chiral Majorana hinge modes become more separated from other modes in energy, as shown in Figs. S3(d) and S3(e) and Figs. S4(c) and S4(d).

In the following, we develop an analytic theory to explain the robustness of helical Majorana hinge modes against the Zeeman field. Following the standard procedure, we first perform a lower-energy expansion of the lattice Hamiltonian in Eq. (1) of the main article around the band-inversion momentum. Without loss of generality, we still consider that the band inversion occurs at the $\Gamma$ point. Accordingly, the low-energy continuum Hamiltonian reads

$$\mathcal{H}(\boldsymbol{k}) = (m + tk^2)\sigma_z s_0 \tau_z + 2\lambda \sigma_x \boldsymbol{k} \cdot \boldsymbol{s} \tau_z - \mu \sigma_0 s_0 \tau_z + \sigma_0 \boldsymbol{h} \cdot \boldsymbol{s} \tau_0 - [\Delta_0 + 2\Delta_s - \frac{\Delta_s}{2}(k_x^2 + k_y^2)]\sigma_0 s_0 \tau_x, \quad (S21)$$

where $m = m_0 - 6t$. For simplicity, we focus on $\mu = 0$ in the following and take $t$ and $\lambda$ to be positive. Let us first derive the low-energy Hamiltonian for the surface states on the $z$-normal surfaces. To proceed, let us consider a semi-infinity sample with $0 \le z < +\infty$. The presence of a boundary breaks the translational symmetry in the $z$ direction, so the $k_z$ in Hamiltonian (S21) needs to be replaced by $-i\partial_z$. Following Ref. [S5], we divide the Hamiltonian into two parts, $\mathcal{H} = \mathcal{H}_1 + \mathcal{H}_2$, where

$$\mathcal{H}_1(k_x, k_y, -i\partial_z) = [m + t(k_x^2 + k_y^2) - t\partial_z^2]\sigma_z s_0 \tau_z - 2i\lambda \partial_z \sigma_x s_z \tau_z,$$

$$\mathcal{H}_2(k_x, k_y, -i\partial_z) = 2\lambda \sigma_x (k_x s_x + k_y s_y)\tau_z + \sigma_0 \boldsymbol{h} \cdot \boldsymbol{s} \tau_0 - [\Delta_0 + 2\Delta_s - \frac{\Delta_s}{2}(k_x^2 + k_y^2)]\sigma_0 s_0 \tau_x. \quad (S22)$$

In the division, we have taken $\mathcal{H}_2$ as a perturbation, which is quite accurate when the pairing amplitude and the strength of Zeeman field are small. By solving the eigenvalue equation $\mathcal{H}_1(k_x, k_y, -i\partial_z)\psi_\alpha(z) = E_\alpha \psi_\alpha(z)$ with the boundary condition $\psi_\alpha(0) = \psi_\alpha(+\infty) = 0$, one can find that there exist four zero-energy solutions. The wave functions

$\psi_\alpha(z)$ can be compactly written as

$$\psi_\alpha(z) = \mathcal{N} \sin(\kappa_1 z) e^{-\kappa_2 z} e^{ik_x x} e^{ik_y y} \chi_\alpha, \tag{S23}$$

with normalization given by $|\mathcal{N}|^2 = |4\kappa_2(\kappa_1^2 + \kappa_2^2)/\kappa_1^2|$. The two parameters $\kappa_1$ and $\kappa_2$ are respectively given by $\kappa_1 = \sqrt{\frac{[t(k_x^2+k_y^2)-m]}{t} - \left(\frac{\lambda}{t}\right)^2}$ and $\kappa_2 = \frac{\lambda}{t}$. The most important information is contained in $\chi_\alpha$. Here $\chi_\alpha$ satisfy $\sigma_x s_z \tau_0 \chi_\alpha = -\chi_\alpha$. We can explicitly choose solutions as

$$\begin{aligned}
\chi_1 &= |\sigma_y = -1\rangle \otimes |s_z = 1\rangle \otimes |\tau_z = 1\rangle, \\
\chi_2 &= |\sigma_y = 1\rangle \otimes |s_z = -1\rangle \otimes |\tau_z = 1\rangle, \\
\chi_3 &= |\sigma_y = -1\rangle \otimes |s_z = 1\rangle \otimes |\tau_z = -1\rangle, \\
\chi_4 &= |\sigma_y = 1\rangle \otimes |s_z = -1\rangle \otimes |\tau_z = -1\rangle.
\end{aligned} \tag{S24}$$

By projecting $\mathcal{H}_2$ into the four-dimensional space expanded by $\chi_\alpha$, we obtain the low-energy Hamiltonian for the surface states on the $z = 0$ surface (bottom surface), which reads

$$\mathcal{H}_{\text{eff}}(k_x, k_y) = 2\lambda(k_y s_x - k_x s_y)\tau_z + M_Z s_z + M_S \tau_x, \tag{S25}$$

where $M_Z = h_z$ denotes the Dirac mass induced by the Zeeman field, and $M_S = \Delta_0 + 2\Delta_s$ in the leading order denotes the Dirac mass induced by the superconductivity. It is easy to see that for the Zeeman field, only the component perpendicular to the concerned surface contributes to the Dirac mass. It is worth noting that the two Dirac mass terms commute with each other, and a closure of the surface gap happens at $M_Z = M_S$. That is, when $M_Z$ becomes larger than $M_S$, the nature of the Dirac mass on the surface will change from a superconductivity-dominated one to a Zeeman-field-dominated one.

With Eq. (S25), now we can explain why the helical Majorana hinge modes are stable against the Zeeman field when its strength is below the critical value. To proceed, let us consider that the $y$ direction also becomes open and a domain wall hosting Majorana helical modes is formed at the boundary. Following the same steps, we divide the Hamiltonian (S25) into two parts, $\mathcal{H}_{\text{eff}} = \mathcal{H}_{\text{eff};1} + \mathcal{H}_{\text{eff};2}$, where

$$\begin{aligned}
\mathcal{H}_{\text{eff};1}(k_x, -i\partial_y) &= -2i\lambda \partial_y s_x \tau_z + M_S(y)\tau_x, \\
\mathcal{H}_{\text{eff};2}(k_x, k_y) &= -2\lambda k_x s_y \tau_z + M_Z s_z.
\end{aligned} \tag{S26}$$

As explained previously, such a division is justified when the Zeeman field, so $M_Z$, is small. From our analysis above, one can immediately read from $\mathcal{H}_{\text{eff};1}$ that the space for the low-energy modes is spanned by $\tilde{\chi}_\alpha$ which satisfy either $s_x \tau_y \tilde{\chi}_\alpha = \tilde{\chi}_\alpha$ or $s_x \tau_y \tilde{\chi}_\alpha = -\tilde{\chi}_\alpha$ [the sign depends on the detail of the profile of $M_S(y)$]. That is, the two-dimensional space for the helical Majorana modes is spanned either by $(|s_x = 1\rangle \otimes |\tau_y = 1\rangle, |s_x = -1\rangle \otimes |\tau_y = -1\rangle)$ or by $(|s_x = 1\rangle \otimes |\tau_y = -1\rangle, |s_x = -1\rangle \otimes |\tau_y = 1\rangle)$. Projecting $\mathcal{H}_{\text{eff};2}$ into these two possible two-dimensional spaces, one can find that the low-energy Hamiltonian on the hinge is given by

$$\mathcal{H}_{\text{hinge}}(k_x) = -2\lambda k_y s_y. \tag{S27}$$

One can immediately see that in the weak field regime (the regime for which the perturbative treatment is justified), the presence of Zeeman field does not alter the helical nature, though the time-reversal symmetry is broken. According to Eq. (S25), we know that a closure of the surface gap happens at $M_Z = M_S$. This indicates that when the Dirac mass term induced by the Zeeman field is equal to the one induced by superconductivity, a surface topological phase transition occurs, accompanying a change of the nature of domain walls.

## V. HELICAL AND CHIRAL MAJORANA HINGE MODES FOR A FINITE-SIZE CUBIC SAMPLE IN THE ABSENCE OF VORTEX LINES

In this section, we show the dispersion of Majorana hinge modes for a cubic geometry in the absence of vortex lines. As shown in Figs. S5(a)-S5(c), one can find that for the second-order topological superconducting phase with helical Majorana hinge modes, the eigenenergies of the helical Majorana hinge modes show a four-fold degeneracy (a factor two is from the helical nature, and the other factor two is from the equivalence between the top and bottom surfaces) and a regular jump. Most importantly, there is no zero-energy mode which is not due to finite size effects. Similar results are also found for the second-order topological superconducting phase with chiral Majorana hinge modes, as shown in Fig. S6.

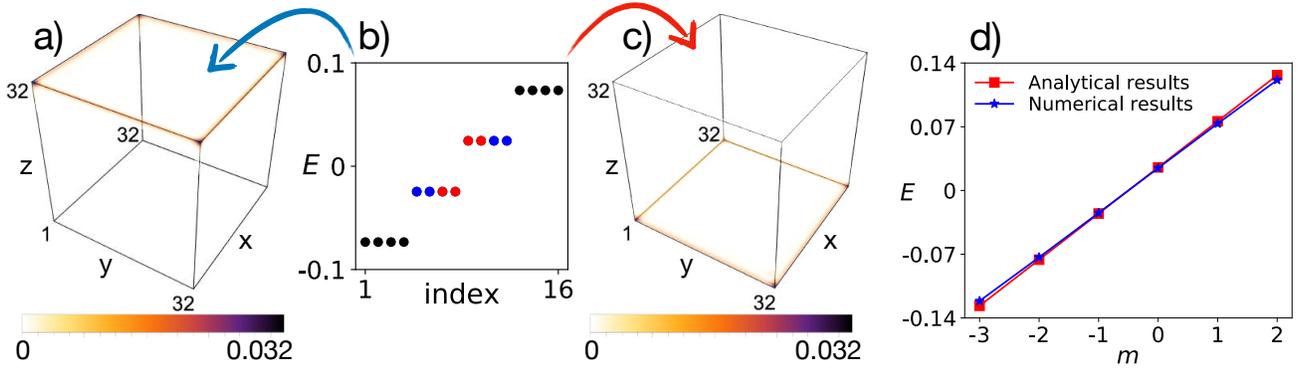

FIG. S5. (Color online) The dispersions of helical Majorana hinge modes for a finite-size system in the absence of vortex lines. Common parameters are $t = 1$, $m_0 = 2.5$, $\lambda = 0.5$, $\mu = 0$, $\Delta_0 = -\Delta_s = 0.25$, and $\boldsymbol{h} = 0$. All directions take open boundary conditions and their lengths contain $L_x = L_y = L_z = 32$ lattice sites. (a) The probability density profile of the hinge states. (b) The 16 energy eigenvalues closest to zero energy. No robust Majorana zero mode exists. (d) Comparison between the analytical (red line) and the numerical (blue line) small eigenvalues. The results show an excellent agreement.

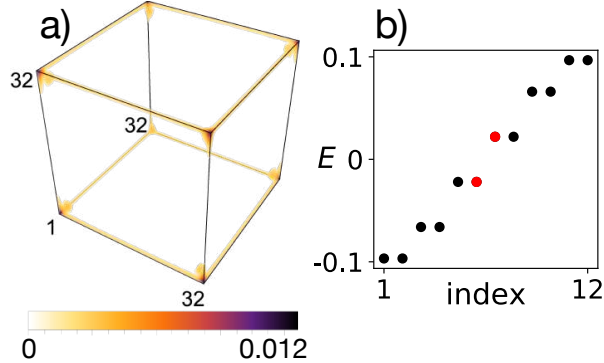

FIG. S6. (Color online) The dispersions of chiral Majorana hinge modes for a finite-size system in the absence of vortex lines. Common parameters are $t = 1$, $m_0 = 2.5$, $\lambda = 0.5$, $\mu = 0$, $\Delta_0 = -\Delta_s = 0.25$, and $\boldsymbol{h} = (h_x, h_y, h_z) = (0, 0, 0.3)$. All directions take open boundary conditions and their lengths contain $L_x = L_y = L_z = 32$ lattice sites. (a) The probability density profile of the hinge states. (b) The 12 energy eigenvalues closest to zero energy. No robust Majorana zero mode exists.

To understand the absence of zero-energy modes (Majorana zero modes), one needs to note that the periodic motion of the helical or chiral Majorana modes along the hinges of the top and bottom $z$-normal surfaces will lead the momentum to be quantized discretely. Accordingly, the dispersion of the helical or chiral Majorana modes will follow $E = \pm v q_m$, with $q_m = (2m + \alpha)\pi/L$, where $L$ denotes the total length of the closed path [here $L = 2(L_x + L_y) - 4$], $m$ is an integer, and $\alpha = 0$ and $1$ refer to periodic and antiperiodic boundary condition, respectively. According to Eq. (S27), we further have $v = 2\lambda$ for the helical case (we do not have an accurate analytical expression of $v$ for the chiral case since the analytical treatment in Sec. IV loses its accuracy when the Zeeman field is strong). Apparently, there exists a big difference between the two kinds of boundary conditions. For the periodical boundary condition, it is readily seen that $q_m$ can take zero value ($q_0 = 0$), so the helical or chiral hinges modes will contain Majorana zero modes. In contrast, for the antiperiodic boundary condition, it is readily seen that $q_m$ cannot take zero value, so there is no Majorana zero mode for the helical or chiral hinge modes when the system size is finite. The absence of Majorana zero modes in Fig. S5(b) and Fig. S6(b) thus indicates that these propagating Majorana modes take an antiperiodic boundary condition. As shown in Fig. S5(d), the analytical results for the helical case, i.e., $E = \pm 2\lambda(2m + 1)\pi/L$, agree very well with the numerical results shown in Fig. S5(b), demonstrating the correctness of our analytical analysis. The antiperiodic boundary condition of the helical or chiral Majorana modes can be understood from the fact that these low-energy Majorana modes originate from the gapless Dirac surface states which has an intrinsic $\pi$ Berry phase.

Interestingly, when a $\pi$-flux vortex line is inserted along the $z$ direction, the closed path that the helical or chiral Majorana modes propagate along will enclose the vortex line, so the helical or chiral Majorana modes will pick up an additional $\pi$ phase from the $\pi$-flux vortex line. Accordingly, the boundary condition of these propagating Majorana

modes will change from an antiperiodic one to a periodic one. Then the helical or chiral Majorana modes will contain Majorana zero modes according to our analysis (see the following section).

## VI. VORTEX LINES FAR AWAY FROM THE HELICAL MAJORANA HINGE MODES

In the main text, based on a local perspective of surface Dirac mass, we have argued that when the vortex line is far away from the hinges, the existence of helical Majorana modes on the hinges should have negligible impact on the vortex line, so the vortex line will host robust Majorana zero modes at its ends in the weakly doped regime. In this section, we provide numerical results to support this argument. In Fig. S7, we show the result for a cubic lattice with size $L_x = L_y = L_z = 32$ lattice sites. A $z$-directional $\pi$-flux vortex line is inserted at the center of the $xy$ plane. According to our analysis in Sec. V, the six modes close to zero energy in Fig. S7(b) correspond to two vortex-end Majorana zero modes and four hinge Majorana zero modes. However, because of the finite-size effect, the modes within the top or bottom surface will hybridize with each other, and the modes of the two surfaces also have finite coupling, so all modes are split away from the zero energy (if the length along the $z$ direction increases to infinity, then because the particle-hole symmetry of the superconductor forces the Majorana zero modes to be created or annihilated in pairs, two of the modes will take exactly zero energy, even though the system size of the $xy$ plane is finite). To show that the increase of the distance between the vortex line and the helical Majorana modes will decrease their hybridization, we further consider a cubic lattice with size $L_x = L_y = 40$ and $L_z = 32$ lattice sites, with the result shown in Fig. S8. Compared to Fig. S7, it is readily seen that the energy of the four red modes becomes more close to zero energy. As these four modes correspond to two hinge modes and two vortex-end modes [see Fig. S7(c)

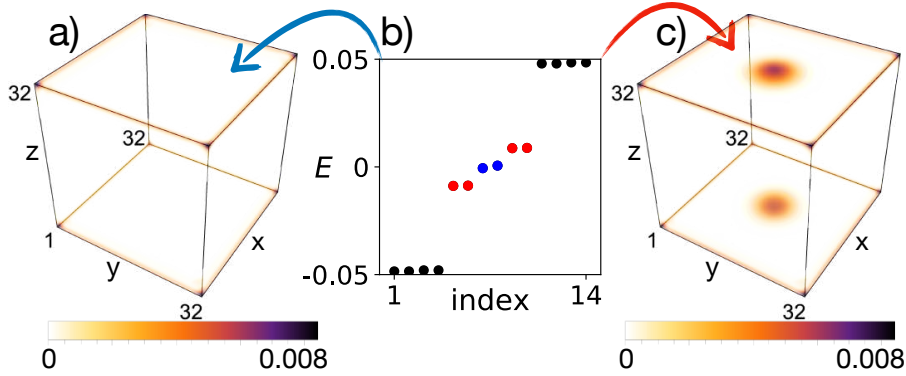

FIG. S7. (Color online) Common parameters are $t = 1$, $m_0 = 2.5$, $\lambda = 0.5$, $\mu = 0$, $\xi = 4$, $\boldsymbol{h} = 0$, and $\Delta_0 = -\Delta_s = 0.25$. All directions take open boundary conditions and their lengths contain $L_x = L_y = L_z = 32$ lattice sites. The vortex line is at the center of $xy$-plane.

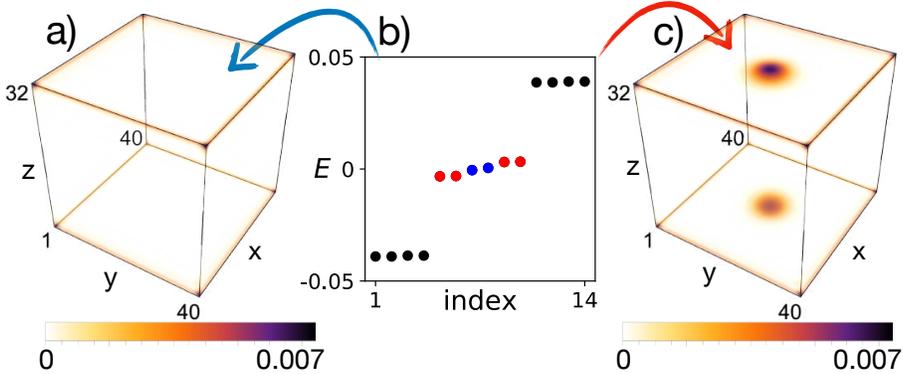

FIG. S8. (Color online) Common parameters are $t = 1$, $m_0 = 2.5$, $\lambda = 0.5$, $\mu = 0$, $\xi = 4$, $\boldsymbol{h} = 0$, and $\Delta_0 = -\Delta_s = 0.25$. All directions take open boundary conditions and their lengths contain $L_x = L_y = 40$, and $L_z = 32$ lattice sites. The vortex line is at the center of $xy$ plane and in the $z$ direction.

and Fig. S8(c)], the decrease of their energy splitting indicates that the hybridization of the vortex-end and hinge Majorana zero modes decreases with the increase of the distance between the vortex line and the hinges. Therefore, it is justified to expect that for vortex lines far away from the hinges, the vortex-end bound states will take exactly zero energy, restoring their exact self-conjugate ($\gamma = \gamma^\dagger$) nature.

## VII. VORTEX LINES CLOSE TO THE HINGES WITHOUT GAPLESS HINGE MODES

In the main text, we have shown when a topological vortex line is moved close to the hinges with helical Majorana modes, their hybridization will trivialize the vortex line. For comparison, in this section we consider that the vortex line is also moved close to the hinges, but the hinges do not host helical Majorana modes. Similar to Fig. 3(e) in the main text, we find that the accidental double degeneracy of the lowest-energy spectrum exhibited in the spectrum of Fig. 3(a) in the main text is lifted when the vortex line is no longer located at the center of the system since the $C_{4z}$ rotation symmetry is broken, as shown in Fig. S9(a). Nevertheless, this lifting of degeneracy does not affect the vortex line topology in the weakly doped regime since the dispersion of the vortex line remains gapped. As expected, in the weakly doped regime, the two lowest-energy modes are strongly localized at the vortex ends, as shown in Figs. S9(b)-S9(d). These results demonstrate that the trivialization of the vortex line for the case with helical Majorana modes does indeed originate from their hybridization.

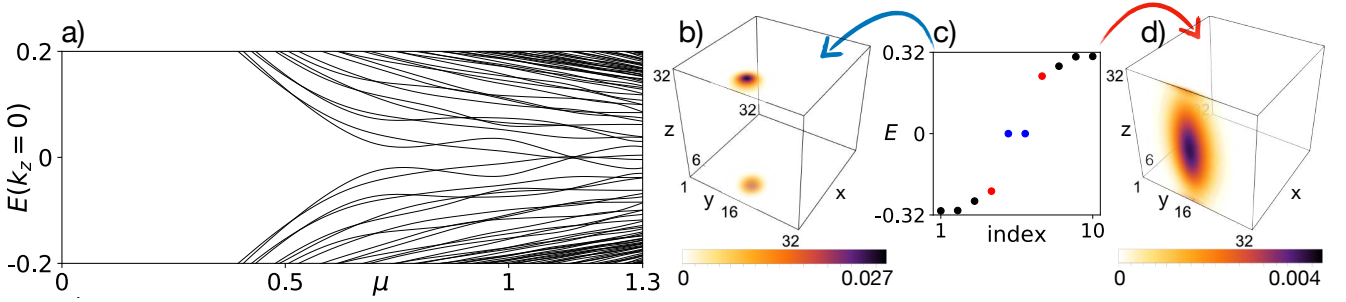

FIG. S9. (Color online) Common parameters are $t = 1$, $m_0 = 2.5$, $\lambda = 0.5$, $\mu = 0$, $\xi = 4$, $\bm{h} = 0$, and $\Delta_0 = \Delta_s = 0.25$. All directions take open boundary conditions and their lengths contain $L_x = L_y = L_z = 32$ lattice sites. The vortex line is at $(x_c, y_c) = (6.01, 16.01)$ point and in the $z$ direction.

## VIII. BOUNDARY FERMI SURFACE AND VORTEX BOUND STATES

In Sec. IV, we have shown that the low-energy effective Hamiltonian on the bottom surface reads (see Eq.(S25))

$$\mathcal{H}_{\text{eff}}(k_x, k_y) = 2\lambda(k_y s_x - k_x s_y)\tau_z + M_Z s_z + M_S \tau_x, \tag{S28}$$

where $M_S = \Delta_0 + 2\Delta_s$ and $M_Z = h_z$. Such a form is derived for the special case with $\mu = 0$. When $\mu \neq 0$, the low-energy effective Hamiltonian becomes

$$\mathcal{H}_{\text{eff}}(k_x, k_y) = 2\lambda(k_y s_x - k_x s_y)\tau_z - \mu\tau_z + M_Z s_z + M_S \tau_x. \tag{S29}$$

Accordingly, the energy spectra are

$$E(k_x, k_y) = \pm\sqrt{4\lambda^2(k_x^2 + k_y^2) + \mu^2 + M_Z^2 + M_S^2 \pm 2\sqrt{\mu^2[4\lambda^2(k_x^2 + k_y^2) + M_Z^2] + M_Z^2 M_S^2}}, \tag{S30}$$

whose gap gets closed at $k_x = k_y = 0$ only when $M_Z^2 = \mu^2 + M_S^2$. As long as $\mu < \sqrt{M_Z^2 - M_S^2}$ (since here only the absolute values of $\mu$ and $h_z$ are relevant, we consider both of them to be positive in this section for the convenience of discussion), no matter whether the superconducting pairing is purely on-site ($\Delta_s = 0$) or extensive ($\Delta_s \neq 0$), the system corresponds to a second-order topological superconductor with chiral Majorana hinge states circling around the top and bottom surfaces because the nature of Dirac masses on these two surfaces is different from that of the side surfaces.

In Fig. 4 of the main text, we have shown the absence of vortex-end Majorana zero modes when the superconductor falls in the chiral second-order superconducting phase. From a surface viewpoint, this result can be simply understood as a consequence of the absence of a boundary Fermi surface in the normal state. The underlying reason is that the vortex-core MZMs must originate from certain low-energy modes, so the absence of a boundary Fermi surface naturally implies the absence of vortex-core MZMs. As the condition for the existence of a boundary Fermi surface before the introduction of superconductivity is $\mu > h_z$, whereas the chiral second-order topological superconducting phase requires $\mu < \sqrt{M_Z^2 - M_S^2} < h_z$, their incompatibility indicates that vortex-end Majorana zero modes will always be absent on the top and bottom surfaces when the system is in the chiral second-order topological superconducting phase.

Focusing on the normal state, in effect, the Zeeman term not only opens a gap of the size $2h_z$ on the top and bottom surfaces, but also reduces the bulk gap. This can be easily inferred from the normal-state Hamiltonian which reads

$$H_N(\boldsymbol{k}) = (m_0 - 2t\sum_{i=x,y,z}\cos k_i)\sigma_z s_0 + 2\lambda\sum_{i=x,y,z}\sin k_i \sigma_x s_i + h_z \sigma_0 s_z, \tag{S31}$$

whose energy spectra are given by

$$E(\boldsymbol{k}) = \pm\sqrt{4\lambda^2(\sin^2 k_x + \sin^2 k_y) + \left(h_z \pm \sqrt{4\lambda^2 \sin^2 k_z + (m_0 - 2t\sum_{i=x,y,z} t\cos k_i)^2}\right)^2}. \tag{S32}$$

According to the above equation, one can easily find that the bulk gap will reduce from $E_g$ (the zero-field value) to $E_g - 2h_z$ when $h_z < E_g/2$. When $h_z > E_g/2$, the normal state will become a Weyl semimetal. Importantly, the increase of surface gap and the decrease of bulk gap imply that bulk and boundary Fermi surfaces will coexist when $h_z > E_g/4$ and $h_z < \mu < E_u$, where $E_u$ denotes the upper limit beyond which the surface states do not exist on the top and bottom surfaces. Therefore, in the regime $h_z > E_g/4$, the low-energy surface Hamiltonian only cannot fully describe the low-energy physics. In the following, we restrict to the regime $h_z < E_g/4$, so that the low-energy surface Hamiltonian can faithfully describe the low-energy physics when the chemical potential only crosses the surface bands.

We first consider that the superconducting pair contains both on-site and extended $s$-wave components, and the configuration of the band inversion surface and the pairing node surface realizes a helical (time-reversal invariant) second-order topological superconductor at $h_z = 0$ and $\mu = 0$. Next, we consider $h_z > |\Delta_0 + 2\Delta_s|$, so that the system realizes a chiral second-order topological superconductor at $\mu = 0$. In Fig. S10(a), the numerical result shows that increasing $\mu$ will induce a gap closure on the top and bottom surfaces when $\mu = \sqrt{h_z^2 - (\Delta_0 + 2\Delta_s)^2}$ (see the dashed purple line). The gap closure signals a change of the topological property on the boundary. When $\mu < \sqrt{h_z^2 - (\Delta_0 + 2\Delta_s)^2}$, robust chiral Majorana modes exist on the hinges, as shown in Fig. S10(b). When $\mu > \sqrt{h_z^2 - (\Delta_0 + 2\Delta_s)^2}$, we find that the hinges previously harboring one chiral Majorana modes now harbor two counter-propagating Majorana modes, as shown in Fig. S10(c). The existence of two counter-propagating Majorana hinge modes in this regime is similar to the helical case at the limit $h_z = 0$. However, because the time-reversal symmetry is broken by a finite $h_z$, the counter-propagating Majorana modes on each hinge are no longer protected by time-reversal symmetry but by certain additional symmetry on the hinge. After inserting a $\pi$-flux vortex line along the $z$ direction, we diagonalize the Hamiltonian under the cubic geometry with lattice sizes $L_x = L_y = L_z = 32$. The numerical results presented in Figs. S10(d)(e) suggest that the lowest-energy modes are located at the hinges. Within the energy window shown in Fig. S10(d), we do not find the signature of vortex bound states. There are two possible reasons for this result. The first possibility is that the vortex line itself is trivial so there is no Majorana zero mode at the vortex-line ends. The second possibility is that the vortex line is topological and harbors Majorana zero modes at its ends, but the hybridization between the vortex-end Majorana zero modes and the low-energy modes on the hinge shift the energies of the vortex-end Majorana zero modes to finite values, similar to the helical case. It is worth noting that the hybridization is caused by the overlap of wave functions in real space, so in principle one can simply distinguish these two possibilities by considering a larger system in which the vortex line is sufficiently far away from the hinges. However, the smallness of the surface gap in Fig. S10(c) implies that the wave functions of hinge modes have a rather long localization length. Under such a condition, a very large system size is required to reduce the hybridization between the vortex-end Majorana zero modes and the hinge modes to a negligible level. To avoid diagonalizing a Hamiltonian with very large size, here we consider another path to distinguish the two possibilities. That is, we further consider the case with pure on-site $s$-wave pairing for comparison.

In order to have a direct comparison, we consider that the value of $\Delta_0$ for the case with pure on-site $s$-wave pairing is equal to the value of $|\Delta_0 + 2\Delta_s|$ for the case with both on-site and extended $s$-wave pairings, so that the superconductivity-induced Dirac masses for these two cases are equivalent. Because of this equivalence, one

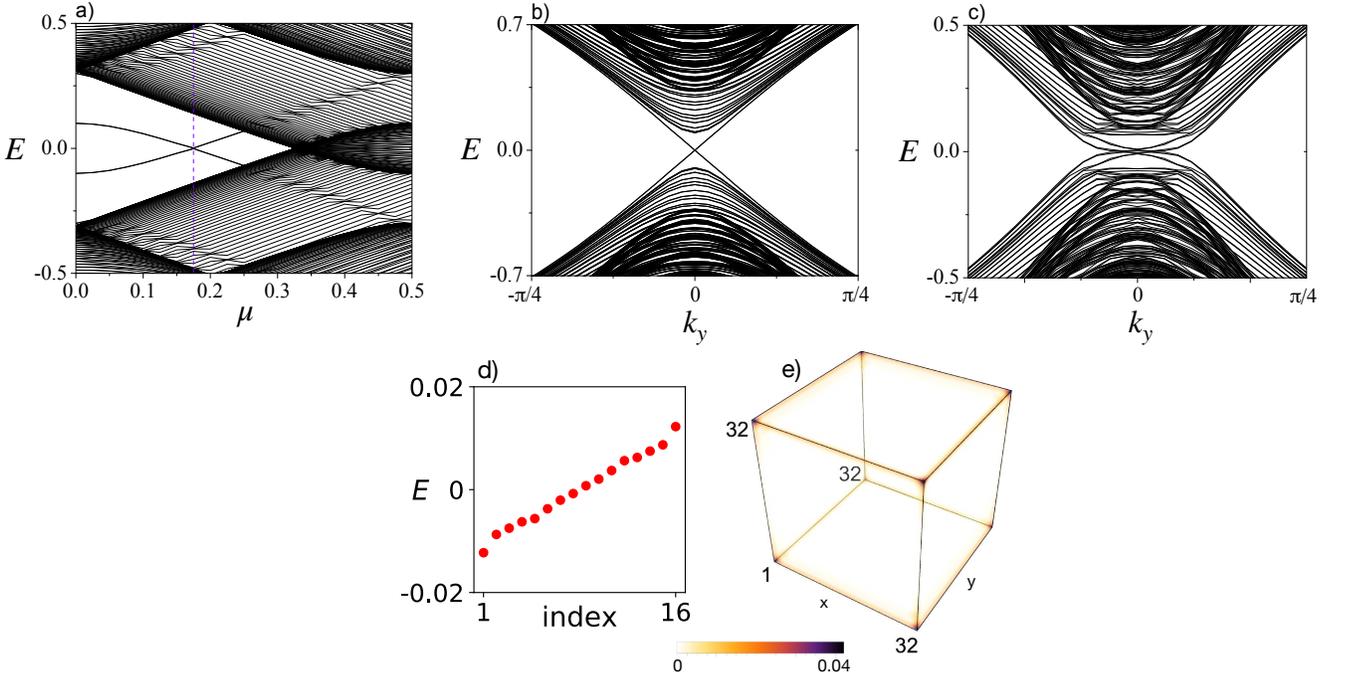

FIG. S10. (Color online) Common parameters are $t = 1$, $m_0 = 2.5$, $\lambda = 0.5$, $\Delta_0 = -\Delta_s = 0.1$, and $\boldsymbol{h} = (h_x, h_y, h_z) = (0, 0, 0.2)$. (a) Energy spectra for a geometry with open boundary directions in the $z$ direction ($L_z = 200$) and periodic boundary conditions in the $x$ and $y$ directions. Only the energy spectra at $(k_x, k_y) = (0,0)$ are shown. The mid-gap spectra correspond to surface states on the $z$-normal surfaces, which show a gap closure at $\mu_c = \sqrt{h_z^2 - (\Delta_0 + 2\Delta_s)^2} = 0.173$. (b)(c) Energy spectra for a geometry with open boundary conditions in the $x$ and $z$ directions ($L_x = L_z = 30$) and periodic boundary conditions in the $y$ direction. $\mu = 0$ for (b) and $\mu = 0.25$ for (c). In (b), the mid-gap spectra have two-fold degeneracy, which indicates that each of the $y$-directional hinges harbors one chiral Majorana mode. In (c), there are eight mid-gap spectra, which indicates that each of the $y$-directional hinges harbors two counter-propagating gapless Majorana modes. (d) The lowest 16 eigen-energies for a geometry with open boundary conditions in all directions. The $z$-directional vortex line is located at the center of the $xy$ plane. (e) The total probability density profiles of the lowest 16 eigen-states shown in (d). There is no signature for the existence of vortex bound states in the given energy window. In (d)(e), $\mu = 0.25$, $\xi = 4$ and $L_x = L_y = L_z = 32$.

can find that the results shown in Figs. S11(a)(b) are almost the same as those in Figs. S10(a)(b). However, when $\mu > \sqrt{h_z^2 - \Delta_0^2}$, the difference emerges. In Fig. S11(c), we show the numerical result for $\mu > h_z$. One can see that there are no gapless Majorana modes on the hinges. Because of the absence of low-energy Majorana modes on the hinges, the hybridization between low-energy vortex bound states and low-energy Majorana hinge modes can be avoided. After the insertion of a $\pi$-flux vortex line, we also diagonalize the Hamiltonian under the cubic geometry with lattice sizes $L_x = L_y = L_z = 32$. The results shown in Figs. S11(d)(e) now clearly demonstrate the appearance of vortex-end Majorana zero modes. As here the only difference is the absence of counter-propagating gapless Majorana hinge modes, we reach the conclusion that the absence of low-energy vortex bound states within the given energy window in Fig. S10(d) originates from the hybridization between the vortex bound states and the counter-propagating gapless Majorana hinge modes.

Now we can establish a complete picture for the interplay of second-order topology and vortex-line topology from a surface perspective. For the low-energy surface Hamiltonian given in Eq. (S29), it is known that it can be mapped to a form describing a chiral $p$-wave superconductor by some basis transformations. When $\mu < h_z$, the boundary Fermi surface is absent and the effective chiral $p$-wave superconductor falls into the so-called strong-pairing phase [S6]. According to the physics in chiral $p$-wave superconductors, it is known that the vortex core does not harbor any Majorana zero mode in the strong-pairing regime [S6]. As here the chiral second-order topological superconducting phase always has $\mu < h_z$, it indicates that vortex-end Majorana zero modes are always absent when the system is in the the chiral second-order topological superconducting phase. In the regime $h_z < E_g/4$ and the chemical potential only crosses the surface bands, the low-energy surface Hamiltonian description is valid, and the effective surface chiral $p$-wave superconductor falls into the weak-pairing regime [S6]. Accordingly, the vortex will bind one robust Majorana zero modes at its core. However, if there exist counter-propagating gapless Majorana modes on the hinges and the vortex is generated to be close to the hinges, then the hybridization between the vortex-core Majorana zero modes

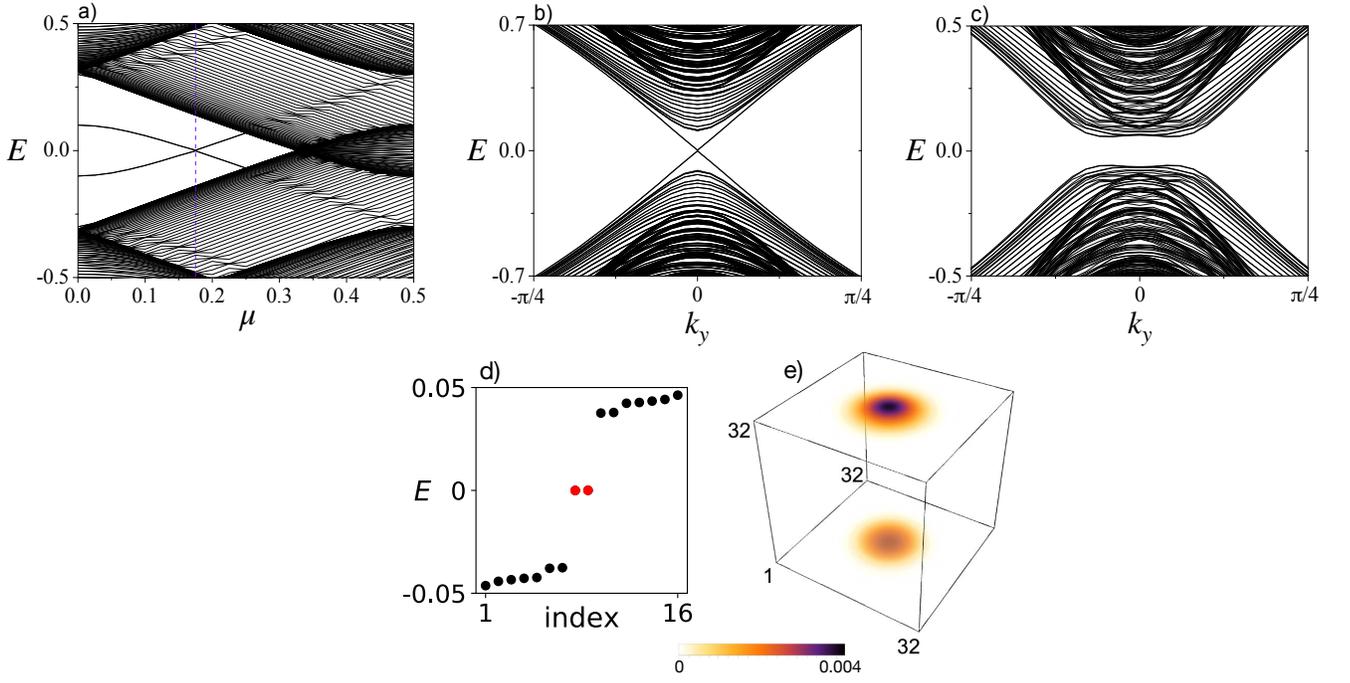

FIG. S11. (Color online) Common parameters are $t = 1$, $m_0 = 2.5$, $\lambda = 0.5$, $\Delta_0 = 0.1$, $\Delta_s = 0$, and $\boldsymbol{h} = (h_x, h_y, h_z) = (0, 0, 0.2)$. (a) Energy spectra for a geometry with open boundary directions in the $z$ direction ($L_z = 200$) and periodic boundary conditions in the $x$ and $y$ directions. Only the energy spectra at $(k_x, k_y) = (0, 0)$ are shown. The mid-gap spectra correspond to surface states on the $z$-normal surfaces, which show a gap closure at $\mu_c = \sqrt{h_z^2 - (\Delta_0 + 2\Delta_s)^2} = 0.173$. (b)(c) Energy spectra for a geometry with open boundary conditions in the $x$ and $z$ directions ($L_x = L_z = 30$) and periodic boundary conditions in the $y$ direction. $\mu = 0$ for (b) and $\mu = 0.25$ for (c). In (b), the mid-gap spectra have two-fold degeneracy, which suggests that each of the $y$-directional hinges harbors one chiral Majorana mode. The result in (c) shows the absence of gapless Majorana modes. (d) The lowest 16 eigen-energies for a geometry with open boundary conditions in all directions. The $z$-directional vortex line is located at the center of the $xy$ plane. (e) The probability density profiles of the lowest two eigen-states shown in (d) (labeled by red dots). The results in (d) and (e) indicate the existence of Majorana zero modes at the vortex ends. In (d)(e), $\mu = 0.25$, $\xi = 4$ and $L_x = L_y = L_z = 32$.

and the counter-propagating gapless Majorana hinge modes will split the vortex-core Majorana zero modes and lead to the absence of robust Majorana zero modes at the vortex core.

## IX. VORTEX LINES IN THE $x$ DIRECTION

In the main article, we have restricted ourselves to vortex lines generated in the $z$ direction in order to be directly comparable with the experiments. Here we provide the results for vortex lines generated in the $x$ direction for completeness. It is worth noting that because the results for vortex lines generated in the $y$ direction are the same due to the $C_{4z}$ rotation symmetry, we will not show them to avoid repetition.

As the vortex line is forced to be trivial when the Zeeman field dominates over the superconductivity, here we focus on the case with vanishing Zeeman field. Similar to the $z$-directional vortex line, a local surface perspective suggests that an $x$-directional vortex line will harbor Majorana zero modes at its ends when its position is far away from the hinges and the doping level is low. Nevertheless, there exists a qualitative difference between the $x$-directional and $z$-directional vortex lines. The difference is that for an $x$-directional vortex line, the closed path that the helical Majorana modes propagate along does not enclose the vortex line. As a result, the helical Majorana modes will not pick up an additional $\pi$ phase from the vortex line after going around the hinges for one circle. In other words, the helical Majorana modes will keep its antiperiodic boundary condition when the vortex line is generated along the $x$ direction, and so the helical Majorana modes will not contain zero-energy modes. The absence of Majorana zero modes on the hinges indicates that the vortex-end Majorana zero modes cannot directly couple with the helical Majorana hinge modes due to the constraint from the intrinsic particle-hole symmetry of superconductors (a Majorana zero mode can get split only when it couples with another Majorana zero mode). Anyway, because the helical Majorana

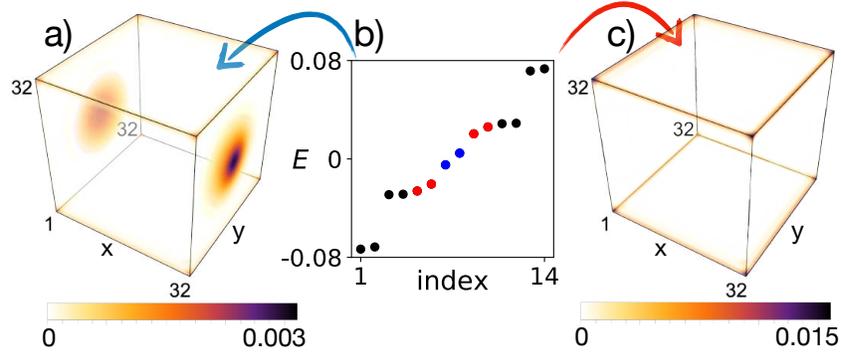

FIG. S12. (Color online) Common parameters are $t = 1$, $m_0 = 2.5$, $\lambda = 0.5$, $\mu = 0$, $\xi = 4$, $\Delta_0 = -\Delta_s = 0.25$, and $\boldsymbol{h} = 0$. All directions take open boundary conditions and their lengths contain $L_x = L_y = L_z = 32$ lattice sites. (a) The wave functions of the two lowest-energy modes [the two blue dots shown in (b)] are mostly localized at the vortex ends. (b) The 14 energy eigenvalues closest to zero energy. (c) The wave functions of the four finite-energy modes in red color are localized along the hinges, indicating that they correspond to the helical Majorana modes.

modes provide some gapless channels which connect the two $x$-normal surfaces, they will enhance the coupling of the two Majorana zero modes at the vortex ends when the vortex line is moved close to the hinges with helical Majorana modes. This enhancement of coupling can also split the vortex-end Majorana zero modes, but now there will be no robust Majorana zero mode on the hinges.

In Fig. S12, we show the results for a geometry with size $L_x = L_y = L_z = 32$ lattice sites. Fig. S12(a) shows that the two lowest-energy modes have most weight at the vortex ends, suggesting the vortex line to be topological. Compared to Fig. S7(b) or Fig. S8(b), the pattern of the near-zero-energy modes in Fig. S12(b) indicates that the helical Majorana modes keep the antiperiodic boundary condition when the vortex line is generated in the $x$ direction.

* yanzhb5@mail.sysu.edu.cn
[S1] Andreas P. Schnyder, Shinsei Ryu, Akira Furusaki, and Andreas W. W. Ludwig, "Classification of topological insulators and superconductors in three spatial dimensions," Phys. Rev. B **78**, 195125 (2008).
[S2] Alexei Kitaev, "Periodic table for topological insulators and superconductors," in *AIP Conference Proceedings*, Vol. 1134 (AIP, 2009) pp. 22–30.
[S3] Alexei Kitaev, "Unpaired Majorana fermions in quantum wires," Physics-Uspekhi **44**, 131 (2001).
[S4] M. Wimmer, "Algorithm 923: Efficient numerical computation of the Pfaffian for dense and banded skew-symmetric matrices," ACM Trans. Math. Softw. **38**, 30:1–30:17 (2012).
[S5] Zhongbo Yan, Fei Song, and Zhong Wang, "Majorana corner modes in a high-temperature platform," Phys. Rev. Lett. **121**, 096803 (2018).
[S6] N. Read and Dmitry Green, "Paired states of fermions in two dimensions with breaking of parity and time-reversal symmetries and the fractional quantum Hall effect," Phys. Rev. B **61**, 10267–10297 (2000).


%========================
\end{document}